\documentclass[11pt]{article}

\title{\vspace{-0.4in}Diagnostics for Respondent-driven Sampling}
\author{Krista J. Gile\footnote{Corresponding Author, Department of Mathematics and Statistics, 
Lederle Graduate Research Tower, Box 34515 University of Massachusetts Amherst,
Amherst, MA 01003-9305, USA, gile@math.umass.edu.}
\\Lisa G. Johnston\footnote{Tulane University School of Public Health \& Tropical Medicine, 1440 Canal Street New Orleans, LA 70112, USA and University of California, San Francisco, Global Health Science, 50 Beale Street, Suite 1200, San Francisco, CA 94105, USA.}
\\Matthew J. Salganik\footnote{Department of Sociology and Office of Population Research, Princeton University, 145 Wallace Hall, Princeton, NJ 08544, USA.}\\Authorship alphabetical; all authors contributed equally to the paper.}

\usepackage{fancyhdr}
\usepackage[compact]{titlesec}  
\titlespacing{\section}{0pt}{*1}{*0}
\titlespacing{\subsection}{0pt}{*1}{*0}
\usepackage{mdwlist}

\usepackage{amssymb,amsmath,amsthm}
\usepackage{graphicx}
\usepackage{mathrsfs} 
\usepackage{natbib}
\usepackage{amsmath}
\usepackage{amssymb}
\usepackage[all]{xy}
\usepackage{color, graphics}
\usepackage{url, multicol}
\usepackage{graphicx}
\usepackage{subfigure}
\usepackage{mathrsfs}
\usepackage{multicol}
\usepackage{threeparttable}
\usepackage{comment}
\usepackage{multirow}
\usepackage{fmtcount}
\usepackage[small]{caption}
\usepackage{lineno} 
\usepackage{booktabs} 

\newcommand{\bi}{\begin{itemize}}
\newcommand{\ei}{\end{itemize}}
\definecolor{Emphcolor}{cmyk}{0,0.89,0.94,0.1}
\definecolor{Netcolor}{rgb}{.8,0,.9}
\definecolor{Diseasecolor}{rgb}{1,.8,.2}
\definecolor{Sampcolor}{rgb}{0,.9,.3}
\definecolor{Black}{rgb}{0,0,0}
\definecolor{Red}{rgb}{1,0,0}
\definecolor{Blue}{rgb}{0,0,1}
\definecolor{Gray}{gray}{.6}

\theoremstyle{plain}

\newcounter{saveenum}

\theoremstyle{definition}

\usepackage{graphicx}
\usepackage{float}
\usepackage{rotating}
\usepackage{natbib}
\usepackage{color}
\usepackage{hypernat}
\usepackage{pdflscape}

\usepackage{graphics}
\usepackage{subfigure}
\usepackage[all]{xy}
\usepackage{amsmath}
\usepackage{amssymb}
\usepackage[all]{xy}
\usepackage{mathrsfs}
\usepackage{multicol}
\usepackage{enumerate}

\definecolor{Emphcolor}{rgb}{.1,.1,.5}
\definecolor{Red}{rgb}{.9,0,.1}
\definecolor{Blue}{rgb}{.1,.1,.5}

\newcommand{\bea}{\begin{eqnarray}}
\newcommand{\eea}{\end{eqnarray}}

\usepackage{ifthen}
\newboolean{draft}
\setboolean{draft}{false}
\newcommand{\knote}[1]{\ifthenelse{\boolean{draft}}{{\bf knote:~}{\it
#1}\relax}{}}
\newcommand{\lisa}[1]{\ifthenelse{\boolean{draft}}{{\bf LISA:~}{\it
#1}\relax}{}}

\newcommand{\bq}{\begin{enumerate}[(A)]
\setcounter{enumi}{\value{saveenum}}
 \setlength{\itemsep}{1pt}
  \setlength{\parskip}{0pt}
  \setlength{\parsep}{0pt}
}

\newcommand{\eq}{\setcounter{saveenum}{\value{enumi}}
\end{enumerate}}

\newenvironment{my_enumerate}{
\begin{enumerate}
  \setlength{\itemsep}{1pt}
  \setlength{\parskip}{0pt}
  \setlength{\parsep}{0pt}}{\end{enumerate}
}

\begin{document}
\maketitle
\thispagestyle{empty}




{\bf Summary:}
Respondent-driven sampling (RDS) is a widely used method for sampling from hard-to-reach human populations, especially groups most at-risk for HIV/AIDS.  Data are collected through a peer-referral process in which current sample members harness existing social networks to recruit additional sample members.  RDS has proven to be a practical method of data collection in many difficult settings and has been adopted by leading public health organizations around the world.  Unfortunately, inference from RDS data requires many strong assumptions because the sampling design is not fully known and is partially beyond the control of the researcher.  In this paper, we introduce diagnostic tools for most of the assumptions underlying RDS inference.  We also apply these diagnostics in a case study of 12 populations at increased risk for HIV/AIDS.  We developed these diagnostics to enable RDS researchers to better understand their data and to encourage future statistical research on RDS.

\vspace{.2in}
{\bf Keywords:} {diagnostics, exploratory data analysis, hard-to-reach populations, HIV/AIDS, link-tracing sampling, non-ignorable design, respondent-driven sampling, social networks, survey sampling\\}

{\bf Acknowledgments:} {We are grateful to Tessie Caballero Vaillant, El Consejo Presidencial del SIDA, Dominican Republic, for allowing us to use these data.  We would also like to thank Maritza Molina Ach\'{e}car, Juan Jose Polanco and Sonia Baez of El Centro de Estudios Sociales y Demograficos, Dominican Republic, for overseeing data collection, and Chang Chung, Sharad Goel, Mark Handcock, Doug Heckathorn, Martin Klein, Dhwani Shah, and Cyprian Wejnert for helpful discussions.  Finally, we would like to thank all those who participated in these studies.  Research reported in this publication was supported by grants from the NIH/NICHD (R01-062366) and the NSF (CNS-0905086). The content is solely the responsibility of the authors.}

\newpage
\setcounter{page}{1}
\pagestyle{plain}

\section{Introduction}

Many problems in social science, public health, and public policy require detailed information about ``hidden'' or ``hard-to-reach'' populations.  For example, efforts to understand and control the HIV/AIDS epidemic require information about the disease prevalence and risk behaviors in the groups most at-risk for the disease: female sex workers (FSW), illicit drug users (DU), and men who have sex with men (MSM)~\citep{magnani_review_2005}.  Respondent-driven sampling (RDS) is a recently introduced link-tracking network sampling technique for collecting such information~\citep{heckathorn_respondent-driven_1997}.  Because of the pressing need for information about the most at-risk groups and the weaknesses of alternatives approaches, RDS has already been used in more than 120 HIV-related studies in 20 countries~\citep{malekinejad_using_2008} and has been adopted by leading public health organizations, such as the US Centers for Disease Control and Prevention (CDC)~\citep{lansky_developing_2007, barbosa_junior_transfer_2011, montealegre_respondent-driven_2012}.  

Collectively, these previous studies demonstrate that RDS is able to generate large samples in a wide variety of otherwise hard-to-reach population.  However, the quality of estimates derived from these data has been challenged in a number of recent papers~\citep{heimer_critical_2005, scott_they_2008, poon_parsing_2009, bengtsson_global_2010, goel_assessing_2010, gile_respondent-driven_2010, mccreesh_evaluation_2012, salganik_commentary:_2012, burt_evaluating_2012}.  A major source of concern is that inference from RDS data requires many assumptions, some of which are widely believed to be incorrect.  Unfortunately, these assumptions are seldom examined in practice. The widespread use of RDS for important public health problems combined with its reliance on untested assumptions, creates a pressing need for exploratory and diagnostic techniques for RDS data.

\begin{figure}
  \centering
   \includegraphics[width=0.9\textwidth]{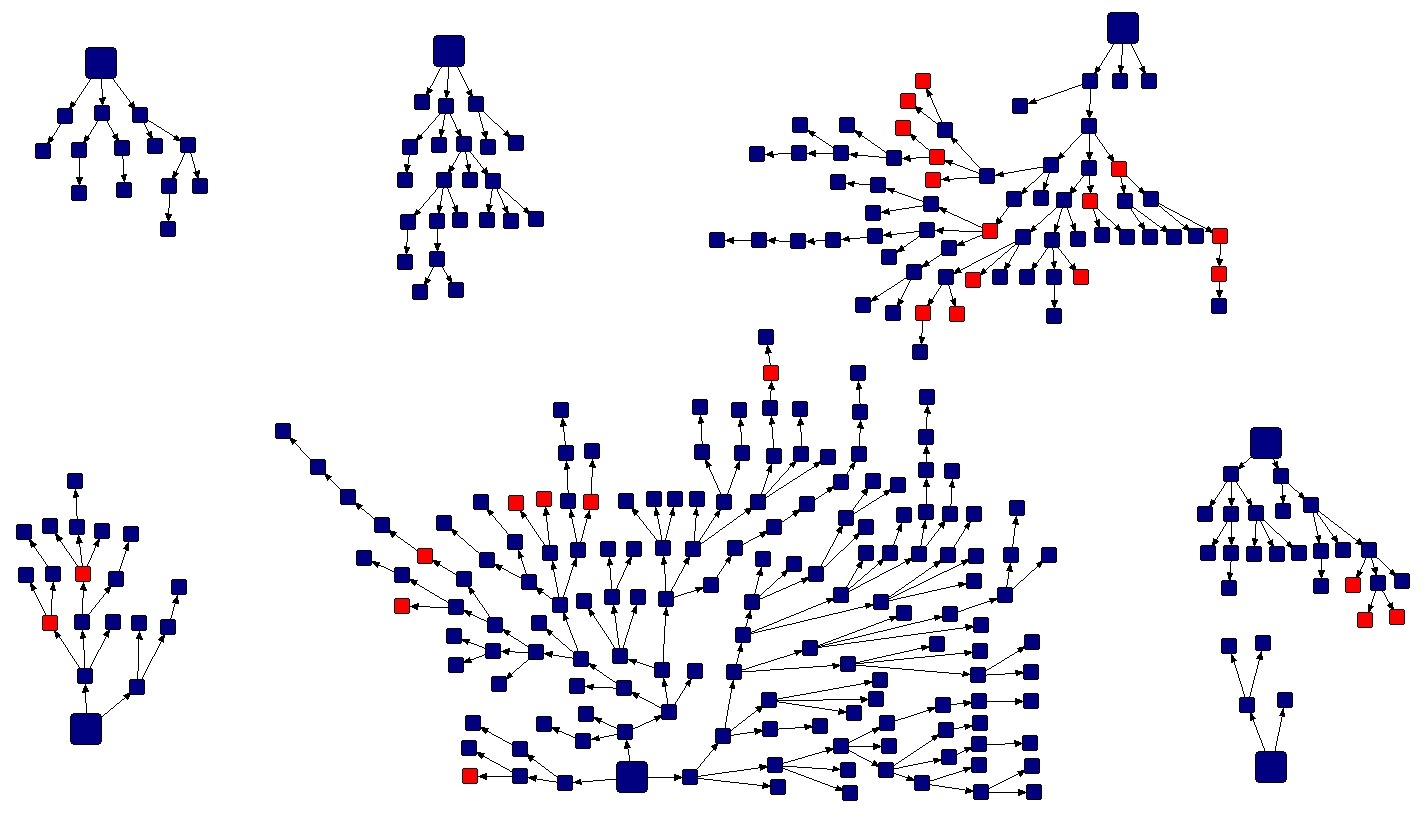}
   \caption{Recruitment Chains Plot from sample of MSM in Higuey.  Red nodes self-identify as ``heterosexual.''}
   \label{fig:chains} 
\end{figure}

RDS data collection begins when researchers select, in an ad-hoc manner, typically 5 to 10 members of the study population to serve as ``seeds.''  Each seed is interviewed and provided a fixed number of coupons (usually three) that they use to recruit other members of the study population.  These recruits are in turn provided with coupons that they use to recruit others.  In this way, the sample can grow through many waves, resulting in recruitment trees like those shown in Fig.~\ref{fig:chains}.  Respondents are encouraged to participate and recruit through the use of financial and other incentives~\citep{heckathorn_respondent-driven_1997}.  The fact that the majority of participants are recruited by other respondents and not by researchers makes RDS a successful method of data collection.  However, the same feature also inherently complicates inference because it requires researchers to make assumptions about the recruitment process and the structure of the social network connecting the study population.  

There are three interrelated approaches to addressing the assumptions underlying inference from RDS data.  First, researchers can identify assumptions whose violations significantly impact estimates, either analytically or through computer simulation (e.g.,~\citet{gile_respondent-driven_2010, tomas_effect_2011, blitzstein_bias-variance_2011, lu_sensitivity_2012}).  Second, researchers can develop new estimators that are less sensitive to these assumptions (e.g.,~\citet{gile_improved_2011, gile_network_2011, lu_respondent-driven_2012}).  Third, researchers can develop methods to detect the violation of assumptions in practice.  This third approach is the primary focus of this paper, but we hope that our results will help motivate and inform research of the first two types.  

This paper makes two main contributions.  First, we review and develop diagnostics for most assumptions underlying statistical inference from RDS data.  One reason for the relative dearth of RDS diagnostics is that the same conditions that complicate inference from RDS data also complicate formal diagnostic tests.  In particular, the potential dependence between recruiters and recruits renders most standard tests invalid.  Therefore, when possible, we develop diagnostic approaches that are intuitive, graphical, and not reliant on statistical testing.  Further, when possible, we emphasize approaches that can be used while data collection is occurring so that problems can be investigated and potentially resolved.  In order to provide these features, our diagnostics frequently take advantage of three specific features of RDS studies that are not typically utilized: information about the time sequences of responses, contact with respondents who visit the study site twice, and the multiple seeds used to begin the sampling process.  The second main contribution of our paper is to deploy these diagnostics in 12 RDS studies conducted in accordance with the national strategic HIV surveillance plan of the Dominican Republic.  We believe that these case studies---which include samples of female sex workers (FSW), drug users (DU), and men who are gay, transsexual, or have sex with men (MSM) in four cities---are reasonably reflective of the way that RDS is used in many countries.  Therefore, we believe that our empirical results have broad applicability for RDS practitioners and researchers who wish to develop improved methods of RDS data collection and inference.

The remainder of the paper is organized as follows: in Section~\ref{sec:assumptions} we briefly review the assumptions underlying RDS estimation and in Section~\ref{sec:data} we describe the data from 12 studies in the Dominican Republic that will be used throughout the paper.  Sections~\ref{sec:with-replacement_sampling} through~\ref{sec:participation} present diagnostics, including extensions of previous approaches as well as wholly new approaches.  In Section~\ref{sec:discussion} we discuss the results and conclude with suggestions for future research.  We also include online Supporting Information with additional results and approaches.

\section{Assumptions of RDS}
\label{sec:assumptions}

Estimation from RDS data requires many assumptions about the sampling process, underlying population, and respondent behavior.  These assumptions are outlined in Table \ref{tab:assred} and described fully in~\citet{gile_respondent-driven_2010}.  In particular, these assumptions are required by the estimator proposed by~\citet{volz_probability_2008}.  Other available estimators require similar assumptions, especially pertaining to respondent behavior.

\begin{table}[h]\caption[Assumptions of the Volz-Heckathorn Estimator]{Assumptions of the Volz-Heckathorn Estimator.  Assumptions in \textbf{\textit{bold-italics}} are considered in this paper, with section numbers given.  A version of this table appeared in~\citet{gile_respondent-driven_2010}.}
\begin{center}
\begin{small}
\begin{tabular}{l||c|c}
& Network Structure  & Sampling Assumptions\\
& Assumptions & \\
\hline
\hline
Random Walk &  Network size large ($N >> n$) & \textbf{\textit{With-replacement sampling (\ref{sec:with-replacement_sampling})}} \\
Model & 
&  Single non-branching chain \\
\hline
Remove Seed &  \textbf{\textit{Homophily weak enough (\ref{sec:detecting_convergence}, \ref{sec:bottlenecks})}} &  \textbf{\textit{Enough sample waves (\ref{sec:detecting_convergence})}} \\
Dependence & \textbf{\textit{Bottlenecks limited (\ref{sec:bottlenecks})}}  & \\
 & Connected graph & \\
\hline
Respondent   & \textbf{\textit{All ties reciprocated (\ref{sec:reciprocation})}} & \textbf{\textit{Degree accurately measured (\ref{sec:degree})}}  \\
Behavior & &  \textbf{\textit{Random referral (\ref{sec:participation})}}   \\
\end{tabular} \label{tab:assred}
\end{small}
\end{center}
\end{table}

Each row of this table includes assumptions according to their roles in allowing for estimation.  The first row (``Random Walk Model'') corresponds to assumptions required to allow the sampling process to be approximated by a random walk on the nodes.  Critically, the random walk model requires with-replacement sampling, while the true sampling process is known to be without-replacement.  We, therefore, first consider diagnostics designed to detect impacts of the without-replacement nature of the sampling (Sec.~\ref{sec:with-replacement_sampling}).

The second row (``Remove Seed Dependence'') contains assumptions required to reduce the influence of the initial sample---the seeds---on the final estimates.  Because the initial sample is usually a convenience sample, RDS is intended to be carried out for many sampling waves through a well-connected population in order to minimize the impact of the seed selection process.  Therefore, we consider diagnostics designed to detect seed bias that may remain due to an insufficient number of sample waves (Secs.~\ref{sec:detecting_convergence} and~\ref{sec:bottlenecks}).

The final row of the table, (``Respondent Behavior,'') contains assumptions related to respondent behavior.  Unlike traditional survey sampling, RDS is characterized by a significant role of respondent decision-making in the sampling process, and, therefore, assumptions about these decisions are needed for estimating sampling probabilities.  In particular, we consider the assumptions that all network ties are reciprocated, that degree (also referred to as number of contacts or personal network size) is accurately reported, and that future participation is random among contacts in the study population (Secs.~\ref{sec:reciprocation},~\ref{sec:degree}, and~\ref{sec:participation}).

\section{Case study: 12 sites in the Dominican Republic}
\label{sec:data}

We employ these diagnostics in a case study of 12 parallel RDS studies conducted in the spring of 2008 using standard RDS methods~\citep{johnston_introduction_2008}.  As part of the national strategic HIV surveillance plan of the Dominican Republic, data were collected from female sex workers (FSW), drug users (DU), and men who are gay, transsexual, or have sex with men (MSM) in four cities: Santo Domingo (SD), Santiago (SA), Barahona (BA), and Higuey (HI). These studies are typical of the way RDS is used in national HIV surveillance around the world.  Eligible persons were 15 years or older and lived in the province under study.  Eligible FSW were females who exchanged sex for money in the previous six months, DU were females or males who used illicit drugs in the previous three months, and MSM were males who had anal or oral sexual relations with another man in the previous six months. Seeds were purposively selected through local non-governmental organizations or through the use of peer outreach workers.  Each city had a fixed interview site where respondents enrolled in the survey. 

\begin{figure}
  \centering
   \includegraphics[width=0.7\textwidth]{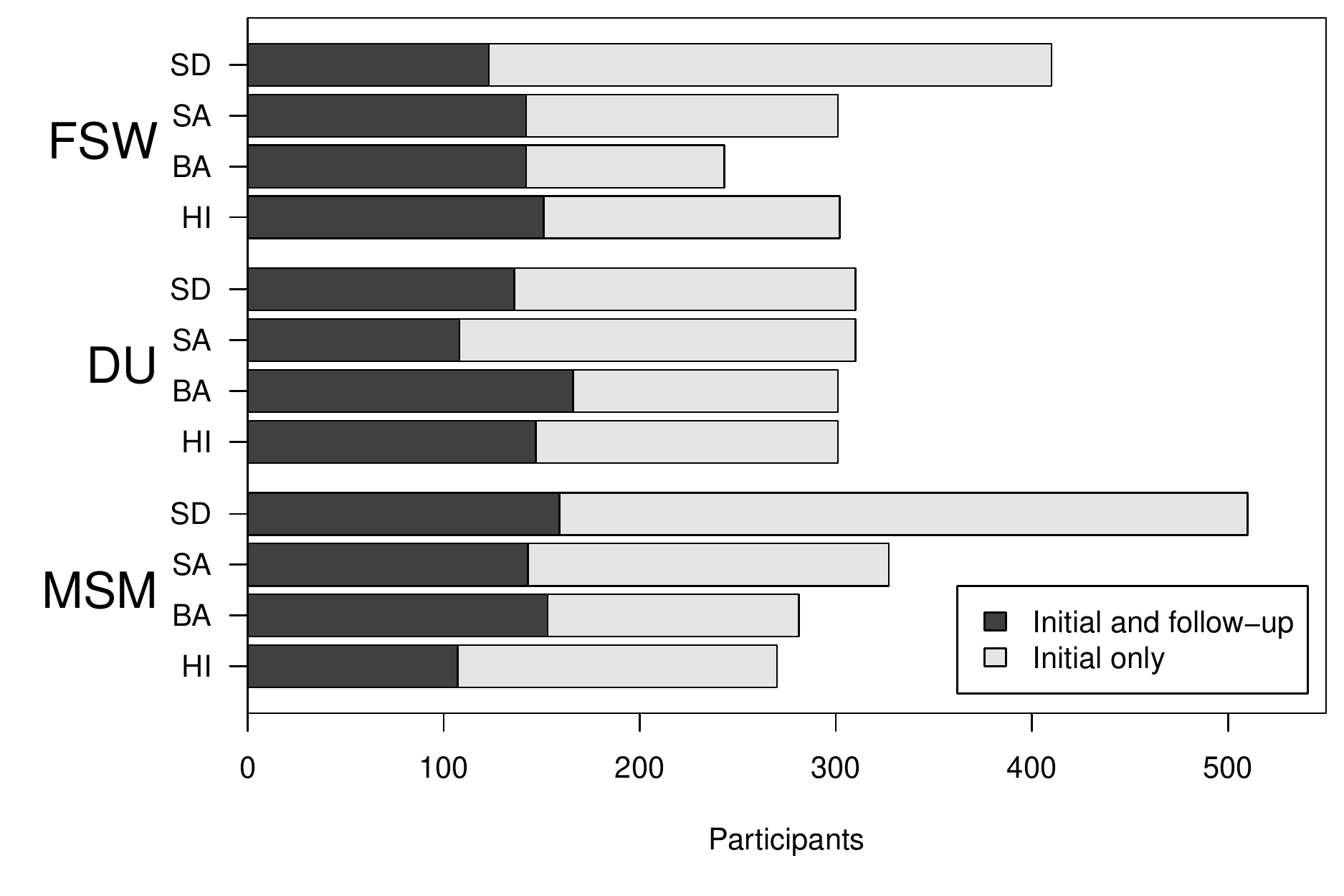}
   \caption{Sample sizes from the 12 studies.  In total, 3,866 people participated, of which 1,677 (43\%) completed a follow-up survey.}
   \label{fig:samplesizes} 
\end{figure}

During the initial visit, consenting respondents were screened for eligibility, completed a face-to-face interview, received HIV pre-test counseling and provided blood samples that were tested for HIV, Hepatitis B and C, and Syphilis.  Before leaving the survey site, respondents were encouraged to set an appointment to return two weeks later for a follow-up visit during which they would receive HIV post-test counseling, collect infection test results and, if necessary, be referred to a nearby health facility for care and treatment.  During the follow up visit respondents also completed a follow-up questionnaire and received secondary incentives for any peers they recruited; respondents were compensated the equivalent of \$9.00 USD for completing the initial survey and \$3.00 USD for each successful recruitment (up to a maximum of three).  To ensure confidentiality, respondents' coupons, questionnaires and biological tests were identified using a unique study identification number;  no personal identifying information was collected. The studies ranged in sample size from 243 to 510 with a total sample size of 3,866 people, of which 1,677 (43\%) completed a follow-up survey (see Fig.~\ref{fig:samplesizes}).

We analyze these data using the estimator introduced in~\citet{volz_probability_2008} because it has been used in most of the recent evaluations of RDS methodology~\citep{wejnert_empirical_2009, goel_respondent-driven_2009, goel_assessing_2010, gile_respondent-driven_2010, tomas_effect_2011, blitzstein_bias-variance_2011, lu_sensitivity_2012, mccreesh_evaluation_2012}.  The estimator of the proportion of the population with a specific trait (e.g., HIV infection) is:

\begin{equation}
\hat{p}= \frac{\sum_{j \in I} \frac{1}{d_j}}{\sum_{j \in S} \frac{1}{d_j}},
\label{vhest}
\end{equation}
where $S$ is the full sample, $I$ is the infected sample members, and $d_j$ is the self-reported ``degree,'' or number of contacts of respondent $j$.  Equation (\ref{vhest}), sometimes called the RDS II estimator or the Volz-Heckathorn (VH) estimator, is a generalized ratio estimator of a population mean, with inverse probability weighting, and sampling weights proportional to degree.

\section{With-replacement Sampling}
\label{sec:with-replacement_sampling}

Many estimators for RDS data are based on the assumption that the sample can be treated as a with-replacement random walk on the social network of the study population.  In particular, respondents are assumed to choose freely which of their contacts to recruit into the study.  In practice, sampling is {\it without} replacement;  respondents are not allowed to recruit people who have already participated.  This restriction may lead to inaccurate estimates of inclusion probabilities and biased estimates, as described in \citet{gile_improved_2011}. 

Indications of the influence of earlier respondents on subsequent sampling decisions would suggest potentially problematic violations of the sampling-with-replacement assumption.  The finite population size may affect sampling in two ways: locally, when members of a small well-connected sub-group are sampled at a high rate, influencing the future referral choices of other sub-group members, and globally, when the study population as a whole is sampled at a high enough rate that later samples are influenced by earlier samples.  If the finite population affects sampling, it is possible this will induce a bias in resulting estimates.

In this section, we examine the with-replacement sampling assumption in several ways.  First, we use three types of evidence to detect local and global finite population effects on sampling.  Next, we assess the impact of global finite population effects on estimates.  Finally, we compare the methods  and conclude with recommendations.

\subsection{Failure to Attain Sample Size}
\label{sec:failed_to_attain_sample_size}

Strong evidence of global finite population effects is provided by failure to attain the target sample size due to the inability of respondents to sample additional members of the study population. This occurred in three of our studies: FSW-BA, MSM-BA, and MSM-HI.
When all final wave respondents lack alters that are eligible, available, and previously un-sampled, the sample was clearly affected by the finite size of the study population.  As a diagnostic, however, this indicator has two primary limitations.  First, it cannot be assessed until the study is complete.  Second, while failure to attain sample size is an indication that finite population effects are present, the absence of such failure is not an indication that those effects are absent.  In our comparison, therefore, this indicator serves to indicate ``true positives,'' but not ``true negatives.''

\subsection{Failed Recruitment Attempts}
\label{sec:failed_recruitment_attempts}

If the sampling process were not influenced by the previous sample, each respondent could distribute coupons without considering whether contacts had already participated in the study.  Therefore, respondents who returned for a follow-up survey were asked
\begin{enumerate}[(A)]
\item How many people did you try to give a coupon but they had already participated in the study?
\label{nfailq}
\eq

\begin{figure}
  \centering
   \includegraphics[width=0.5\textwidth]{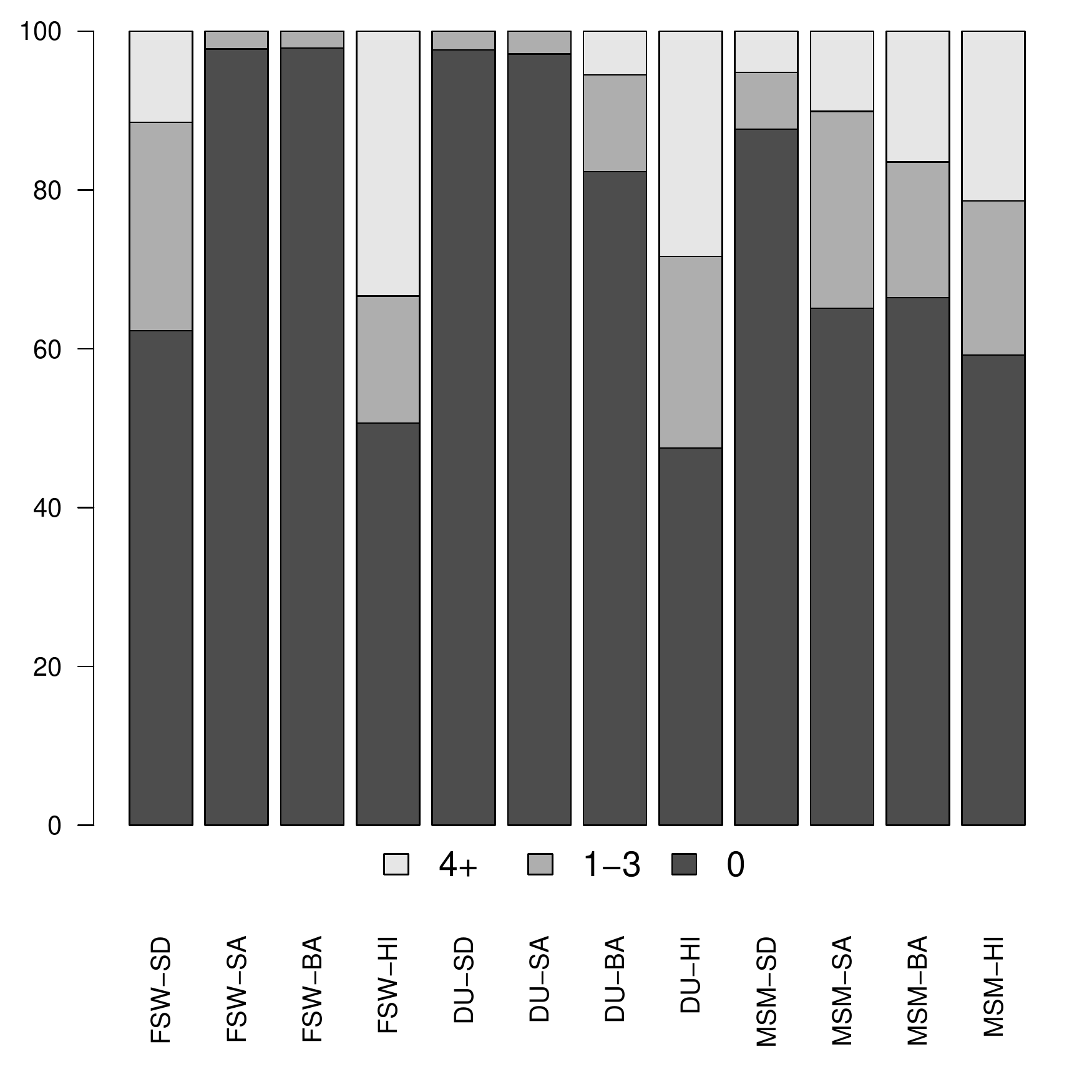}
   \caption{Percent of respondents reporting 0, 1-3, or 4+ failed recruitment attempts.  In 6 sites, at least 25\% of respondents reported at least one failed recruitment attempt.}
   \label{fig:allfailed} 
\end{figure}

Responses to this question are summarized in Fig.~\ref{fig:allfailed}.  Rates of failed coupon distributions varied widely by site, with the most failures among drug users in Higuey, with over half of follow-up respondents reporting a failed attempt to distribute a coupon, and the fewest failures being among DU in Santo Domingo and Santiago, and among FSW in Santiago and Barahona, with 3\% or fewer respondents reporting failed coupon distributions. In six of the 12 sites, at least 25\% of respondents participating in follow-up interviews indicated they had attempted to give coupons to at least one person who had already participated in the study (Table~\ref{replacesummary}).  Where present, these reported failures provide direct evidence that respondents' recruiting decisions were affected by earlier parts of the sample.  Where absent, they can either indicate a lack of such influence or accurate knowledge of which alters have already participated in the study.

\subsection{Contacts Participated}
\label{sec:contacts_participated}

Respondents' coupon-passing choices could also be influenced by the contacts they know have already participated in the study.  To assess this possibility, respondents were asked the following question (see also \citet{mccreesh_evaluation_2012}):
\bq
\item How many other MSM/DU/FSW do you know that have already participated in this study, without counting the person who gave a coupon to you?
\label{knowalreadyq}
\eq
Across all 12 datasets, only 30\% of respondents answered ``0,'' with 
mean proportion of alters reported to have already participated 36\%.  This result suggests that previously sampled population members may indeed impact the alters available for the passing of coupons.  Note that about 10\% of respondents (347 out of 3,866) reported knowing more people who had already participated than they reported knowing in (\ref{sizeq3}).  It is possible that the distinction is due to the fact that the group in (\ref{sizeq3}) was limited to ``people you know and they know you,'' while (\ref{knowalreadyq}) applies to all ``people you know'', however this is more likely due to inaccurate reporting.  Throughout this section, we truncate responses at one less than the reported number of people known.

If this phenomenon is uniform across the sampling process, it may be partially explained by measurement error or low-level local clustering with minimal connection to global finite population effects.  An increase in this effect over the course of the sample, however, suggests the population is becoming increasingly depleted, such that previously sampled alters constrain the choices of later respondents more than those of earlier respondents.  In looking for evidence of a time trend, we fit a simple linear model relating the sample order to the proportion of alters who already participated.  To serve as a conservative flagging criterion, in a setting where formal testing is likely invalid, we flag any cases with positive trends over time.
We find positive trends in probability of having been previously sampled for increasing survey order, in eight of the 12 populations (DU-SD, DU-SA, FSW-SA, FSW-BA, FSW-HI, MSM-SA, MSM-BA, MSM-HI), suggestive of potential finite population effects.  In the Supporting Information, we consider two approaches to visualizing these effects.  Results were very similar when more complex models models were applied.

\subsection{Assessing Finite Population Effects on Estimates}
\label{sec:ss}

The results in Sections~\ref{sec:failed_to_attain_sample_size} to~\ref{sec:contacts_participated} focused on detecting finite population effects on sampling.  Next, we turn to detecting global finite population effects on estimates using an approach that requires knowing or estimating the size of the study population.  If the study population is very large compared to the sample size, then global exhaustion is unlikely to be of concern.  If the study population is small, however, then a bias may be induced, but the magnitude of estimator bias will depend on the relative degree distributions of the groups of interest (such as infected and uninfected people): the greater the systematic difference in degrees, the greater the potential bias in estimate~\citep{gile_respondent-driven_2010, gile_improved_2011}.  Finite population biases can be mitigated by using estimators designed to account for finite population effects, such as the estimator based on successive sampling  (SS) introduced in \citet{gile_improved_2011} and implemented in the {\bf R}~\citep{r_core_team_r:_2012} package {\tt RDS} \citep{handcock_rds:_2009}.  Further, a comparison between the results of the SS estimator and the VH estimator can serve as a sensitivity analysis to global finite population effects because these two estimators differ only in that the former corrects for finite population effects.  If the two estimators are nearly identical for reasonable estimates of the population size, then global exhaustion is likely not inducing bias into estimates.  

In order to undertake this sensitivity analysis, and as described in greater detail in the Supporting Information, we estimated the size of our study populations using two different approaches: 1) drawing on meta-analysis of related studies and 2) the approach introduced in~\citet{handcock_estimating_2012} and implemented in the package {\tt size} \citep{handcock_size:_2011}, which uses information in the degree sequence in the RDS sample.  Using these estimated population sizes, we then compared the SS and VH estimators in all 12 study populations for all characteristics described in Section \ref{sec:detecting_convergence}.  In most cases, the two estimates were within $0.01$ of each other (Fig.~\ref{fig:ssvhdif}); Table~\ref{tab:ssvhdif} lists all traits with differences larger than 0.01.  Overall, therefore, this analysis suggests that there were not large finite population effects on the VH estimator in these studies.  Note that we also studied the degree sequence directly, as summarized in the Supporting Information.  Surprisingly, direct evaluation of the trend in degree over time suggested little evidence of finite population effects on sampling.

\begin{figure}
  \centering
     \includegraphics[width=\textwidth]{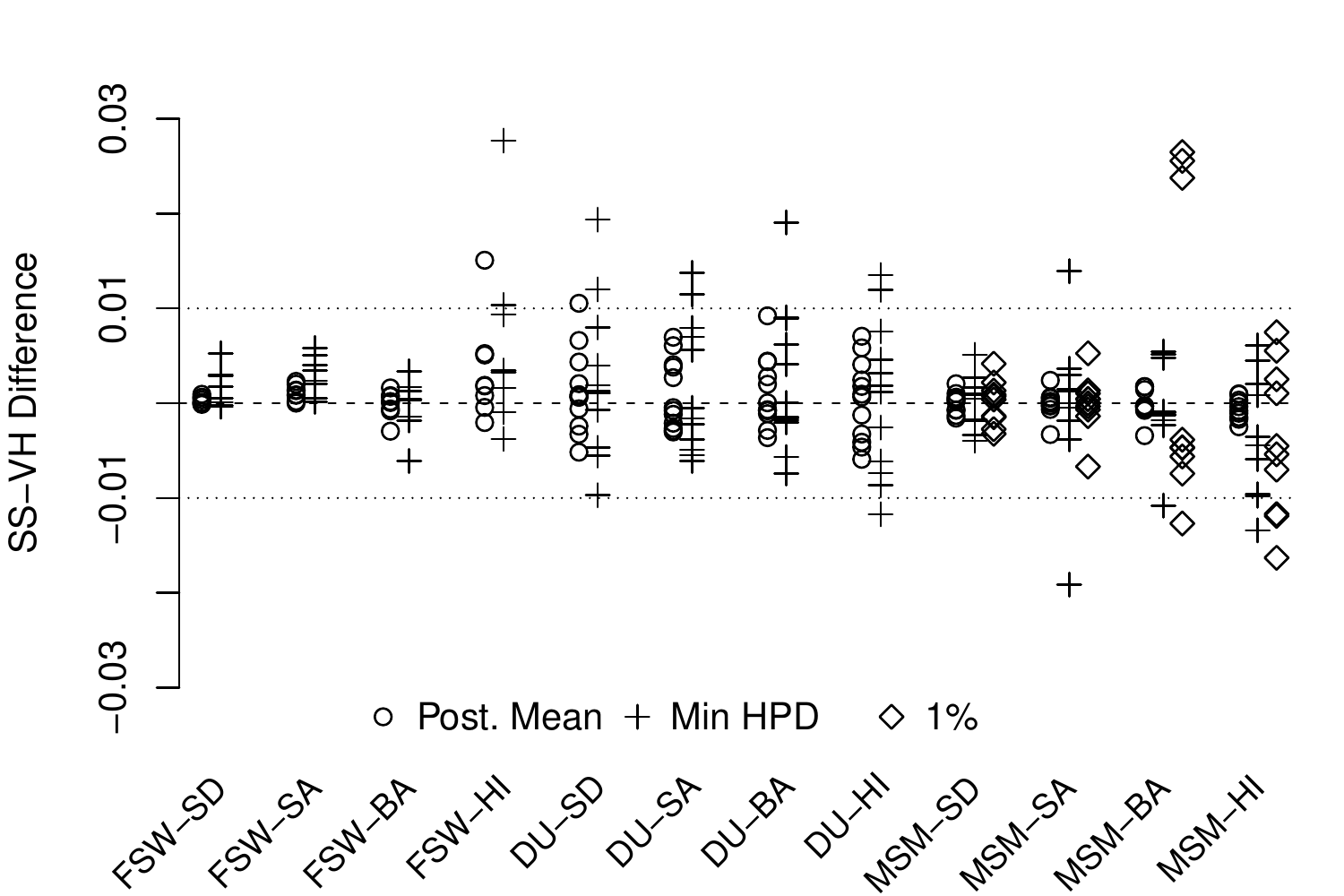}
   \caption{Difference between Successive Sampling and Volz-Heckathorn estimators, over many traits. Three population size estimates are considered:  the posterior mean using the method in \citet{handcock_estimating_2012}  (a ``best guess''), the lower bound (of the HPD interval) from this method, and, for the MSM, a lower bound from the literature at 1\% of the relevant target population. One point (``had HIV Test'' among MSM in BA, $1\%$ population size) had a difference of -.054, and is not shown.}
  \label{fig:ssvhdif} 
\end{figure}

\subsection{Comparison of Approaches and Current Recommendations}

Table \ref{replacesummary} summarizes all of the sampling process indicators across survey sites.  Failure to attain the desired sample size (FSW-BA, MSM-BA, MSM-HI) is a clear indication that the earlier samples impacted the later sampling decisions.  Consistent with this result, the MSM sites in Barahona and Higuey showed evidence of without-replacement sampling effects on all three of these proposed indicators.  Nearly all sites, however, had evidence of finite population effects on at least one indicator.  Together, these indicators show that finite population effects on sampling were frequent and that reasonable diagnostic approaches for detecting them can produce different results.  These differences between indicators can either be the result of random variation or the result of different indicators reflecting different features of the underlying process.  

The most effective diagnostic of global effects on estimates is the comparison of the VH and SS estimators.  Unlike the other indicators, this indicator measures the direct effect on the estimate.  It is possible that global finite population effects influence sampling (as indicated by one of the earlier indicators), but do not induce bias in the estimator because of other features of the network, such as similar degree distributions between the two sub-populations of interest.   This is the case, for example, among FSW in Barahona, and MSM in Higuey, which do not exhibit worrisome finite population effects on estimates, despite failing to reach their intended sample sizes.  Among MSM in Barahona, however, the large sample fraction may well be influencing estimates.  One challenge in implementing this diagnostic is that the SS estimator requires an estimate of the size of the study population, and these size estimates can be difficult to construct~\citep{unaids_guidelines_2010, bernard_counting_2010, salganik_assessing_2011, handcock_estimating_2012}.

In future studies, questions about failed recruitments and numbers of known participants (Questions \ref{nfailq} and \ref{knowalreadyq}), can be helpful in diagnosing local effects, and should be collected and studied further to determine the extent to which local clustering may impact inference.  Further, when diagnostics suggest large finite population effects on sampling, researchers should use estimators that do not depend on the sampling with replacement assumption (e.g.,~\citet{gile_improved_2011, handcock_estimating_2012}), or minimally these estimators should be used for sensitivity analysis as in Section \ref{sec:ss}.  Methods for inference in the presence of local finite population effects are not yet available.  

\begin{table}[ht]\caption{\small Summary of indicators of violations of the with-replacement sampling assumption.  First row indicates sites which were not able to attain the intended sample sizes (Sec.~\ref{sec:failed_to_attain_sample_size}).  The second indicates at least 25\% of follow-up respondents reporting they attempted to give coupons to at least one person who had already participated in the study (Sec.~\ref{sec:failed_recruitment_attempts}).  The third indicates a positive coefficient of sample order in the linear regression model for probability an alter is in the study (Sec.~\ref{sec:contacts_participated}).}
\begin{center}
\begin{small}
\begin{tabular}{lcccccccccccc}
& \multicolumn{4}{c}{FSW} & \multicolumn{4}{c}{DU} & \multicolumn{4}{c}{MSM}\\
\cmidrule(rl){2-5} 
\cmidrule(rl){6-9}
\cmidrule(rl){10-13}
 & SD & SA & BA & HI & SD & SA & BA & HI & SD & SA & BA & HI\\
\hline
Failed to Attain Sample Size & & & X & & & & & & & & X & X\\
Failed Attempts $> 25$\% & X & & & X & & & & X & & X & X & X\\ 
Increasing Participants Known & & X & X & X & X & X & & X & & X & X & X\\
\end{tabular}
\end{small}
\label{replacesummary}
\end{center}
\end{table}

\section{Detecting convergence}
\label{sec:detecting_convergence}

In RDS studies the initial sample members (``seeds'') are not selected from a sampling frame, but are instead an ad-hoc convenience sample.  In general, the seed selection mechanism has not concerned RDS researchers because of asymptotic results showing that the choice of seeds does not effect the final estimate~\citep{heckathorn_respondent-driven_1997, heckathorn_respondent-driven_2002, salganik_sampling_2004}.  However, these asymptotic results only hold as the sample size goes to infinity, and in practice samples are far from infinite.  Therefore, a natural question is whether a given sample is large enough to overcome the potential biases introduced during seed selection. 

There are some apparent similarities between the current problem and the monitoring of convergence of computer-based Markov-chain Monte Carlo (MCMC) simulations. 
  Standard MCMC methods, unfortunately, cannot be directly applied here.  First, single chain methods, such as~\citet{raftery_how_1992}, are not applicable because we have multiple chains (as highlighted in Section~\ref{sec:bottlenecks}).  Further, multiple chain methods, such as~\citet{gelman_inference_1992}, are not directly applicable because RDS chains are of different lengths.  Finally, these standard approaches typically rely on far longer sample chains than are available in RDS data; for example, the longest chain in these studies is 16 respondents long.

The currently used diagnostic for assessing whether the RDS sample is big enough is to compare the length of the longest chain to the calculated number of waves required for the sampling process to approximate its stationary distribution under a first-order Markov chain model on group membership~\citep{heckathorn_extensions_2002}.  This approach is now standard in the field~\citep{johnston_implementation_2008, malekinejad_using_2008, montealegre_respondent-driven_2012}, but is based on a different model for the sampling process than is assumed in most RDS estimators, does not address sampling past the point of ``convergence,''  and has generated a great deal of confusion (see for example,~\citet{ramirez-valles_networks_2005, heimer_critical_2005, ramirez-valles_fit_2005, wejnert_web-based_2008}).  Here we propose a more direct and interpretable approach to assess convergence.  Rather than focusing on the \emph{simulated} dynamics of the \emph{sample composition}, we focus on the \emph{actual} dynamics of the \emph{RDS estimate}.  Roughly, the more the estimate changes as we collect more data, the more concern we should have that the choice of seeds is still influencing the estimate (see also~\citet{bengtsson_implementation_2012} for a similar approach).

More concretely, let $\hat{p}_t$ be the estimated trait prevalence using the first $t$ observations (where we exclude all seeds).  To assess the possible lingering impact of seed selection, we plot $\hat{p}_1, \hat{p}_2, \ldots, \hat{p}_n$ and see if the estimates seem to stabilize.  Fig.~\ref{fig:dynamics_bad} shows a \emph{Convergence Plot} for the proportion of DU in Barahona that report using drugs every day.  The estimate is increasing over time suggesting that the seeds and early samples were atypical in their drug use pattern.  This constant and sharp increase in estimates actually under-represents the differences between the early and late parts of the sample because the estimate is cumulative.  For example, based on the first 50 respondents we would estimate that 8\% of the population use drugs every day, but from the final 50 respondents we would estimate that 67\% use drugs every day.  Compare these dynamics with Fig.~\ref{fig:dynamics_good} which plots the estimated proportion of DU in Barahona that reported engaging in unprotected sex in the last 30 days.  This estimate appears to be stable for the second half of the sample.  Note that both of these estimates arise from the same sample and, therefore, highlight the fact that convergence is a property of an estimate not a sample.

\begin{figure}
  \centering
   \subfigure[DU-BA, Use drugs every day]{
    \label{fig:dynamics_bad} 
     \includegraphics[width=0.45\textwidth]{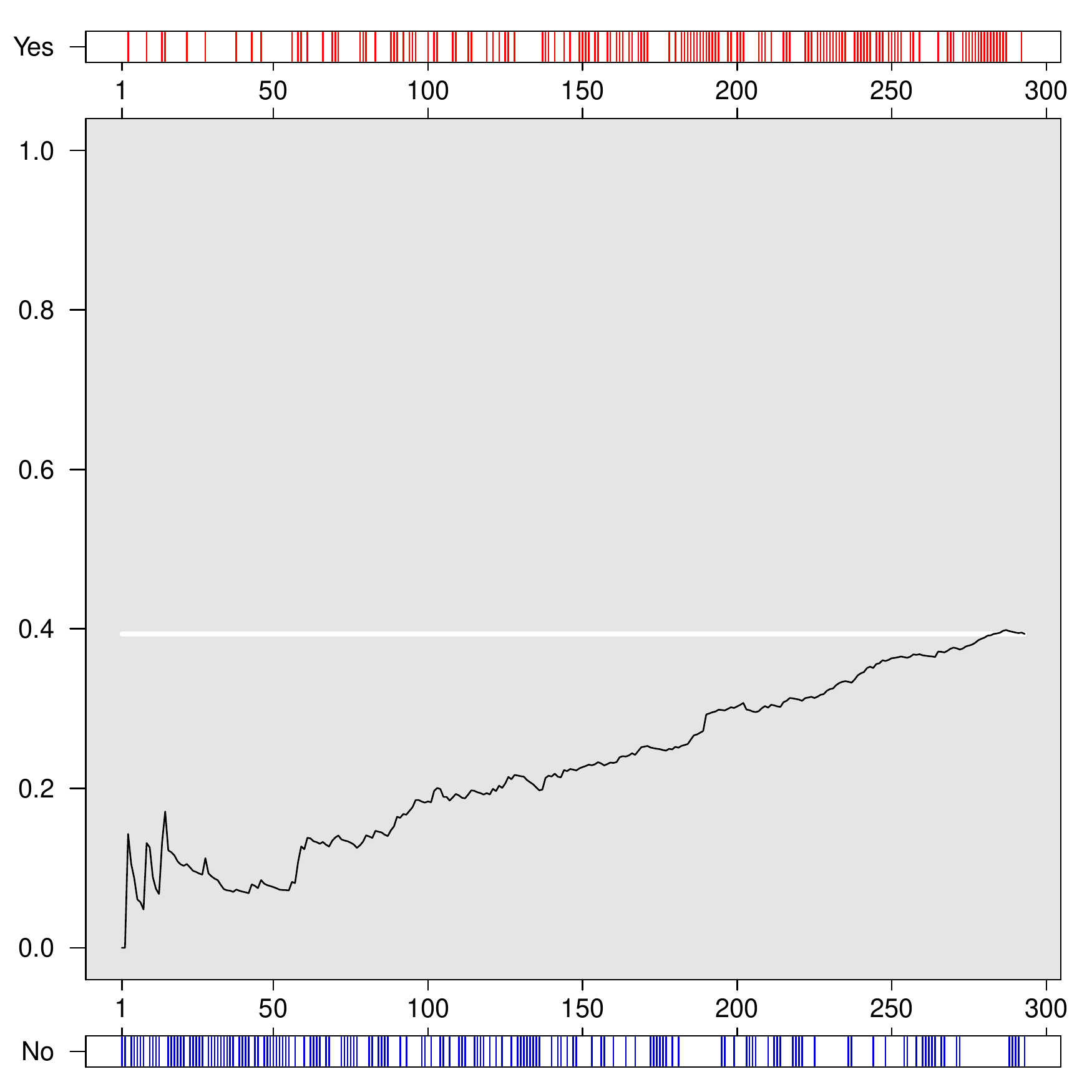}}
  \hspace{0in}
  \subfigure[DU-BA, Risky sex]{
  \label{fig:dynamics_good} 
   \includegraphics[width=0.45\textwidth]{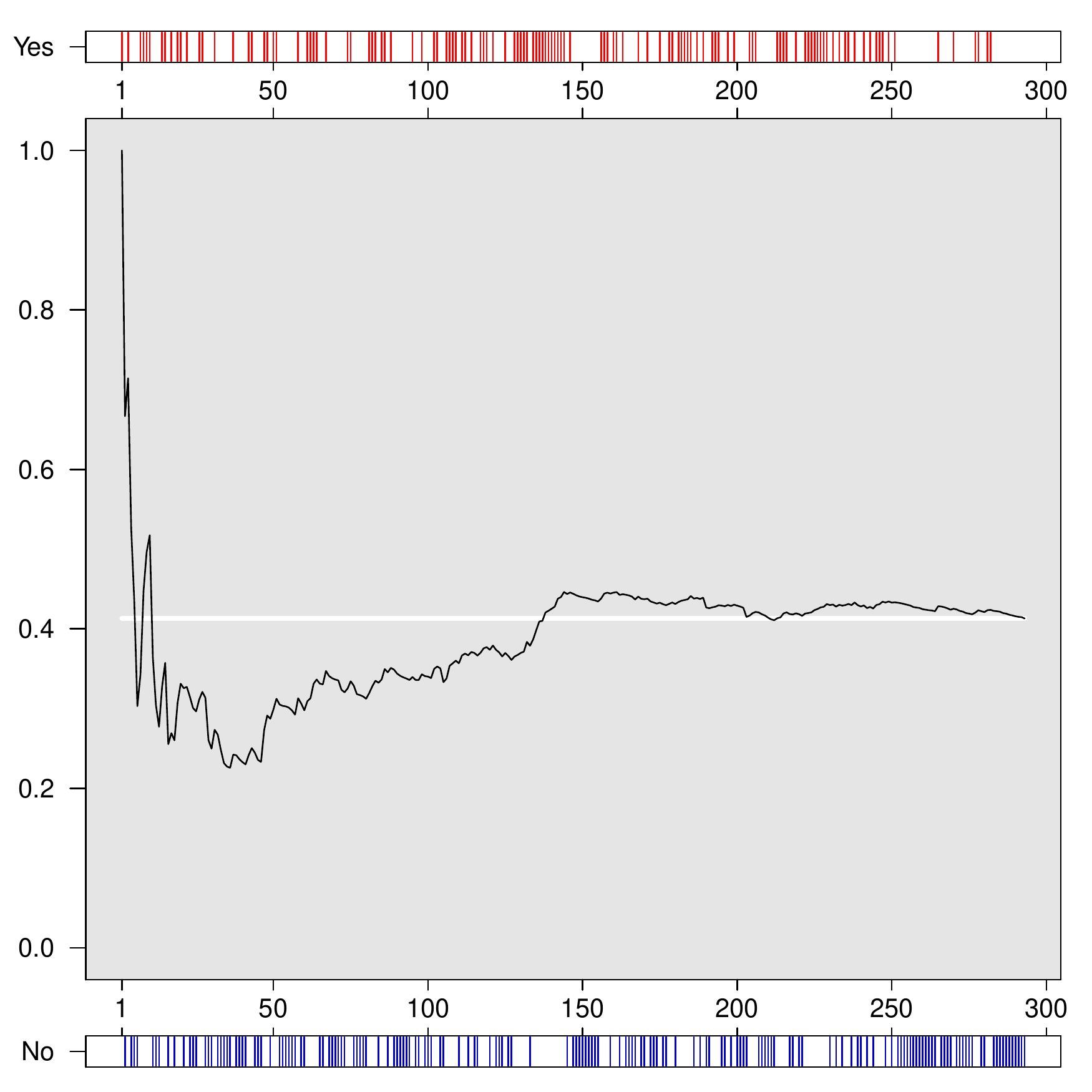}}
  \caption{{\it Convergence Plots} showing $\hat{p}_1, \hat{p}_2, \ldots, \hat{p}_n$.  The headers and footers plot the sample observations with and without the trait.  The white line shows the estimate based on the complete sample ($\hat{p}_n$).}
     \label{fig:convergence_plots} 
\end{figure}

We recommend visual inspection of Convergence Plots rather than a formal decision rule, but in cases where there are many study sites and many traits of interest, it may be difficult to monitor all of these plots.  Therefore, traits can be flagged for further inspection if the estimates seem to be changing at the end of the sample.  That is, a trait should be flagged if  
\begin{equation}
\mbox{there exists} \quad t < \tau \mbox{ such that } \mid \hat{p}_{(n-t)}-\hat{p}_{(n)} \mid \  > \ \epsilon
\label{eq:convergence}
\end{equation}
where $\tau$ is a parameter that sets how much of the trace will be examined and $\epsilon$ represents the maximum allowable difference between the estimate at time $t$ and the final estimate.  We suspect that the desired values of $\tau$ and $\epsilon$ will vary from study to study, but in this case we set $\tau=50$ and $\epsilon=0.02$.  In other words, we ask whether there are any of the final 50 estimates that have a difference of more than 0.02 from the final estimate.  We run this procedure on 120 group $\times$ trait $\times$ city combinations shown in Fig.~\ref{fig:convergence_results}, and we find the most convergence problems in MSM data: 37.5\% of traits were flagged, as compared with 25\% of traits for DU and 22\%  for FSW.  Increasing $\epsilon$ to 0.05 results in flagging only two traits, both in MSM populations:  Bisexual in Santiago and Use Drugs in Higuey.  The convergence problems that we detected could be caused by the network structure in the population, the method of seed selection, and the interaction between the two.

\begin{figure}
  \centering
   \includegraphics[width=\textwidth]{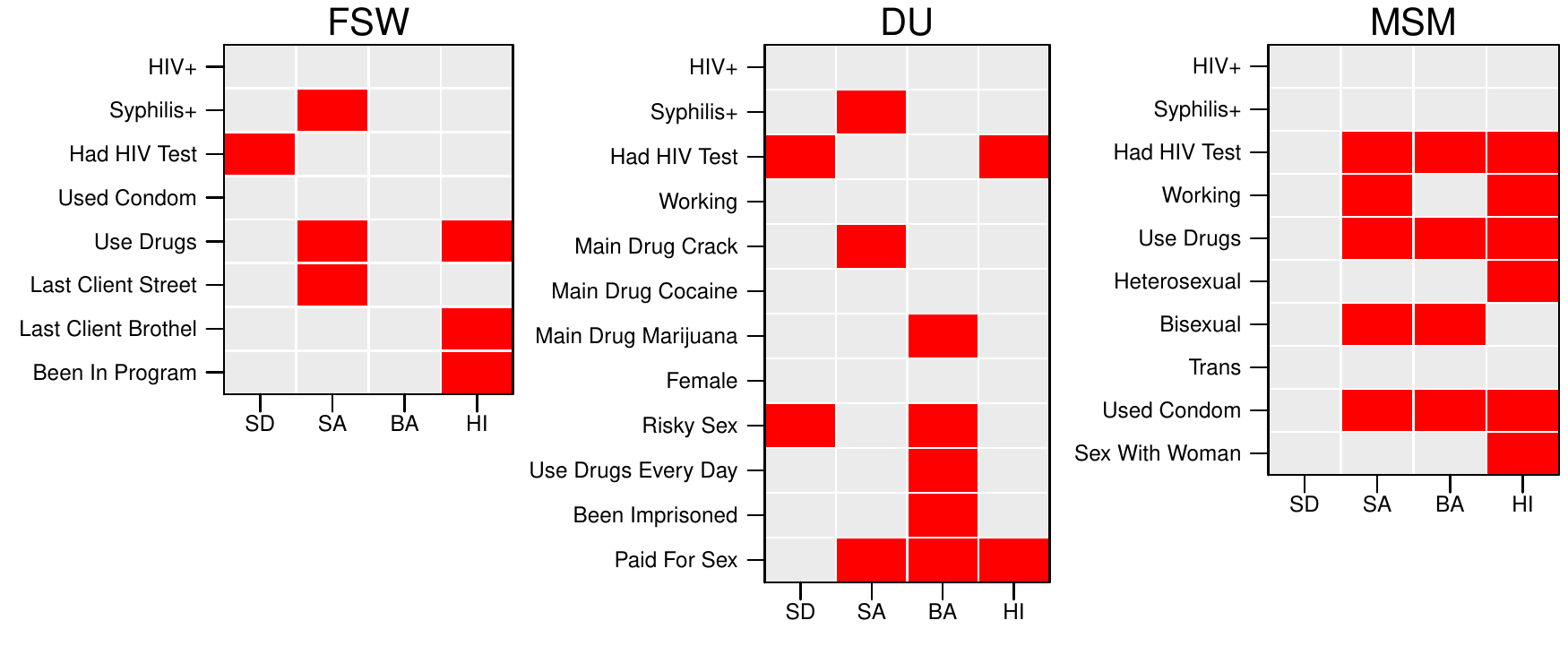}
   \caption{Convergence test results for $\tau=50$ and $\epsilon=0.02$.  Red cells represent traits flagged for possible lack of convergence.   }
   \label{fig:convergence_results} 
\end{figure}

\subsection{Current Recommendations}

We recommend creating Convergence Plots for all traits of interest during data collection.  Evidence of unstable estimates (e.g., Fig.~\ref{fig:dynamics_bad}) should be taken as an indication that results may be suspect and that more data should be collected.  If additional data collection is not possible, researchers may need to use more advanced estimators that are designed to correct for features such as seed bias (e.g.,~\citet{gile_network_2011}).  If it is not possible to create Convergence Plots during data collection, they should still be made, used to consider alternative estimators, and, if unusual, presented with published results.  

We wish to emphasize that there are cases where the Convergence Plot could fail to detect a real problem.  For example, we could imagine cases where the estimates appear stable (Fig.~\ref{fig:dynamics_good}), but then the sample could move to a previously unexplored part of the study population yielding very different estimates.  Researchers can therefore gain some, but not perfect, confidence by looking at how the estimate changes over time.  

\section{Detecting bottlenecks}
\label{sec:bottlenecks}

RDS can perform poorly in populations that divide into ``communities'', if those communities differ in their prevalence of specific traits~\citep{goel_respondent-driven_2009}.  For example, imagine a city with street-based sex workers and brothel-based sex workers where there are many social connections within these groups, but few connections between these groups.  Further, imagine that brothel-based sex workers use condoms regularly, whereas street-based sex workers do not.  This situation will be problematic for RDS because the network ``bottleneck'' between the two groups will prevent the sample from exploring the entire population and could lead to inaccurate estimates about both sex worker type (i.e., brothel-based vs.\ street-based) and condom usage.

The standard method for detecting bottlenecks along a single trait is to create a cross-group recruitment table and calculate a measure referred to as ``homophily,'' which summarizes the tendency for respondents to recruit people who have the same trait as themselves~\citep{heckathorn_respondent-driven_2002}.  However, this approach can be misleading because bottlenecks anywhere in the network can cause problems for estimates, even if the bottlenecks are not primarily based on the trait being estimated~\citep{goel_respondent-driven_2009}.  Returning to the example above, even though there might be little homophily by condom usage, the bottleneck between street-based and brothel-based sex workers still degrades the estimates of condom usage.  

To detect bottlenecks we propose a more holistic approach that uses the different recruitment trees originating at each seed as a type of natural experiment.  Roughly, we ask if the trees seem to be getting stuck in distinct communities.  We assess this visually by creating \emph{Bottleneck Plots} that show the dynamics of the estimates from each seed individually.  For example, Fig.~\ref{fig:bottleneck_plot_bad} shows a Bottleneck Plot for the estimated proportion of MSM in Santo Domingo that use drugs every day.  The plot reveals that none of the seven seeds use drugs (see left panel of plot) and that the sample was dominated by three large trees, one of which produced a completely different estimate.  This pattern suggests that the MSM population in Santo Domingo consists of several communities with different drug use behaviors.  On the other hand, Fig.~\ref{fig:bottleneck_plot_good} shows a Bottleneck Plot for the estimated proportion of MSM in Barahona that are employed.  In this case there are three large trees, each of which produces estimates that were similar to the overall estimate of about 70\% employment rate.  This suggests that there are not large communities where employment is either extremely common or extremely rare.  

\begin{figure}
  \centering
  \subfigure[MSM-SD, Use drugs everyday]{
   \label{fig:bottleneck_plot_bad} 
   \includegraphics[width=0.4\textwidth]{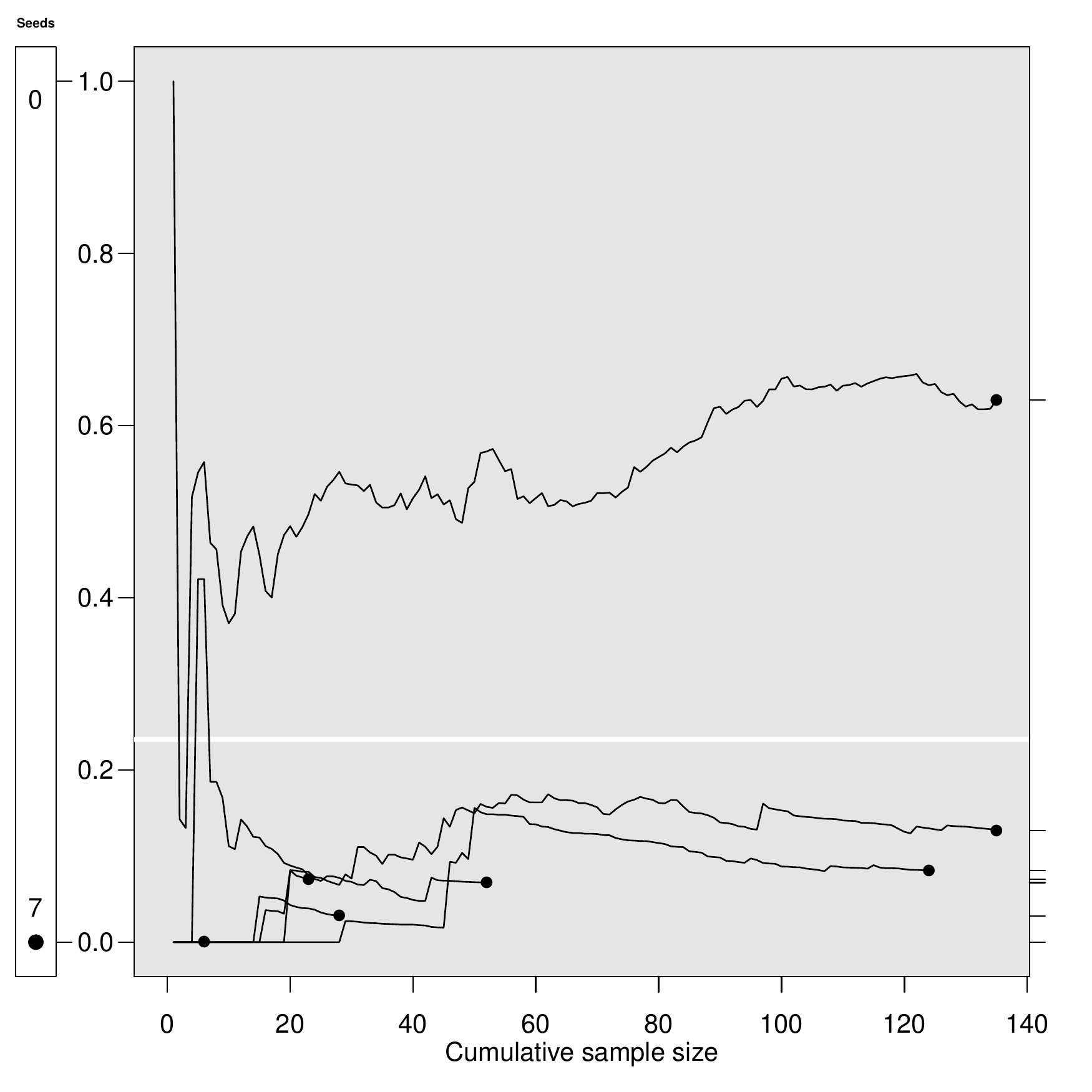}}
  \hspace{0.25in}
  \subfigure[MSM-BA, Employed]{
  \label{fig:bottleneck_plot_good} 
  \includegraphics[width=0.4\textwidth]{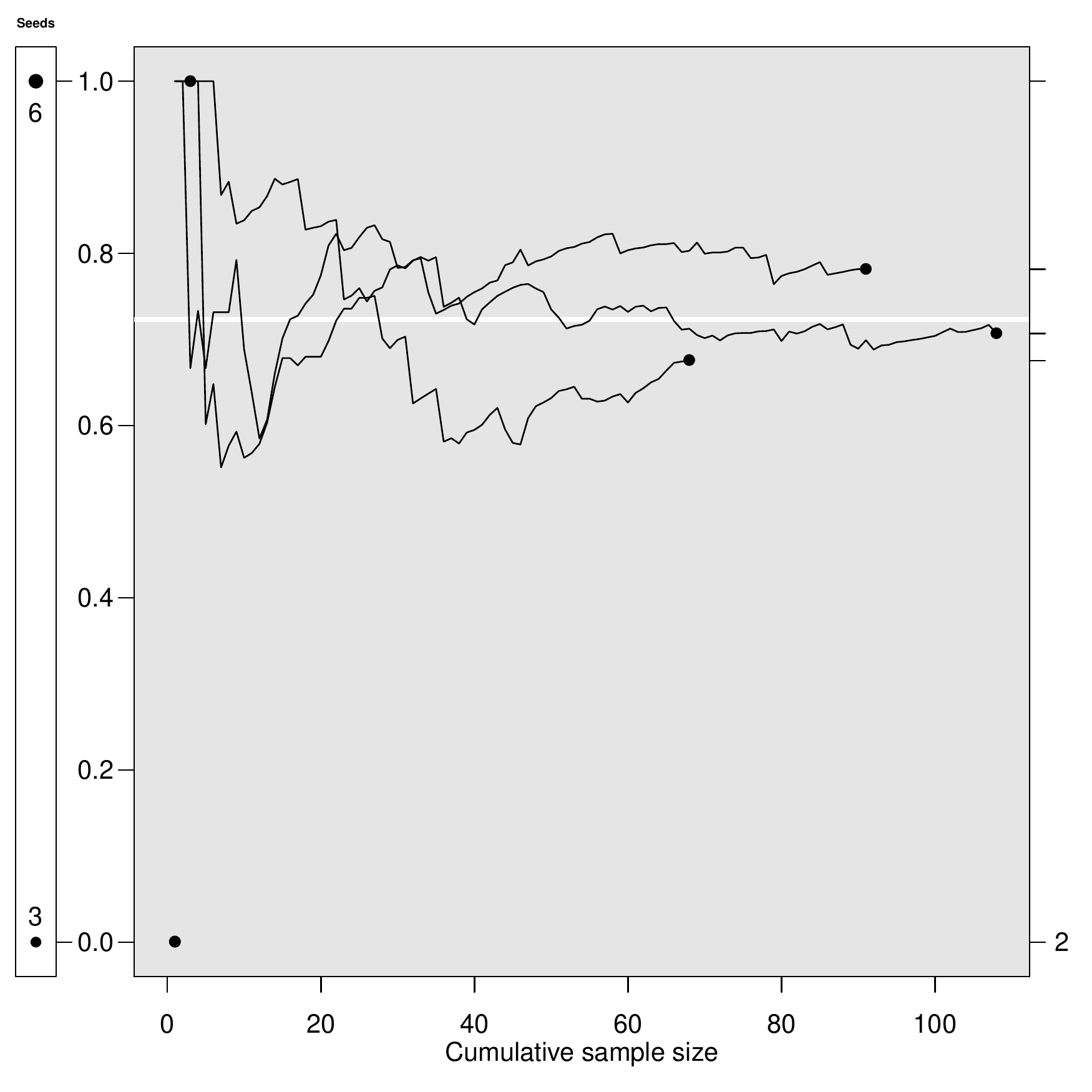}}
  \subfigure[MSM-BA, HIV+]{
  \label{fig:bottleneck_msm_ba_hiv+} 
  \includegraphics[width=0.4\textwidth]{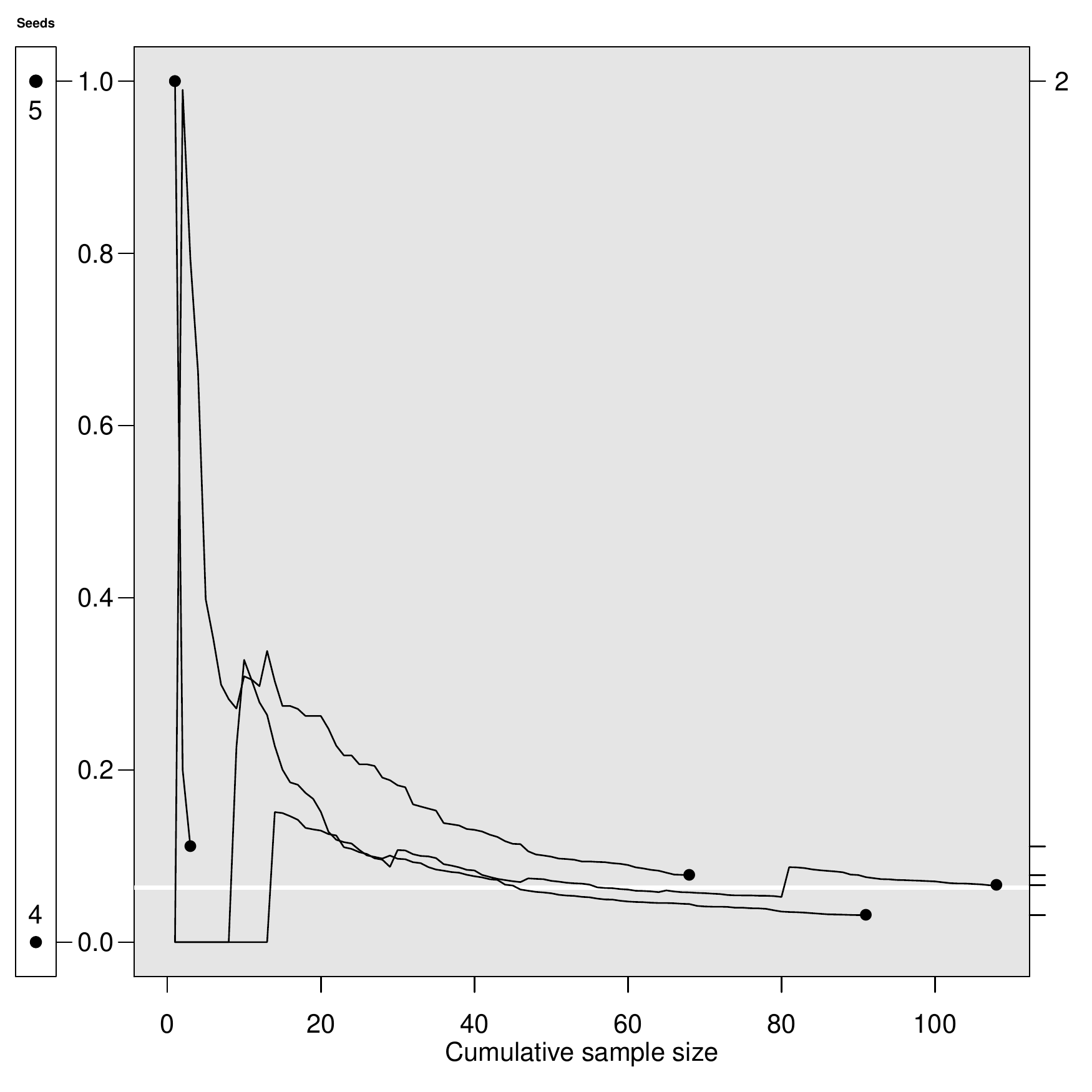}}
  \hspace{0.25in}
  \subfigure[MSM-SA, HIV+]{
  \label{fig:bottleneck_msm_sa_hiv+} 
  \includegraphics[width=0.4\textwidth]{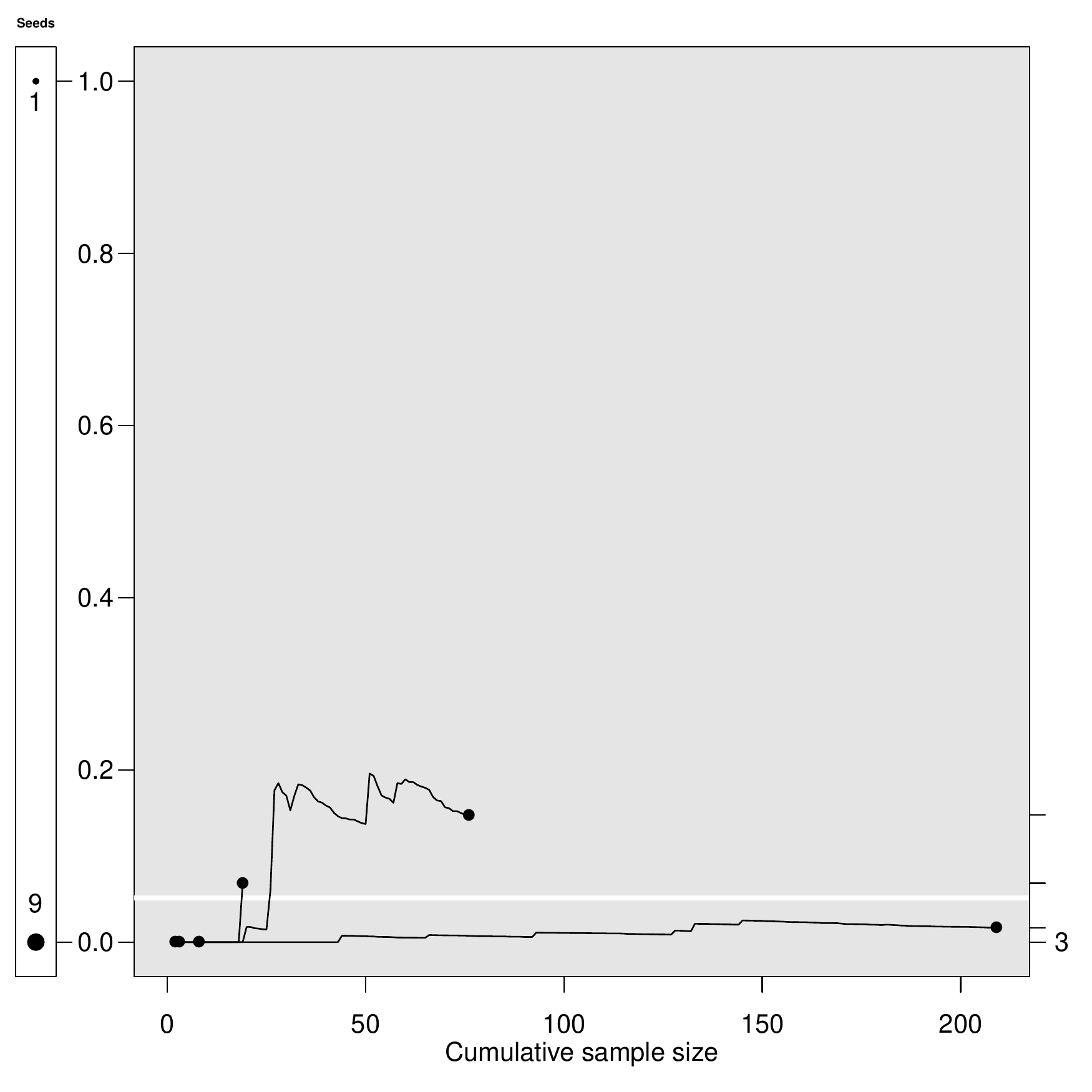}}
  \caption{{\it Bottleneck Plots:}  The left panel in each plot reports the composition of the seeds and the tick marks on the right axis show the final estimates.  If there is more than one tree with the same final estimate, that number is also shown on the right axis (see (c) and (d)).}
     \label{fig:bottleneck_plots} 
\end{figure}

Again, we recommend visual inspection of plots.  In order to aid inspection, researchers may also consider a weighted squared deviation:
\begin{equation}
WSD = \sum_s n_s \cdot (\widehat{p}_{s} - \widehat{p})^2
\label{eq:test_statistic}
\end{equation}
where $\widehat{p}_{s}$ is the estimate from the tree originating at seed $s$ and $n_s$ is the size of the sample resulting from seed $s$ (not including the seed itself and not including cases with missing data on the trait of interest or degree).  In order to assess whether this statistic is unusual, we perform a permutation procedure where the chain lengths and weights within the chain are fixed, but the traits are permuted.  We then calculate the WSD for the permuted data, and we repeat this procedure 10,000 times.  We flag a trait for further investigation if the observed WSD is greater than 90\% of the permuted WSD values; this threshold can be adjusted for desired sensitivity.
 
We run this procedure on the same 120 group $\times$ trait $\times$ city combinations examined in Section~\ref{sec:detecting_convergence} and found that the rates of flagging were highest among FSW (41\%) followed by MSM (30\%) and then DU (23\%) (Fig.~\ref{fig:bottleneck_plots_results}).  Although no trait was flagged in all four cities, these results suggest that likely sources of bottlenecks for FSW are based on sources of clients (e.g., brothel vs.\ street), drug use, and disease status (HIV and Syphilis); for DU based on type of drug used (Marijuana), employment status, and gender; and for MSM based on self-identification (e.g., bi-sexual and transsexual).  These results also suggest that bottlenecks can occur across traits that are not visible to respondents (e.g., disease status) possibly because these traits are correlated with other traits (e.g., age or risky behavior) that do affect social tie formation.  Finally, it is important to note that some study populations (e.g., MSM is Santo Domingo) appear to have bottlenecks along many traits.  

\begin{figure}
  \centering
   \includegraphics[width=\textwidth]{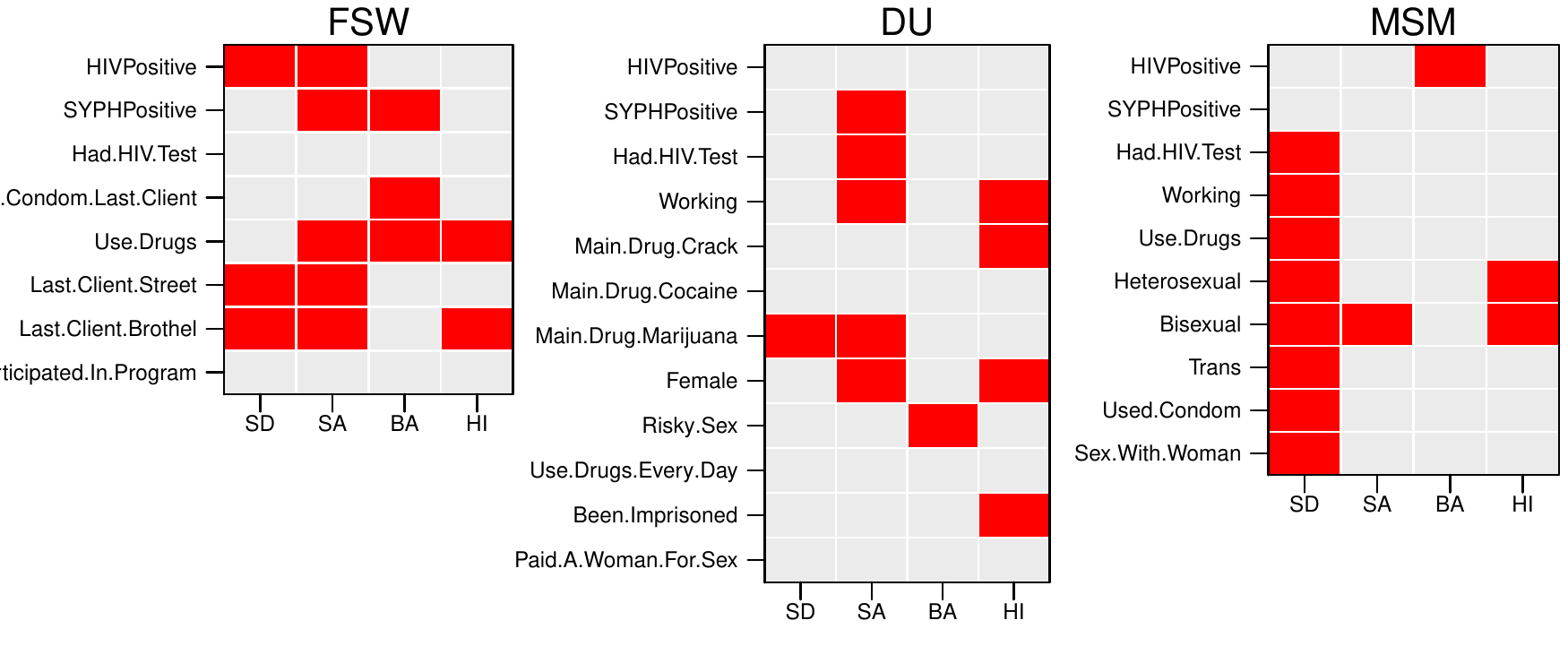}
   \caption{Bottleneck Plots test results. Red cells represent traits flagged for possible bottlenecks. 
   }
   \label{fig:bottleneck_plots_results} 
\end{figure}

\subsection{Current Recommendations}

In addition to looking for bottlenecks during formative research~\citep{simic_exploring_2006, johnston_formative_2010}, we recommend creating Bottleneck Plots for all traits of interest during data collection and conducting the permutation procedure on the weighted squared deviation (Equation~\ref{eq:test_statistic}) whenever there are too many plots to examine by hand.  Evidence of bottlenecks (e.g., Fig.~\ref{fig:bottleneck_plot_bad}) should be taken as an indication that estimates may be unstable and that more data should be collected.  If additional data collection is not possible, researchers should consider presenting estimates for each tree individually rather than trying to combine them into an overall estimate and researchers should be aware that standard RDS confidence intervals will be too small~\citep{salganik_variance_2006, goel_assessing_2010}.

Four caveats are needed when interpreting Bottleneck Plots.  First, the results of our flagging procedure may not always match the intuition of experienced RDS researchers.  For example, our procedure flags HIV status for MSM in Barahona (Fig.~\ref{fig:bottleneck_msm_ba_hiv+}) although this is caused by two chains of length 1, and is therefore probably not cause for concern.  On the other hand, our procedure does not flag HIV status for MSM in Santiago (Fig.~\ref{fig:bottleneck_msm_sa_hiv+}) even though a review of the plot seems to call for further investigation into the difference between the long chain with approximately 15\% estimated prevalence to the other chains with close to zero estimated prevalence.  Second, as with many of the other diagnostics proposed in this paper, lack of evidence of a problem does not mean that a problem does not exist.  For example, if there is a strong bottleneck between brothel-based and street-based sex workers and all the seeds are brothel-based, the sample may never include street-based sex workers and the Bottleneck Plots would not be able to alert researchers to this problem.  Finally, the statistical properties of this approach are currently unknown because of the unknown dependence structure in these data; we recommend, therefore, considering this flagging procedure as a useful heuristic rather than a formal statistical test.

\section{Reciprocation}
\label{sec:reciprocation}

Most current RDS estimators use self-reported degree to estimate sampling probabilities based on the assumption that all ties are reciprocated.  Current best practice monitors this feature by asking respondents during their initial visit about their relationship with the person who recruited them,  typically choosing from a set of categories (e.g., acquaintance, friend, sex partner, spouse, other relative, stranger, or other)~\citep{heckathorn_respondent-driven_2002}.  Here, we present responses to a slightly different question, and in the Supporting Information (Section \ref{sec:reciprocationA}), we present further discussion of additional approaches aimed at studying the reciprocation patterns in the broader social network.

On the follow-up questionnaire, for each coupon given out, respondents were asked:  
\bq
\item Do you think that the person to whom you gave a coupon would have given you a coupon if you had not participated in the study first?
\label{ques:recip}
\eq

\begin{table}[ht]\caption{Percent of affirmative responses to the question \ref{ques:recip}, ``Do you think that the person to whom you gave a coupon would have given you a coupon if you had not participated in the study first?''}
\begin{center}
\begin{tabular}{lcccccccccccc}
& \multicolumn{4}{c}{FSW} & \multicolumn{4}{c}{DU} & \multicolumn{4}{c}{MSM}\\
\cmidrule(rl){2-5} 
\cmidrule(rl){6-9}
\cmidrule(rl){10-13}
 & SD & SA & BA & HI & SD & SA & BA & HI & SD & SA & BA & HI\\
\hline
Percent  Reciprocated & 87 & 98 & 87 & 89 & 86 & 96 & 74 & 79 & 87 & 98 & 91 & 91\\
\end{tabular}
 \label{tab:pctsrecipgive}
\end{center}
\end{table}

Table~\ref{tab:pctsrecipgive} shows the results of this question, separated by population and site.  Overall, about 88\% of responses indicated reciprocation,\knote{is there newer work - Whipple? - showing non-reciprocation at lower rates causes problems?} but there are notable differences across the populations and sites.  Reciprocation rates in Santiago were considerably higher than the other cities, and reciprocation rates of DU were lower than other populations.  The reciprocation rates among DU were especially low in Higuey, and also in Barahona, where participants may have been selling coupons (for more on coupon-selling, see~\citet{scott_they_2008}, also~\citet{broadhead_notes_2008, ouellet_cautionary_2008}).  

\subsection{Current Recommendations}

The reciprocity assumption requires both that the recruiter and recruit are known to each other and that both people would be willing to recruit each other.  Therefore, we recommend that on the initial survey researchers should collect information about the relationship between the recruiter and recruit (see e.g.,~\citet{heckathorn_respondent-driven_2002}) and information directly assessing the possibility of recruitment (similar to question \ref{ques:recip}).  Researchers should calculate reciprocity rates as defined by both questions during data collection.  Low rates of reciprocation by either measure could be used to improve field procedures (e.g., training respondents about how to recruit others) and alert researchers to potential problems (e.g., coupon-selling).  Further, high-rates of non-reciprocation may require alternative RDS estimators.  See \citet{lu_respondent-driven_2012} for one such approach.

\section{Measurement of Degree}
\label{sec:degree}

The Volz-Heckathron estimator (VH, \cite{volz_probability_2008}) weights respondents based on their self-reported degree (see Equation~\ref{vhest}).  The fact that the estimates can depend critically on self-reported degree has troubled RDS researchers~\citep{frost_respondent-driven_2006, wejnert_empirical_2009, iguchi_simultaneous_2009, goel_respondent-driven_2009, bengtsson_global_2010} because of the well-documented problems with self-reported social network data in general~\citep{bernard_problem_1984, marsden_network_1990, brewer_forgetting_2000}.  However, despite the widespread concern about degree measurement, the issue is rarely explored empirically in RDS studies (for important exception, see~\citet{wejnert_web-based_2008, wejnert_empirical_2009, mccreesh_evaluation_2012}).  Here we present several methods of assessing the measurement of degree and the resulting effects on estimates.

In this study, respondents were asked a series of four questions to measure degree~\citep{johnston_implementation_2008} (DU versions, others analogous)::

\bq
\item How many people do you know who have used illegal drugs in the past three months?\label{sizeq1}
\item How many of them live or work in this province? \label{sizeq2}
\item How many of them [repeat response from \ref{sizeq2}] are 15 years old or older? \label{sizeq3}
\item How many of them [repeat response from \ref{sizeq3}] have you seen in the past week? \label{sizeq4}
\eq

The response to the fourth question (\ref{sizeq4}) was the degree used for estimation.  Respondents were also asked:

\bq
\item If we were to give you as many coupons as you wanted, how many of these drug users\footnotemark[\value{footnote}] (repeat the number in \ref{sizeq3}) do you think you could give a coupon to by this time tomorrow? \label{ncupday}
\item If we were to give you as many coupons as you wanted, how many of these drug users\footnotemark[\value{footnote}] (repeat the number in \ref{sizeq3}) do you think you could give a coupon to by this time next week?  \label{ncupweek}
\eq

During the follow-up visit, the series of four main degree questions (\ref{sizeq1}, \ref{sizeq2}, \ref{sizeq3}, and \ref{sizeq4}) was repeated, and respondents were also asked how quickly they distributed each of their coupons.  We use these responses, along with data on the number of days between recruiter and recruit interviews to evaluate three features of the degree question: validity of the ``one week'' time frame used in question (\ref{sizeq4}), test-retest reliability of responses, and the possible effect of inconsistent reporting on estimates. 

\subsection{Validity of Time Window}
\label{sec:degree_time_window}

A time frame of one week was used in the key degree question (\ref{sizeq4}) because that was thought to best approximate the probability that a respondent would be selected, based on previous experience with RDS.  We looked to the data for information on the validity of this time window.  
In the Supporting Information, therefore, we provide detailed examination of recruitment time dynamics and conclude that: respondents reported that a high proportion (92\%) of their alters could be reached within one week (Fig.~\ref{fig:propgiveboth}),  respondents reported distributing most (95\%) of their coupons within one week (Fig.~\ref{fig:daysgiven}), and the number of days between the interview of the recruiter and recruit was usually less than one week (79\% of the time, Fig.~\ref{fig:bargapdata}).  We conclude, therefore, that for these studies, the one-week time window was reasonable.

\subsection{Test-retest reliability}
\label{sec:degree_test_retest}

For participants who returned to the survey site for a second time (about half the participants, see Fig.~\ref{fig:samplesizes}), we have a measure of the consistency, but not accuracy, of their degree responses.  The median difference between degree at the initial and follow-up visits was 0 (Fig.~\ref{fig:12sites_delta_degree_box}) suggesting that there was nothing systematic about the two visits that led to different answers on the questionnaire (e.g., different location, different length of interview, etc.).  However, the responses of many individuals differed, in some cases substantially.  The association between the measurements is affected by a small number of outliers, so we use the more robust Spearman's rank correlation to measure the association between the visits.  The rank correlations range from 0.17 to 0.47 with a median correlation for FSW of 0.33 and a median correlation for DU and MSM of 0.41 (Fig.~\ref{fig:12sites_spearman_P204}). As expected, the reliability of the degree question was relatively low.

\subsection{Effect on estimates}
\label{sec:estimates}

Finally, we studied the robustness of our estimates by calculating disease prevalence estimates using degree as measured in the initial and follow-up interviews.  Fig.~\ref{fig:12sites_4disease_2estimates_2column_degreebothtimes} shows estimates for people who participated in both interviews, thus ensuring that we compare the same groups of respondents.  The differences in disease prevalence estimates are generally small in an absolute sense, ranging from 0 to 0.08 (8\%) with a median difference of 0.01.  When broken down by disease, HIV had the largest median absolute difference, 0.031, followed by Syphilis, 0.017.  Hepatitis B and C had median absolute differences in prevalence of essentially 0, possibly driven by the fact that these diseases are very rare in these populations.   Note that these differences probably slightly over-estimate the sensitivity of RDS estimates, as these estimates are restricted to respondents who completed both surveys.

\begin{figure}
  \centering
   \includegraphics[width=\textwidth]{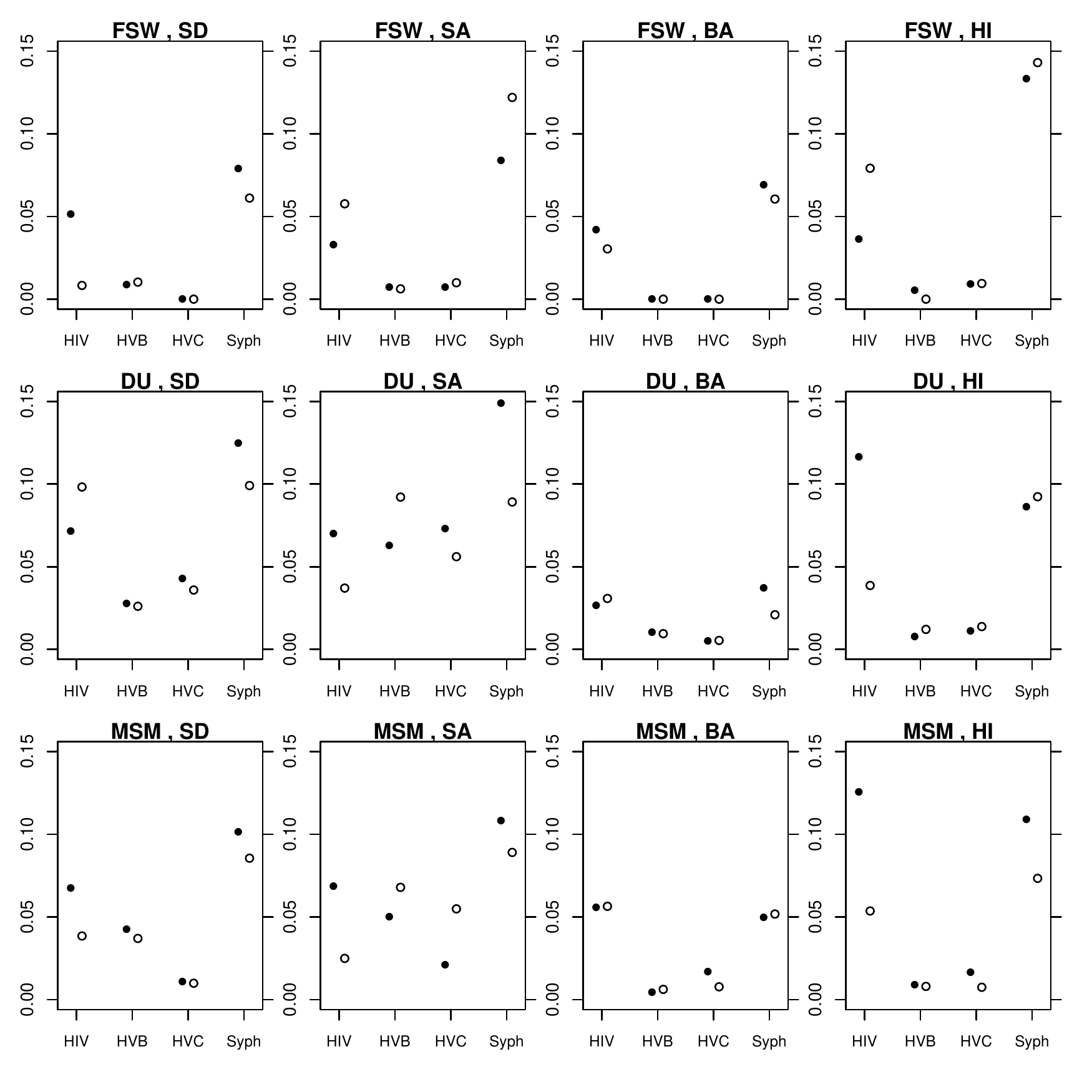}
   \caption{Disease prevalence estimates from 12 studies for 4 diseases using Question \ref{sizeq4} (solid circle) and \ref{sizeq4} at follow-up (hollow circle).  The plot includes only people who participated in test and retest (see Fig.~\ref{fig:samplesizes} for sample sizes).}
   \label{fig:12sites_4disease_2estimates_2column_degreebothtimes} 
\end{figure}

In addition to comparing these differences in absolute units, we also consider the differences in relative units, $(\mid \hat{p} - \hat{p}^\prime \mid) / \hat{p}$.  The difference between the two estimates is more than 50\% of the original estimate in about a quarter of the cases.  Nevertheless, in public health disease surveillance, an estimated increase in disease prevalence of 50\% is likely cause for concern, even if the estimated prevalences themselves were quite low.  These data show that measurement error with respect to degree could introduce a change this large when prevalence is low.  

\subsection{Current Recommendations}

When collecting data, the time period used to elicit self-reported degree should be reflective of the time in which coupons are likely to be distributed.  These results suggest that the one-week period used in these studies was reasonable, but this should be checked in future studies with different populations.  We also recommend that researchers collect degree at both the initial and follow-up visits to assess test-retest reliability.  Further, when making estimates, we recommend that researchers compare the unweighted sample mean to the estimates using each of the degree questions in the study (as was done in~\citet{wejnert_empirical_2009} and~\citet{mccreesh_evaluation_2012}).  To the extent that these estimates are similar, one might be less concerned about the measurement of degree; however, we emphasize that consistency of estimates does not ensure accuracy of estimates.  Finally, in future studies, when considering which measure of degree to use, it is important to recall these measures are being used to approximate the relative probability of selection of respondents.  To the extent that there are other things about a respondent, such as social class or geographic location, that make him or her more or less likely to participate, the probability of selection is no longer proportional to degree.  Any other features thought to be related to the probability of participation should also be collected.

\section{Participation Bias}
\label{sec:participation}

RDS estimation relies on the assumption that recruits represent a simple random sample from the contacts of each recruiter. Limited ethnographic evidence, however, suggests that recruitment decisions can be substantially more complex than is assumed in standard RDS statistical models~\citep{scott_they_2008, ouellet_cautionary_2008, broadhead_notes_2008, bengtsson_global_2010, kerr_selective_2011, mccreesh_evaluation_2012}.  For example, a study of MSM in Brazil found that some people tended to recruit their riskiest friends because they were thought to need safe sex counseling~\citep{mello_assessment_2008}.  Further, the same study found that some MSM refused to participate when recruited because they were worried about revealing their sexual orientation.  Such selective recruitment and participation could lead to non-response bias.  

We find it helpful to consider the process of a new person entering the sample as the product of three decisions:
\begin{my_enumerate}
\item Decision by recruiter to pass coupons (how many and to whom)
\item Decision by recruit to accept coupon
\item Decision by recruit to participate in study given that they have accepted a coupon
\end{my_enumerate}
Biases at any of these steps could result in systematic over or under representation of certain subgroups in the sample, resulting in biased estimates. We assess these possible biases in four ways.  The first two, recruitment effectiveness and recruitment bias, address the cumulative effects of all three decisions on the quantity and characteristics of recruits; the third addresses two forms of non-response corresponding to steps (2) and (3); and the final analysis examines a respondent's motivation for participation, related to their decisions to accept a coupon and participate in the study.

\subsection{Recruitment Effectiveness}

Systematic differences in recruitment effectiveness can lead to biased estimates under some conditions~\citep{tomas_effect_2011}.  For example, if respondents with HIV have systematically more recruits and are also more likely to have contact with others with HIV, then people with HIV will be over-represented in the sample.  In Fig.~\ref{fig:hivkids}, we present the mean numbers of recruits by HIV status for each site.  We call this plot a {\it Recruitment Effectiveness Plot}.  In a single study, paired bars might represent differential recruitment effectiveness by many traits.  In these 12 studies, the most dramatic difference is among FSW in Higuey, where respondents with HIV recruit at only half the rate of those without HIV.  

\begin{figure}
  \centering
     \includegraphics[width=\textwidth]{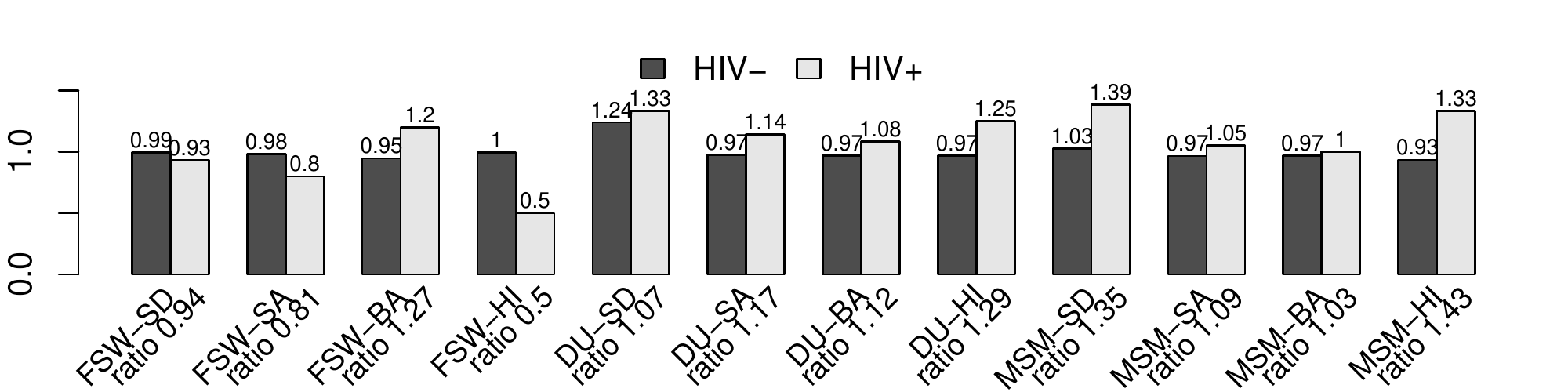}
   \caption{{\it Recruitment Effectiveness Plot:}  Average recruits for HIV+ and HIV- respondents by site.  The ratio is provided under the bars.  Differential recruitment effectiveness can lead to bias in the RDS estimates under some conditions~\citep{tomas_effect_2011}.}
  \label{fig:hivkids} 
\end{figure}

\subsection{Recruitment Bias}

Recruitment bias---when a respondent's contacts have unequal probabilities of selection---can result in a pool of recruits that is systematically different from the pool of contacts of respondents.  Because existing inferential methods assume recruits are a simple random sample from among contacts, these systematic differences may bias resulting estimates \citep{tomas_effect_2011, gile_respondent-driven_2010}.  To examine the effects of such biases on the sample composition of a specific trait, employment status, we introduced the following questions in the DU questionnaires:
\bq
  \item  
  How many of them (repeat number of contacts in Question \ref{sizeq3}) are currently working? \label{empl1}
  \item 
  (follow-up questionnaire):  Do the persons to whom you gave the coupons have work? (asked separately for each of 1 to 3 persons).\label{empl2}
  \item 
  Are you actually working? (we consider responses given by recruits of each respondent.)\label{empl3}
\eq
Overall, then, these questions, in order, should measure the employment characteristics of the pool of potential recruits, the employment characteristics of those who were chosen for referral by the respondents and accepted coupons, and the employment characteristics of those who then chose to return the coupons and enroll in the study.  The difference between the characteristics reported in the first (\ref{empl1}) and second (\ref{empl2}) questions reflect the joint effects of the decisions to pass and accept coupons, while the difference in characteristics between the second (\ref{empl2}) and third (\ref{empl3}) questions reflect the effect of the decision to participate in the interview.  

\begin{figure}
  \centering
   \includegraphics[width=3in]{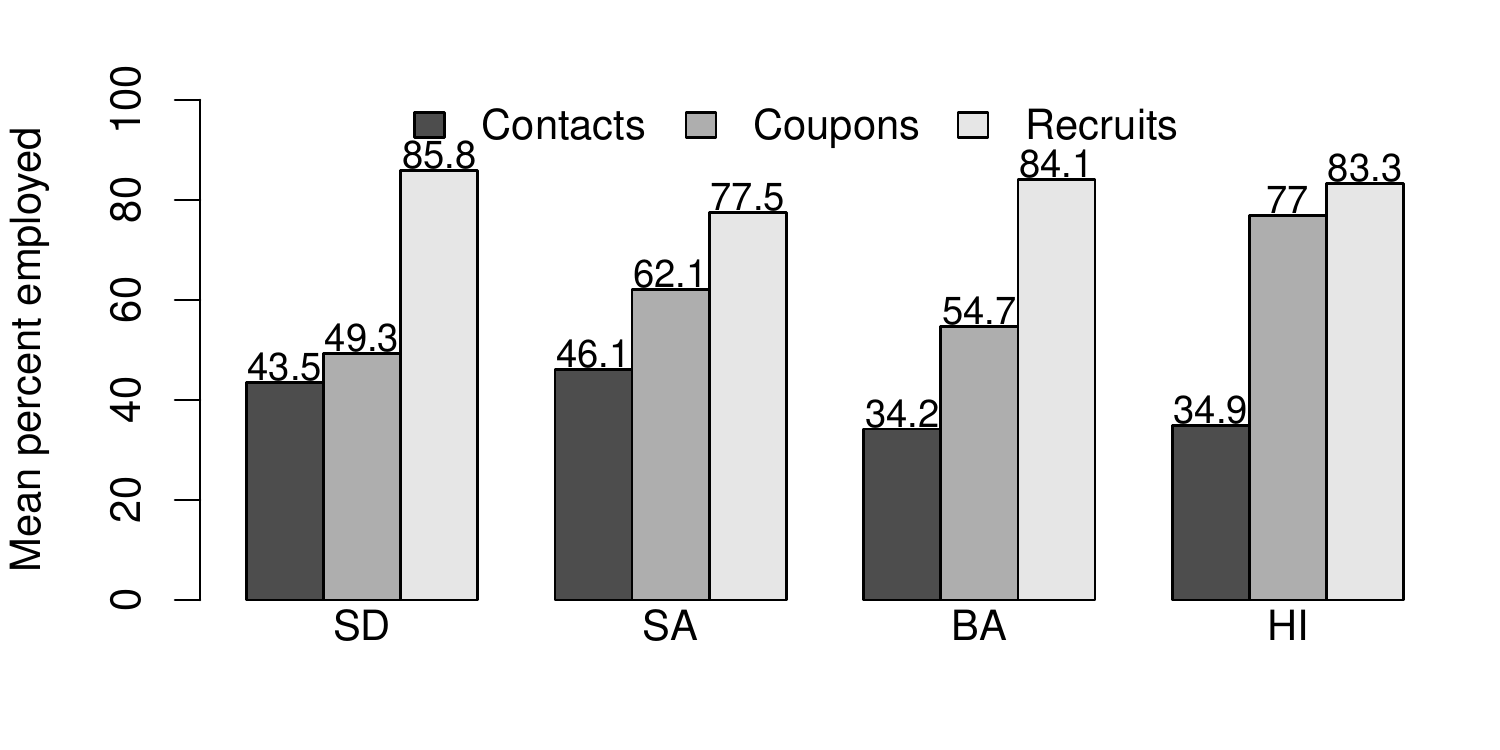}
   \caption{{\it Recruitment Bias Plot:}  Percent employed by location and question.}
   \label{fig:refbiasplot} 
\end{figure}  

Fig.~\ref{fig:refbiasplot}, which we call a {\it Recruitment Bias Plot} provides a summary of the responses to these questions.  This plot compares the composition of comparable sets of respondents' social contacts, of coupon recipients, and of recruits.  To do this, we restrict analysis to the set $S$ of recruiters with data available on all three levels, and then calculate the average percent of contacts, coupon recipients, and recruits who are employed as follows:
\[
\frac{1}{|S|}\sum_{i \in S} \frac{\ref{empl1}_i}{\ref{sizeq3}_i}, ~~~~~ \frac{1}{|S|} \sum_{i \in S} \frac{\sum_{j=1}^{n^c_i} \ref{empl2}_{ij}}{\sum_{j=1}^{n^c_i} 1},   ~~~~\textrm{and} ~~~~ \frac{1}{|S|}\sum_{i \in S} \frac{\sum_{j \in Recruits(i)} \ref{empl3}_j}{\sum_{j \in Recruits(i)} 1}, 
\]
where $\ref{sizeq3}_i, \ref{empl1}_i,$ and $\ref{empl3}_i$ refer to respondent $i$'s response to questions $\ref{sizeq3}, \ref{empl1}$, and $\ref{empl3}$, $\ref{empl2}_{ij}$ is a binary indicator of $i$'s report of the employment status of the person receiving his or her j$^{th}$ coupon, and $n^c_i$ is the number of coupons $i$ reported distributing. 

In every site, there is a marked increase in the reported rate of employment for each stage in the referral process.  These data suggest that respondents distributing coupons are more likely to give them to those among their contacts who are employed, and that among those receiving coupons, those who are employed are more likely to return them.

These results are a provocative suggestion of aberrant respondent behavior, and could belie a dramatic over-sampling of employed DU.  These particular results, however, should be seen in light of other possible explanations, in particular the possibility of survey response bias.  The succession of questions, reflecting increased proportions of reported employment, also correspond to increasing social closeness to the respondent.  Because it is possible ``having work'' is a desirable status, a response bias based on social desirability would also explain the results in this section.  

Other researchers (e.g., \citet{heckathorn_extensions_2002}, \citet{wang_respondent-driven_2005}, \citet{wejnert_web-based_2008}, \citet{iguchi_simultaneous_2009}, and \citet{rudolph_subpopulations_2011}) have introduced and used statistical tests assessing the assumption of random recruitment.  We address these, and introduce a new test, in the Supporting Information.

\subsection{Non-Response}

Non-response, where intended respondents do not participate, is a problem in most surveys.  If non-responders differ systematically from responders, estimates will suffer from non-response bias.  Non-response and non-response bias are particularly challenging to measure in RDS studies because non-responders are contacted by other participants rather than by researchers and because non-response can arise in two ways---by refusing a coupon or by failing to return the coupon to participate in the study.

In order to better understand non-response, respondents were asked during the follow-up interview:
\bq
\item How many coupons did you distribute?\label{ques:numcup}
\item How many people did not accept a coupon you offered to them?\label{ques:numrefuse}
\eq

We estimate the {\it Coupon-Refusal Rate} by comparing responses to question (\ref{ques:numcup}) and question (\ref{ques:numrefuse}), and we estimate the {\it Non-Return Rate} by comparing responses to question (\ref{ques:numcup}) to the number of survey participants presenting coupons from each respondent.  Finally, comparing the number of respondents to the number of attempted eligible coupon-distributions (refused and distributed) we estimate the {\it Total Non-Response Rate}.  Specifically, these rates are respectively computed as follows:
\[
\frac{\sum_{i \in S} n_i^r}{\sum_{i \in S} n_i^r + \sum_{i \in S} n_i^c} ~~~~~ 1-\frac{\sum_{i \in S} \vert Recruits(i) \vert}{\sum_{i \in S} n_i^c} ~~~~~ 1- \frac{\sum_{i \in S} \vert Recruits(i) \vert}{\sum_{i \in S} n_i^r + \sum_{i \in S} n_i^c},
\]
where $S$ is again restricted to those with data on all relevant questions, $n_i^c$ is the number of coupons distributed by $i$, $n_i^r$ is the number of refused coupons reported by $i$, and $\vert Recruits(i) \vert$ represents the number of successful recruits of $i$.  All three rates are summarized in Table \ref{tab:refrates}.  Coupon Refusal rates ranged from more than 50\% (FSW-SD) to almost none (DU-SD), and the Total Non-Response rates varied from  62.3\% (FSW-SD) to 26.3\% (DU-SD).  For comparison, the University of Michigan's Consumer Survey of Attitudes, a high quality telephone survey, has a Refusal Rate of about 20\%~\citep{curtin_changes_2005}.  It is more difficult to compare the RDS Non-Response Rate to the non-response rate of traditional surveys because the ``non-contact rate'' is not clearly defined for RDS (AAPOR, 2011)\nocite{the_american_association_of_public_opinion_researchers_standard_2011}.   

\renewcommand{\tabcolsep}{3pt}

\addtocounter{footnote}{1}
\begin{table}[h]\caption{{\it RDS Non-Response Rates.}  Coupon refusal rate is the total number of reported coupon refusals to eligible alters divided by that number plus the number of reported coupons distributed.  Coupon Non-return is the percent of coupons that were not returned (among accepted coupons). Total Non-Response rate is the percent of attempted recruitments of eligible alters not resulting in survey participation.  All computed based only on recruits of respondents completing the return survey.}
\begin{center}
\begin{small}
\begin{tabular}{lrrrrrrrrrrrr}
& \multicolumn{4}{c}{FSW} & \multicolumn{4}{c}{DU} & \multicolumn{4}{c}{MSM}\\
\cmidrule(rl){2-5} 
\cmidrule(rl){6-9}
\cmidrule(rl){10-13}
Rate & SD & SA & BA & HI & SD & SA & BA & HI & SD & SA & BA & HI\\
\hline
Coupon Refusal &56.5 & 45.3 & 7.5 & 28.0 & 0.4 & 15.9 & 11.3 & 41.3 & 7.7 & 16.5 & 25.4 & 29.2 \\ 
Non-Return & 13.4 & 43.9 & 43.0 & 41.4 & 26.1 & 35.3 & 44.6 & 33.9 & 29.4 & 23.6 & 39.7 & 31.9 \\
\hline
  Total Non-Response & 62.3 & 69.3 & 47.2 & 57.8 & 26.3 & 45.6 & 50.9 & 61.2 & 34.8 & 36.3 & 55.0 & 51.8 \\
\hline
  Number of Recruiters & 123 & 136 & 141 & 151 & 126 & 105 & 164 & 141 & 153 & 128 & 152 & 102 \\
\end{tabular}
\end{small}
\end{center}\label{tab:refrates}
\end{table}

\renewcommand{\tabcolsep}{6pt}

To know whether this non-response could induce non-response bias, we would need to know if the people who refused were different than those who participated.  We could not collect information about non-responders directly so we asked recruiters why their non-respondents had refused coupons, as has been done in previous studies~\citep{stormer_analysis_2006, johnston_effectiveness_2008, iguchi_simultaneous_2009}.  For each of up to 5 refusals, the return survey asked:
\bq
\item What is the principal reason why these persons did not accept a coupon?
\label{ques:whyrefuse}
\eq
Responses to (\ref{ques:whyrefuse}) are summarized in the {\it Coupon-Refusal Analysis} in Table \ref{tab:whyrefuse}.  The most common reason given for refusal was aversion to being identified as a member of the study population (26.6\%).  Many refusers also reported fear of test results (especially HIV test results:  16.3\%).  Some were ``uninterested'' (22.0\%).  Interestingly for study organizers, among the reasons for ``other,'' 5.2\% of MSM refusers reportedly did not trust the study or did not believe the incentive was true.  

\renewcommand{\tabcolsep}{3pt}

\begin{table}[h]\caption{{\it Coupon-Refusal Analysis:}  Responses to the question ``What is the principal reason why these persons did not accept a coupon?''}
\begin{center}
\begin{small}
\begin{tabular}{lrrrrrrrrrrrr}
& \multicolumn{4}{c}{FSW} & \multicolumn{4}{c}{DU} & \multicolumn{4}{c}{MSM}\\
\cmidrule(rl){2-5} 
\cmidrule(rl){6-9}
\cmidrule(rl){10-13}
Response & SD & SA & BA & HI & SD & SA & BA & HI & SD & SA & BA & HI\\
  \hline
Too Busy &7.3 & 80.0 & 10.0 & 0.8 & 0.0 & 10.3 & 0.0 & 3.0 & 0.0 & 30.8 & 4.5 & 12.1 \\
Fear being identified & 31.4 & 0.0 & 63.3 & 21.3 & 100.0 & 17.9 & 22.4 & 20.5 & 4.2 & 30.8 & 30.3 & 31.3 \\
Incentive low/location far & 3.6 & 0.0 & 0.0 & 2.5 & 0.0 & 2.6 & 0.0 & 1.8 & 0.0 & 0.0 & 0.8 & 1.0 \\
Not interested & 26.3 & 0.0 & 10.0 & 38.5 & 0.0 & 2.6 & 10.2 & 15.1 & 0.0 & 30.8 & 30.3 & 19.2 \\ 
Fear HIV/other results & 15.3 & 0.0 & 6.7 & 28.7 & 0.0 & 15.4 & 20.4 & 10.2 & 75.0 & 7.7 & 16.7 & 4.0 \\
Fear giving blood &0.7 & 0.0 & 0.0 & 0.8 & 0.0 & 33.3 & 0.0 & 22.9 & 0.0 & 0.0 & 0.0 & 7.1 \\ 
Fail Eligibility & 0.0 & 0.0 & 10.0 & 0.8 & 0.0 & 0.0 & 4.1 & 6.0 & 0.0 & 0.0 & 3.8 & 2.0 \\
Already got coupon & 1.5 & 20.0 & 0.0 & 0.0 & 0.0 & 0.0 & 2.0 & 0.6 & 0.0 & 0.0 & 10.6 & 0.0 \\
 Other & 13.9 & 0.0 & 0.0 & 6.6 & 0.0 & 17.9 & 40.8 & 19.9 & 20.8 & 0.0 & 3.0 & 23.2 \\ 
 \hline 
 Total Reasons Reported & 137 & 5 & 30 & 122 & 1 & 39 & 49 & 166 & 24 & 13 & 132 & 99 \\ 
\end{tabular}
\end{small}
\end{center}\label{tab:whyrefuse}
\end{table}

\renewcommand{\tabcolsep}{6pt}

\subsection{Decisions to Accept Coupon and Participate in Study}

In addition to exploring reasons for not participating in the study, we also asked about each respondent's reason for participating, as in~\citet{johnston_effectiveness_2008}:  
\bq
\item What is the principle reason why you decided to accept a coupon and participate in this study?
\label{ques:whyaccept}
\eq
Responses are reported in Table \ref{tab:whyparticipate}. In every site, a substantial majority reported participating in the interest of receiving HIV test results.  Further, we go beyond previous researchers and assess whether the motivation for participation is associated with important study outcomes.  For example, we found that the odds of having HIV among those who expressed motivation based on the HIV test was 0.43 (MSM-HI) to  2.03 (MSM-SA) times the odds for those who did not.  We summarize these results in the {\it Motivation-Outcome Plot} in Fig.~\ref{fig:oddsratioHIV}.  Similar relationships hold when analyses are restricted to those who have not had an HIV test in the last 3 months or last 6 months.  Again, the unknown dependence structure, does not allow for formal statistical testing, however, note that to the extent that the probability of participation is associated with participant motivation and participant motivation is associated with an outcome of interest, bias will be introduced into the estimates even if these associations are not statistically significant.

\renewcommand{\tabcolsep}{3pt}

\begin{table}[h]\caption{Responses to the question (\ref{ques:whyaccept}), ``What is the principle reason why you decided to accept a coupon and participate in this study?''  The ``Other'' category includes: ``I have free time'', ``To stop using'' (DU only), and ``Other.''}
\begin{center}
\begin{small}
\begin{tabular}{lrrrrrrrrrrrr}
& \multicolumn{4}{c}{FSW} & \multicolumn{4}{c}{DU} & \multicolumn{4}{c}{MSM}\\
\cmidrule(rl){2-5} 
\cmidrule(rl){6-9}
\cmidrule(rl){10-13}
Response & SD & SA & BA & HI & SD & SA & BA & HI & SD & SA & BA & HI\\
\hline
Incentive & 2.9 & 5.0 & 3.3 & 1.3 & 11.6 & 16.5 & 6.0 & 5.0 & 5.7 & 5.8 & 1.8 & 5.9 \\ 
For HIV test & 88.5 & 77.7 & 90.1 & 86.4 & 51.3 & 63.2 & 65.1 & 70.1 & 71.5 & 71.9 & 81.5 & 83.3 \\
Other/all test & 1.0 & 2.3 & 0.4 & 2.6 & 18.4 & 0.6 & 4.7 & 3.0 & 1.4 & 1.2 & 5.0 & 1.9 \\
Recruiter & 1.7 & 5.0 & 1.6 & 2.0 & 3.9 & 10.3 & 6.3 & 17.6 & 7.7 & 5.8 & 4.6 & 4.8 \\ 
Study interest & 4.6 & 10.0 & 4.1 & 7.3 & 10.0 & 8.7 & 17.9 & 4.3 & 11.3 & 14.7 & 5.0 & 3.7 \\
Other & 1.2 & 0.0 & 0.4 & 0.3 & 4.8 & 0.6 & 0.0 & 0.0 & 2.4 & 0.6 & 2.1 & 0.4 \\ 
\hline
 Total & 410 & 301 & 243 & 302 & 310 & 310 & 301 & 301 & 505 & 327 & 281 & 269 \\ 
\end{tabular}
\end{small}
\end{center}\label{tab:whyparticipate}
\end{table}

\renewcommand{\tabcolsep}{6pt}

\begin{figure}
  \centering
   \includegraphics[width=3in]{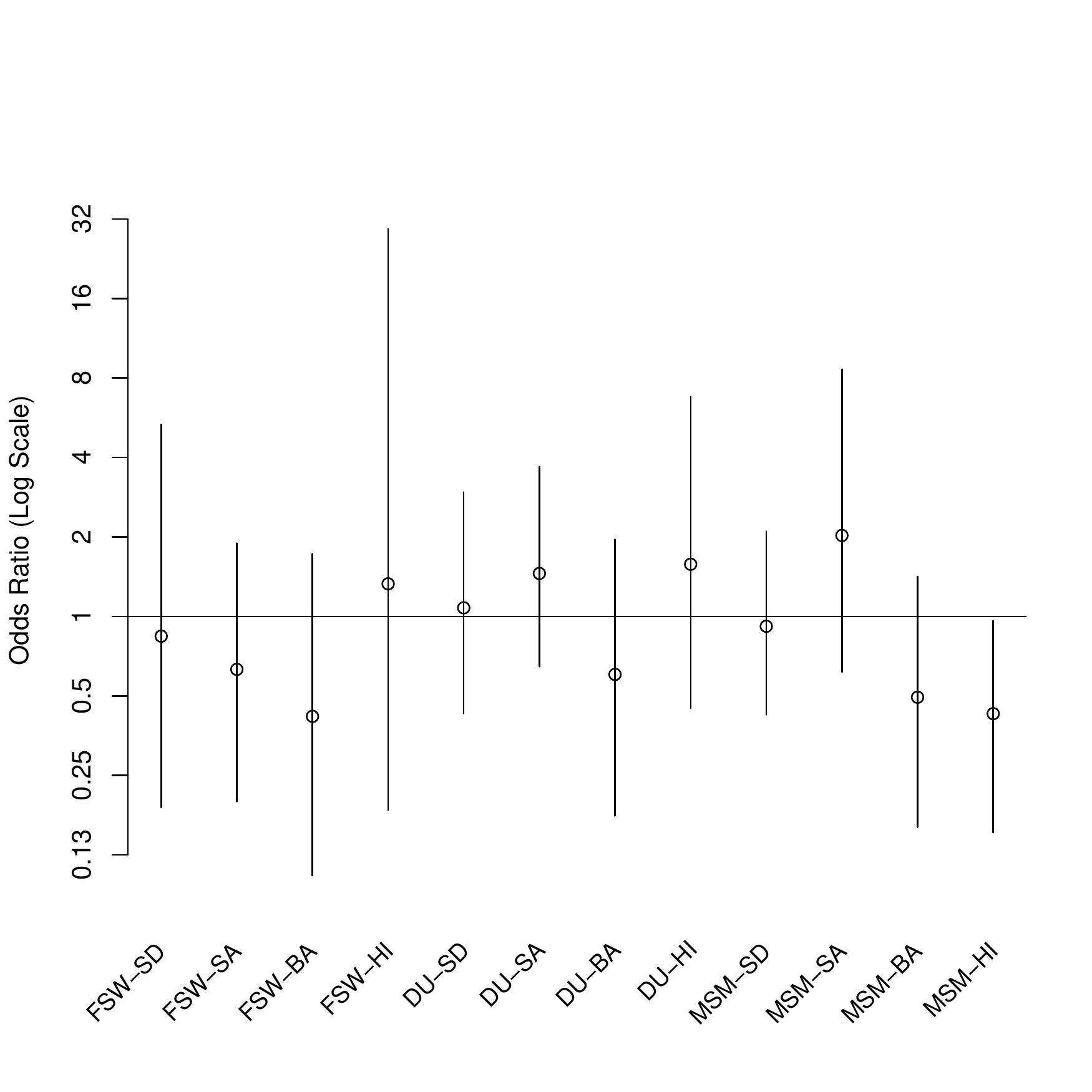}
   \vspace{-.2in}
   \caption{{\it Motivation-Outcome Plot:} Odds ratios of having HIV given HIV test motivation for study participation.  Ratios greater than 1 indicate those participating for HIV test results more likely to have HIV.  For reference, nominal 95\% intervals are based on the inversion of Fisher's exact test (these would be confidence intervals if the data were independent identically distributed).} 
   \label{fig:oddsratioHIV} 
\end{figure}

\subsection{Current Recommendations}

The approaches in this section do not directly indicate the extent to which estimates may be impacted.  Instead, we present approaches for measuring and monitoring potential sources of participation bias, in the interest of (1) adjusting the sampling process, (2) informing the choice of an estimator, or (3) informing the development of new approaches to inference.  Ideally, the quantitative survey-based approaches presented here should be paired with qualitative evaluation of decision-making associated with recruitment and participation~\citep{scott_they_2008, broadhead_notes_2008, ouellet_cautionary_2008, mello_assessment_2008, kerr_selective_2011, mccreesh_evaluation_2012}.  

Differential recruitment effectiveness is possible to evaluate using data readily available in all RDS studies, and is directly actionable in terms of estimators.  Recruitment Effectiveness Plots should be made to study the relationship between recruitment and key study variables, both during and after data collection.  Where differences are found, qualitative study or discussion with survey staff may reveal areas for improvement in the sampling process.  Further, these findings may influence the choice of estimators.  \citet{tomas_effect_2011} show that the estimator in \citet{salganik_sampling_2004} is more robust to differential recruitment effectiveness than other estimators.  The newer estimator of \citet{gile_network_2011} allows researchers to adjust for differential recruitment effectiveness by wave of the sample. 

Recruitment Biases are more difficult to evaluate, in part because they require more specialized data-collection. The particular characteristics of interest in a given study, such as employment status, will be study-specific and require researchers to be very familiar with the population and sampling process.  Any characteristic that may be associated with increased participation should be measured for respondents, potential respondents (i.e., contacts of respondents), and coupon-recipients so that researchers can create Recruitment Bias Plots (Fig.~\ref{fig:refbiasplot}).  The collection of such data may also inspire further development of statistical inference for RDS data.  The relationship between drug user employment and participation is a good example.  If employed alters are indeed more likely to be sampled, and if this tendency can be measured (as in these data), methods may be developed to adjust inference for this tendency.  The estimator in \cite{gile_network_2011} is particularly conducive to this kind of adjustment.  

A thorough evaluation of non-response bias requires a follow-up study of non-responders.  Despite the obvious logistical challenges (see~\citet{mello_assessment_2008, kerr_selective_2011, mccreesh_evaluation_2012}), we recommend such a study whenever researchers have special concerns about non-response.  Absent such a study, computing RDS Non-Response Rates could alert researchers to possible problems with non-response.  A Coupon-Refusal Analysis could then help researchers adjust their studies to remove barriers to participation.   Further, the results of a Coupon-Refusal Analysis could suggest individual characteristics that might be related to non-response and which, therefore, should be measured.  For example, if distance to the survey site seems burdensome, researchers could introduce an additional survey site, or, minimally, collect data on a measure of distance-burden to either adjust estimators or to monitor recruitment bias. 

Participant motivations should also be measured in all studies, as an indication of potential differential valuation of incentives by different sub-populations.  Motivation-Outcome relationships can be studied between any combinations of expressed motivations and relevant respondent characteristics.  Mechanisms to adjust inference for biases introduced by measurable differential incentives to participation, such as those due to interest in HIV test results, are not yet developed.  The precise quantification of these effects, and their impacts on inference are an important area for future research.

\section{Discussion}
\label{sec:discussion}

RDS is designed to enact a near statistical miracle:  beginning with a convenience sample, selecting subsequent samples dependent on previous samples, then treating the final sample as a probability sample with known (or estimable) sampling probabilities.  This is in stark contrast to traditional survey samples, in which all steps of sampling are conducted within well-defined sampling frames according to carefully designed sampling procedures fully controlled by the researcher.

Miracles do not come for free, and where alternative workable strategies are available, RDS is often not advisable.  Unfortunately, alternative approaches are unavailable for many populations of interest.  Therefore, researchers need to be aware of two main costs of RDS: large variance of estimates (see, for example~\citet{goel_assessing_2010, szwarcwald_analysis_2011, wejnert_estimating_2012}) and many assumptions, including those assessed in this paper.

For researchers planning future data collection, we will briefly summarize our current recommendations for the quantitative diagnostics of RDS data.  If possible, we strongly recommend that these analyses be combined with qualitative analysis.  When constructing the questionnaires, both initial and follow-up, we recommend that researchers include all questions analyzed in this paper and two additional questions on the initial questionnaire to measure reciprocation: a question similar to (\ref{ques:recip}) and a question about the relationship between the recruiter and recruit (see e.g.,~\citet{heckathorn_respondent-driven_2002}). We also recommend that researchers adjust questions (\ref{empl1})-(\ref{empl3}) to reflect the most likely sources of recruitment bias in their study population.

During data collection, we recommend that for all traits of interest, researchers should make Convergence Plots, Bottleneck Plots, and All Points Plots (which are described in the Supporting Information (Section~\ref{sec:allpoints})).  Further in order to understand the recruitment processes taking place, we recommend making Recruitment Effectiveness Plots (Fig.~\ref{fig:hivkids}), making Recruitment Bias Plots (Fig.~\ref{fig:refbiasplot}), calculating the reciprocation rate (Table~\ref{tab:pctsrecipgive}), calculating the non-response rates (Table~\ref{tab:refrates}), and conducting a coupon refusal analysis (Table~\ref{tab:whyrefuse}).  Conducting these analyses during data collection could provide valuable insight into the sampling process while it can still be corrected.  If real-time analysis is not possible, we recommend that these analyses be done at the conclusion of the study.

After data collection is complete, we recommend a Motivation-Outcome analysis (Fig.~\ref{fig:oddsratioHIV}), checking for finite population effects in data collection and estimates (Sec.~\ref{sec:with-replacement_sampling}), checking the validity of the time frame used in the degree question (Sec.~\ref{sec:degree_time_window}), calculating test-retest reliability of the degree question (Sec.~\ref{sec:degree_test_retest}), and calculating the unweighted sample mean in addition to estimates using all available degree questions as weights.  

Finally, we emphasize that these diagnostics should continue to be refined and improved as more is learned about RDS sampling and as new estimators are developed.  For now, however, we hope that these suggestions will provide researchers using RDS a better understanding of their sampling processes.   We also hope that it will spur future methodological developments.

{\small
\bibliographystyle{chicago}

\bibliography{rds}
}

\clearpage

\appendix

\setcounter{figure}{0}
\setcounter{section}{0}
\setcounter{table}{0}
\renewcommand{\thefigure}{S\arabic{figure}}
\renewcommand{\thesection}{S\arabic{section}}
\renewcommand{\thetable}{S\arabic{table}}

\vspace{2in}
\begin{center}
{\LARGE Supporting information:\\}
{\Large Diagnostics for Respondent-driven Sampling}
\end{center}


\section{With-replacement Sampling}
\label{sec:with-replacement_samplingA}

\subsection{Multiple Connections to Survey Participants}

In Section~\ref{sec:contacts_participated} we presented results about the proportion of respondents contacts who had already participated in the study.  It may also be of interest to visualize these trends.  Figs.~S\ref{fig:already_bad} and S\ref{fig:already_good} show the reported proportions that already participated for each respondent, by seed, over time.  In Fig.~S\ref{fig:already_bad}, we can see that within seed, particularly seed 1, periods of low proportion already sampled are often followed by periods of higher proportion already sampled.  This may be indicative of the exhaustion of local subgroups.  Fig.~S\ref{fig:already_good} shows less evidence of a positive trend in proportion already sampled over time.  Finally, Fig.~\ref{fig:already_all}  shows the fitted linear trends for all 12 sites.

\begin{figure}[h]
  \centering
   \subfigure[]{
    \label{fig:already_bad} 
     \includegraphics[width=0.45\textwidth]{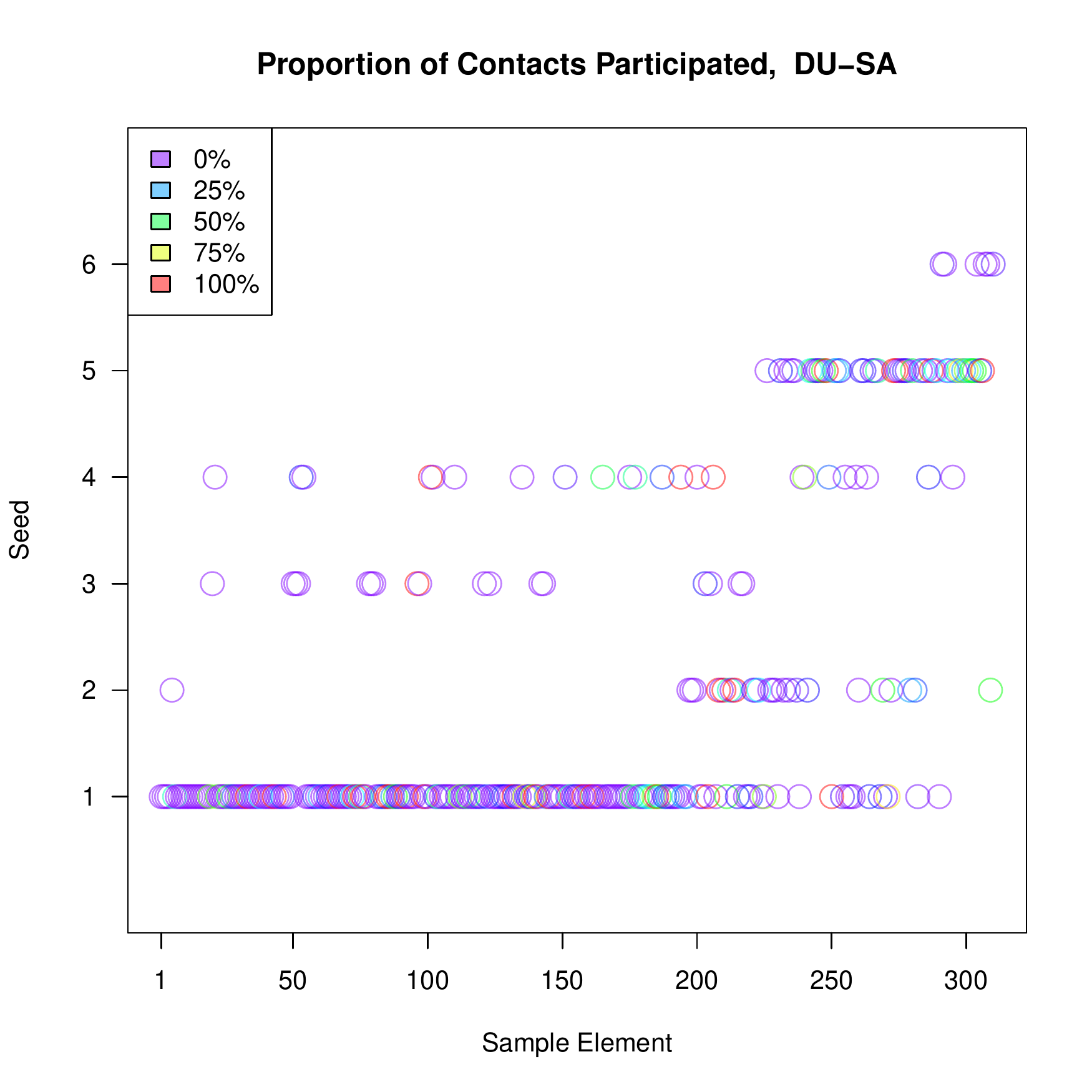}}
  \hspace{0in}
  \subfigure[]{
  \label{fig:already_good} 
   \includegraphics[width=0.45\textwidth]{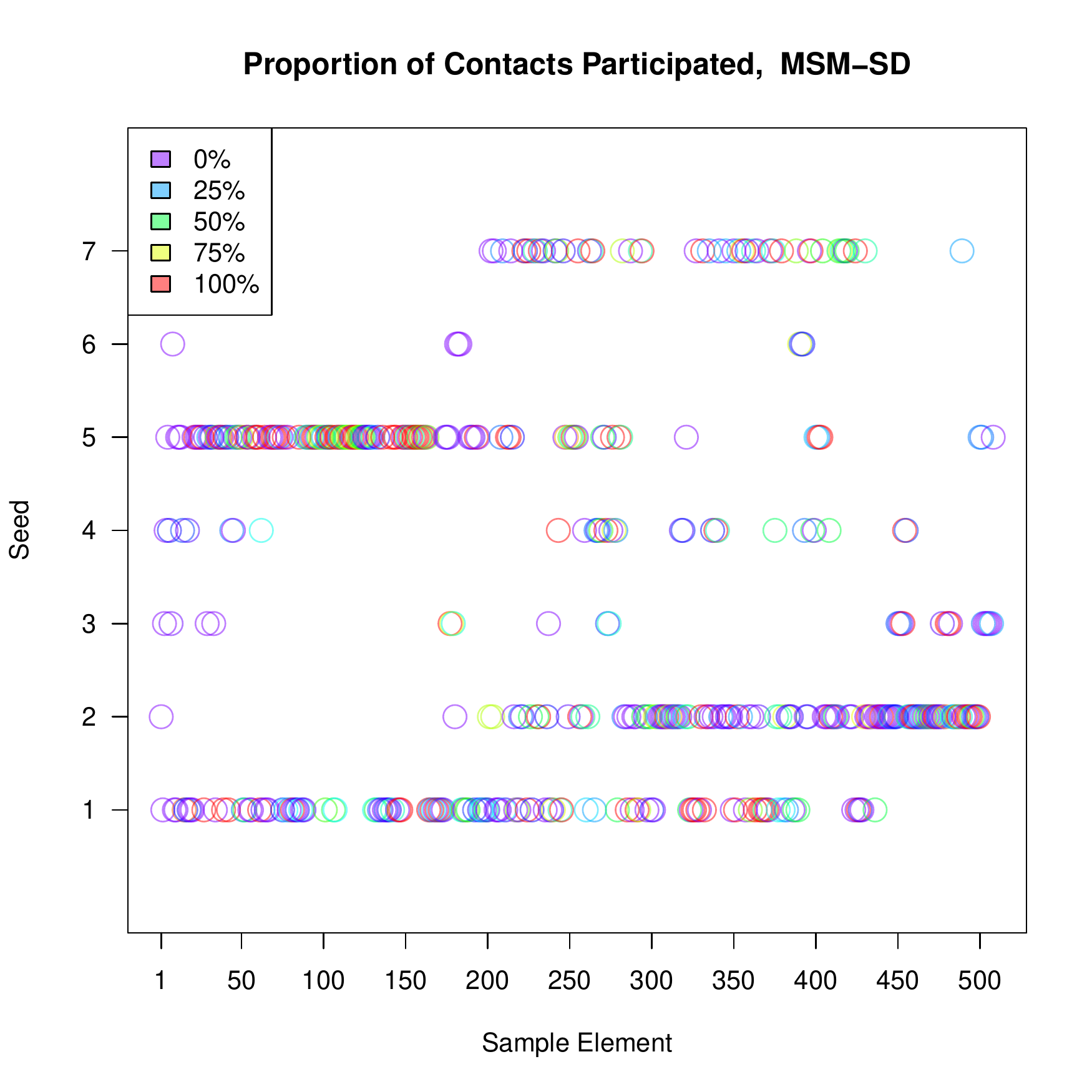}}
  \caption{Proportion of alters already participated, by seed.}
     \label{fig:already} 
\end{figure}

\begin{figure}[h]
  \centering
  \includegraphics[width=0.5\textwidth]{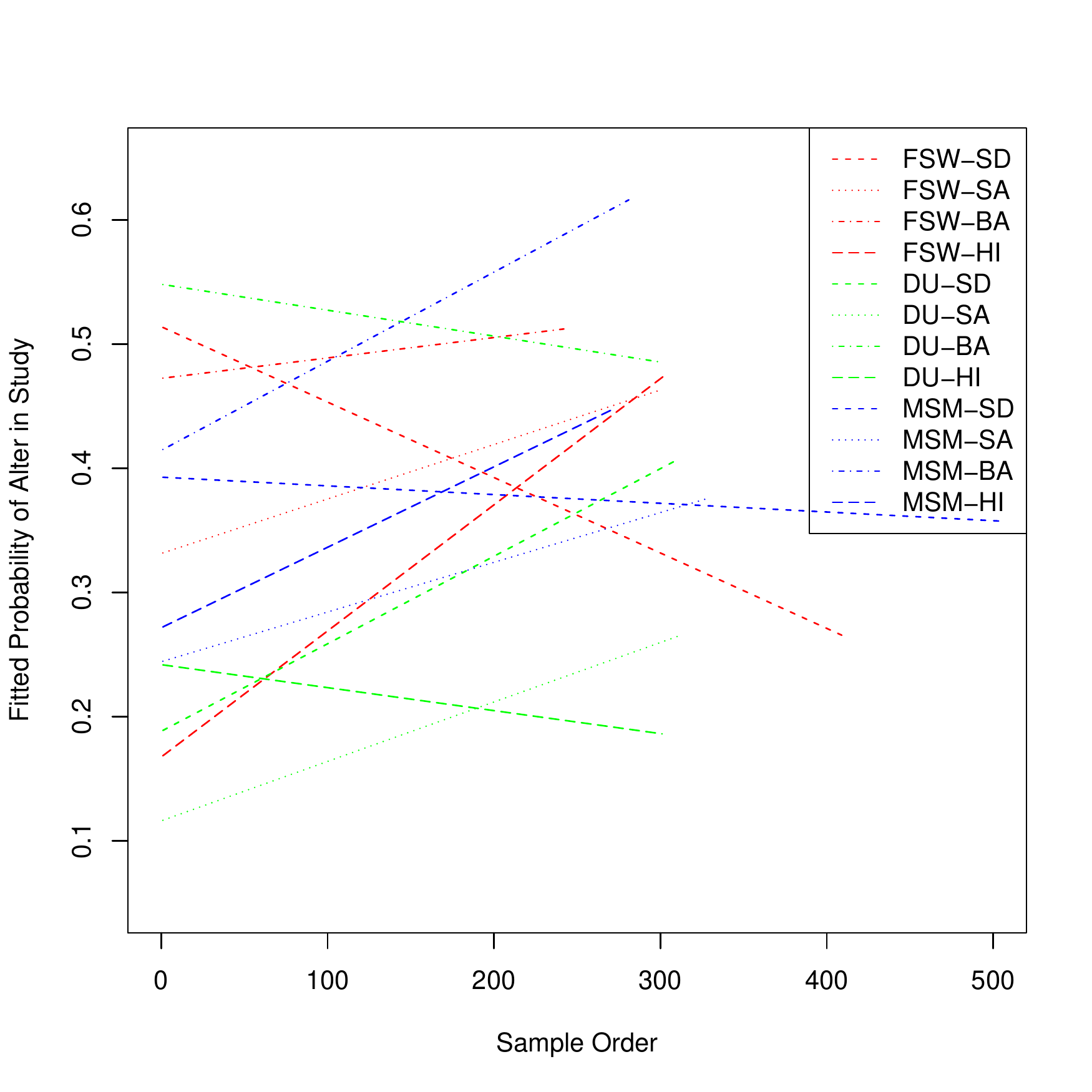}
  \caption{Fitted linear trends for 12 sites for the proportion of respondents contacts who had already participated in the study.}
  \label{fig:already_all} 
\end{figure}

\subsection{Decreasing Degree over Time in Sample}

Under a broad range of assumptions, link-tracing samples result in higher draw-wise sampling probabilities for people with higher degrees~\citep{gile_improved_2011}.  Thus, as the sample begins to deplete the study population, we would expect higher-degree nodes to be sampled earlier, followed by lower-degree nodes, suggesting that a decreasing trend in degree over time could be an indication of finite population effects on sampling.  We compared several options for evaluating the trend of degree over time.\footnote{We used time-order in the study to measure time in these analyses, although results were robust to using survey date.}  These approaches grouped roughly into two families: those sensitive to a small number of high outliers (linear regression, Poisson regression), and those robust to a small number of high outliers (regression on log degree, robust regression approaches (least trimmed squares, M regression, median regression), and rank-based methods (Kendall's Tau and Spearman's Rho); approaches within each family tended to produce similar results.  Because of the dependence in the data structure, we considered only the sign of the coefficient of time in each model.  Surprisingly, we find little evidence of decreasing degree over time with either the non-robust (5 of 12 flagged for the linear model) or robust methods (1-3 of 12 flagged).

Fig.~\ref{fig:degs} illustrates the fitted linear relationship between degree and sample order, as well as the linear relationship fitted to log degree for three sites.  In Fig.~S\ref{fig:degs_bad} (MSM-SA), both approaches found a negative relationship between degree and sample order; in Fig.~S\ref{fig:degs_good} (FSW-BA) both approaches found a positive relationship; and in Fig.~S\ref{fig:degs_weird} (MSM-SD) the two approaches found differing trends, likely driven by the few high responses early in the sample.   

Because we have more faith in the more robust methods, we conclude that this indicator clearly suggests finite population effects for MSM-SA (flagged by all indicators) and perhaps MSM-SD (flagged by most robust indicators).  It is surprising to us that all the other populations, including the three known to have not reached their target sample size (FSW-BA, MSM-BA, MSM-HI), suggested positive or null trends in sample degree over time.  Because we have strong theoretical reason \citep{gile_improved_2011} to expect negative trends in these cases, we hope future research, with other data sets, will help explain this phenomenon.

\begin{figure}
  \centering
   \subfigure[]{
    \label{fig:degs_bad} 
     \includegraphics[width=0.3\textwidth]{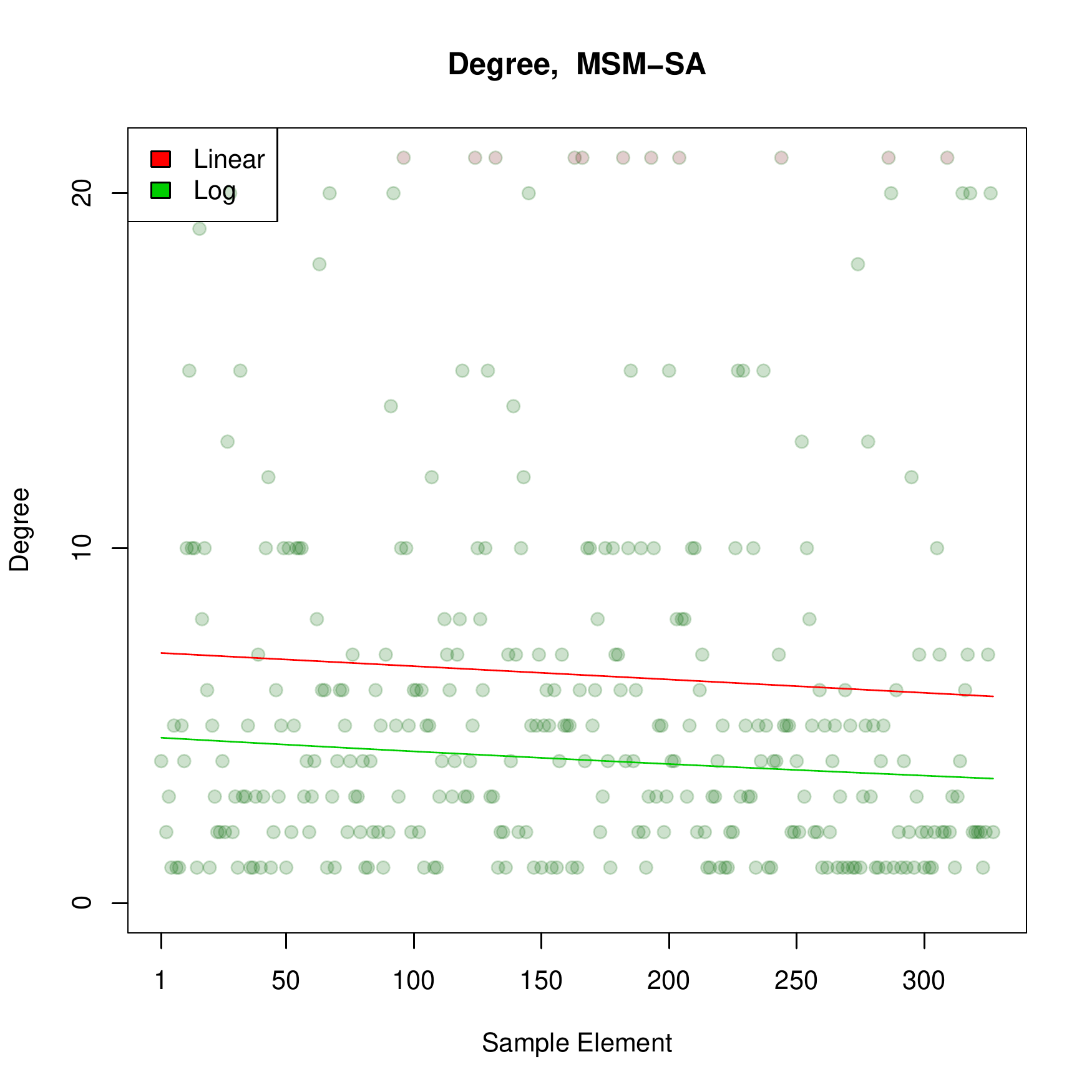}}
  \hspace{0in}
  \subfigure[]{
  \label{fig:degs_good} 
   \includegraphics[width=0.3\textwidth]{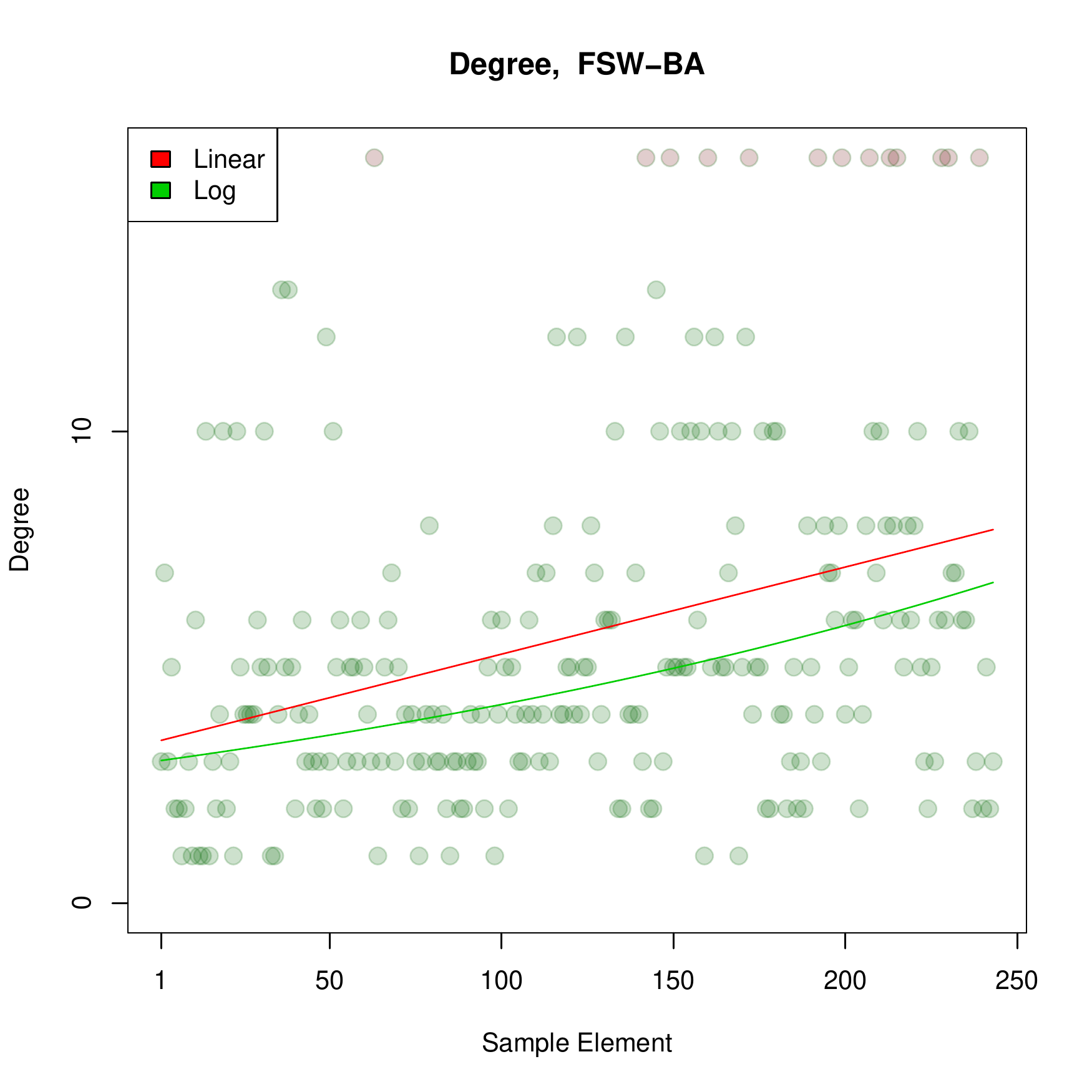}}
  \hspace{0in}
  \subfigure[]{
  \label{fig:degs_weird} 
   \includegraphics[width=0.3\textwidth]{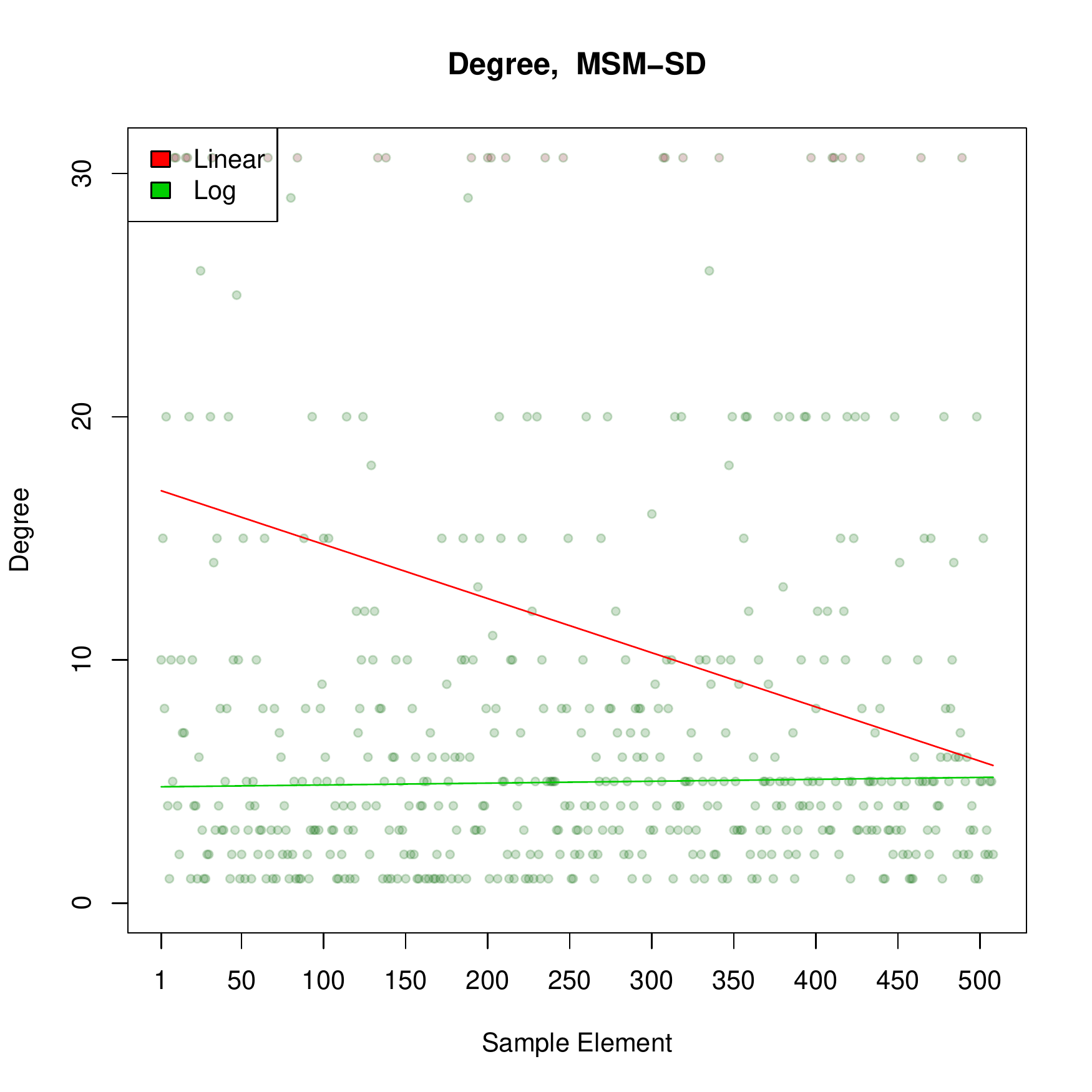}} 
   \caption{Degree of respondents over time, with fitted linear model and linear model for log of degree. For visualization, the highest responses were truncated and represented in red at the tops of the plots.}
     \label{fig:degs} 
\end{figure}

\subsection{Successive Sampling Estimation of Finite Population Bias}

If researchers have an estimate of the size of the study population, they can compare the SS estimator~\citep{gile_improved_2011} to the VH estimator~\citep{volz_probability_2008} in order to assess finite population effects on estimates.  As is typically the case, however, there were no existing estimates of the sizes of our study populations.   Therefore, we use the RDS data itself in order to estimate the sizes of our study populations using the approach introduced in~\citet{handcock_estimating_2012} and implemented in the \texttt{R}~\citep{r_core_team_r:_2012} package {\tt size}~\citep{handcock_size:_2011}.

The method of~\citet{handcock_estimating_2012} requires specifying a prior distribution for the size of the population.  To specify the prior distribution for populations of MSM, we drew on a meta-analysis of~\citet{caceres_estimating_2006}, which provides broad bounds on the proportion of men who have had sex with another man in the past year.  The estimate for the Dominican Republic (and all of Latin America) is 1-8\% of the sexually active adult male population, which we assume to constitute 15-64 year olds.  Combining this information with information on the number of males between 15-64 in each city from the Dominican Republic's National Statistical Office~\citep{oficina_nacional_de_estadistica_poblacion_2009}, we created a conservative upper and lower bound for the size of the MSM population in each city.  These bounds are then used to define the lower and upper quartiles of a prior distribution.  For DU and FSW, no comparable meta-analyses existed so we used broad ranges, consisting of 1-10\% of the 15-64 year old total population (DU), or population of women (FSW).  Again, we used these ranges, combined with information from the Dominican Republic's National Statistical Office~\citep{oficina_nacional_de_estadistica_poblacion_2009} to create prior distributions.  When setting the priors in this manner, the method of~\citet{handcock_estimating_2012} results in posterior mean MSM population size estimates within the original range for SD, SA, and HI, and just above the higher end of the range in Barahona.  For DU and FSW, this procedure produced 6 estimates consistent with the ranges specified in the prior, one (FSW in BA) higher than the 10\% number, and two (DU in SD and SA) lower than 1\%. 

When using the SS estimator, therefore, we used three plausible low population sizes:
\bi
\item The posterior mean (best point estimate from the population size estimation)
\item The lower bound of the posterior highest probability density region (lowest plausible estimate from the population size estimation)
\item For MSM populations, 1\% of the 15-64 year old male population (lower bound of the plausible region from the meta-analysis of~\citet{caceres_estimating_2006}).
\ei
Using each of these estimates of population size, we estimate prevalence of each of the characteristics described in Section~\ref{sec:detecting_convergence}.  A plot of all differences is given in the main text (Fig.~\ref{fig:ssvhdif}).  All items with difference greater than $.01$ are summarized in Table \ref{tab:ssvhdif}.  

\begin{table}[ht]\caption{Prevalence estimates based on Successive Sampling and Volz-Heckathorn estimators for each trait with maximum absolute difference greater than $.01$.}\label{tab:ssvhdif}
\begin{center}
\begin{tabular}{lllccccc}
  \hline
 &  & Trait & VH & Max HPD & Post. Mean & Min HPD & 1\% \\ 
  \hline
 FSW & HI & Last Client Brothel & 0.306 & 0.316 & 0.321 & 0.334 & - \\ 
 FSW & HI & Been In Program & 0.345 & 0.349 & 0.351 & 0.356 & - \\ 
\hline
 DU & SD & Main Drug Crack & 0.263 & 0.267 & 0.270 & 0.275 & - \\ 
 DU & SD & Use Drugs Every Day & 0.378 & 0.385 & 0.388 & 0.397 & - \\ 
\hline
 DU & SA & Use Drugs Every Day & 0.360 & 0.364 & 0.367 & 0.374 & - \\ 
 DU & SA & Been Imprisoned & 0.370 & 0.374 & 0.376 & 0.382 & - \\ 
\hline
 DU & BA & Use Drugs Every Day & 0.391 & 0.397 & 0.400 & 0.410 & - \\ 
\hline
 DU & HI & Main Drug Cocaine & 0.422 & 0.418 & 0.416 & 0.410 & - \\ 
 DU & HI & Use Drugs Every Day & 0.406 & 0.410 & 0.413 & 0.419 & - \\ 
 DU & HI & Been Imprisoned & 0.259 & 0.263 & 0.265 & 0.271 & - \\ 
\hline
 MSM & SA & Had HIV Test & 0.434 & 0.434 & 0.436 & 0.448 & 0.439 \\ 
 MSM & SA & Bisexual & 0.612 & 0.612 & 0.609 & 0.593 & 0.605 \\ 
\hline
 MSM & BA & HIV+ & 0.087 & 0.087 & 0.086 & 0.085 & 0.074 \\ 
 MSM & BA & Had HIV Test & 0.331 & 0.330 & 0.328 & 0.321 & 0.277 \\ 
 MSM & BA & Working & 0.711 & 0.712 & 0.712 & 0.716 & 0.735 \\ 
 MSM & BA & Use Drugs & 0.607 & 0.608 & 0.609 & 0.613 & 0.633 \\ 
 MSM & BA & Sex With Woman & 0.858 & 0.859 & 0.859 & 0.863 & 0.884 \\ 
\hline
 MSM & HI & Had HIV Test & 0.503 & 0.503 & 0.501 & 0.493 & 0.491 \\ 
 MSM & HI & Used Condom & 0.790 & 0.790 & 0.787 & 0.776 & 0.773 \\ 
 MSM & HI & Sex With Woman & 0.834 & 0.834 & 0.833 & 0.825 & 0.823 \\ 
   \hline
\end{tabular}
\end{center}
\end{table}

\section{All Points Plot}
\label{sec:allpoints}

One challenge in interpreting the the Convergence Plots (Section~\ref{sec:detecting_convergence}) and the Bottleneck Plots (Section~\ref{sec:bottlenecks}) is that each obscures some information: the Convergence Plots do not show how the data differ across trees and the Bottleneck Plots do not show how the data vary over time.  For that reason, we suggest an additional plot which we call the \emph{All Points Plot}.  This plot shows all respondents' trait values by seed and sample order.  To demonstrate how these plots can work together, Fig.~\ref{fig:allthree} plots the estimated proportion of MSM in Higuey that self-identify as heterosexual, a key ``bridge group'' because they can spread infection between the high-risk MSM group and the larger heterosexual population.  The Convergence Plot (Fig.~\ref{fig:MSM_HI_hetero_dynamics}) shows that there were no self-identified heterosexuals in the first 100 observations (header), but over time the sample started to reach people who identified as heterosexual.  Since the estimate has not clearly stabilized, we should be worried that the final estimate of $\hat{p}=0.12$ might be unduly influenced by the choice of seeds. Further, the Bottleneck Plot shows that the self-identified heterosexuals were reached only with certain trees suggesting a possible problem with bottlenecks (Fig.~\ref{fig:MSM_HI_hetero_parallchains}).  Finally, the All Points Plot (Fig.~\ref{fig:MSM_HI_hetero_allpoints}) shows that self-identified heterosexuals were unusual in that they both arrived in the sample late and arrived only in a small number of trees, a fact that is difficult to infer from the previous two plots.

\begin{figure}
  \centering
  \subfigure[Convergence Plot]{
  \label{fig:MSM_HI_hetero_dynamics} 
     \includegraphics[width=0.3\textwidth]{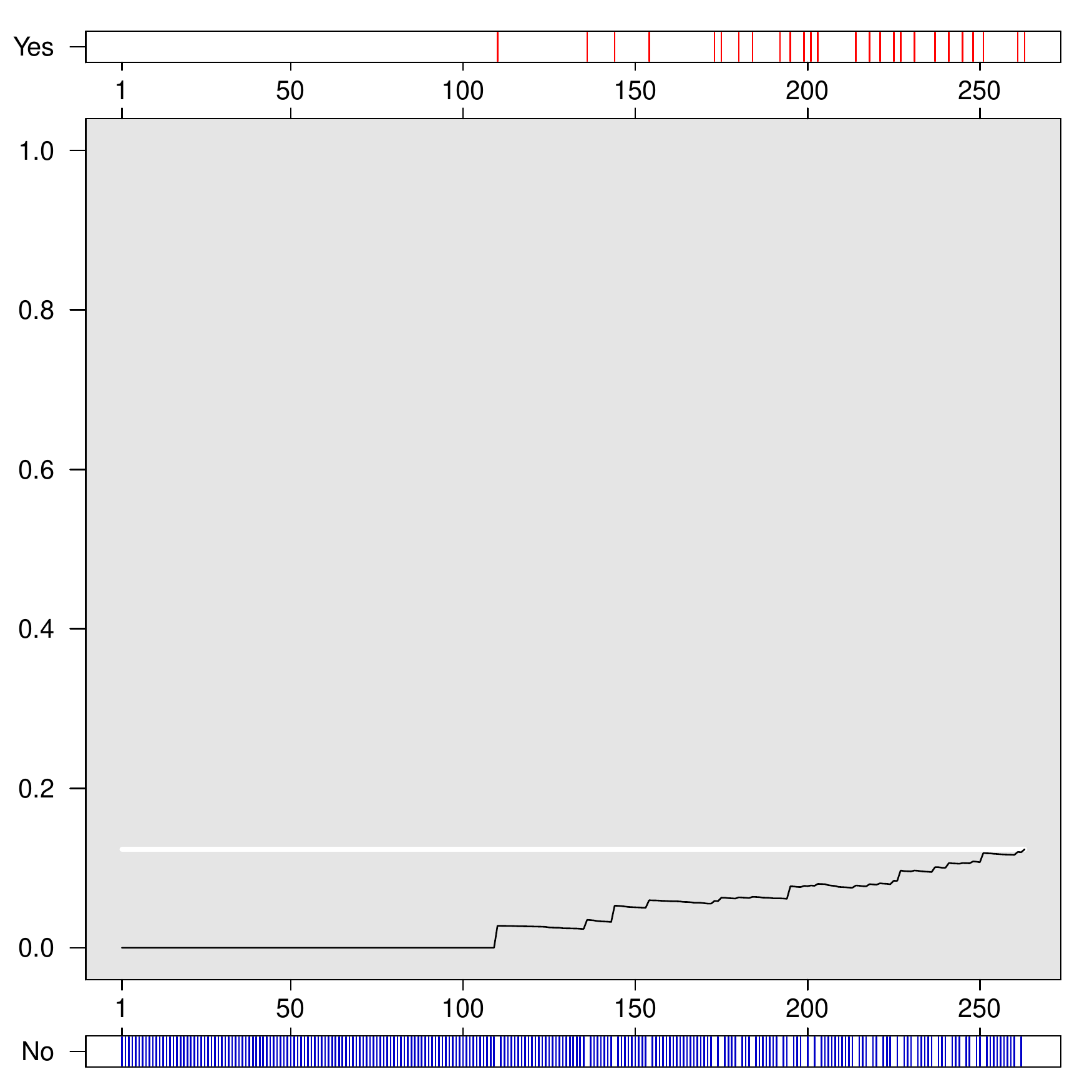}}
  \hspace{0in}
  \subfigure[Bottleneck Plot]{
    \label{fig:MSM_HI_hetero_parallchains} 
     \includegraphics[width=0.3\textwidth]{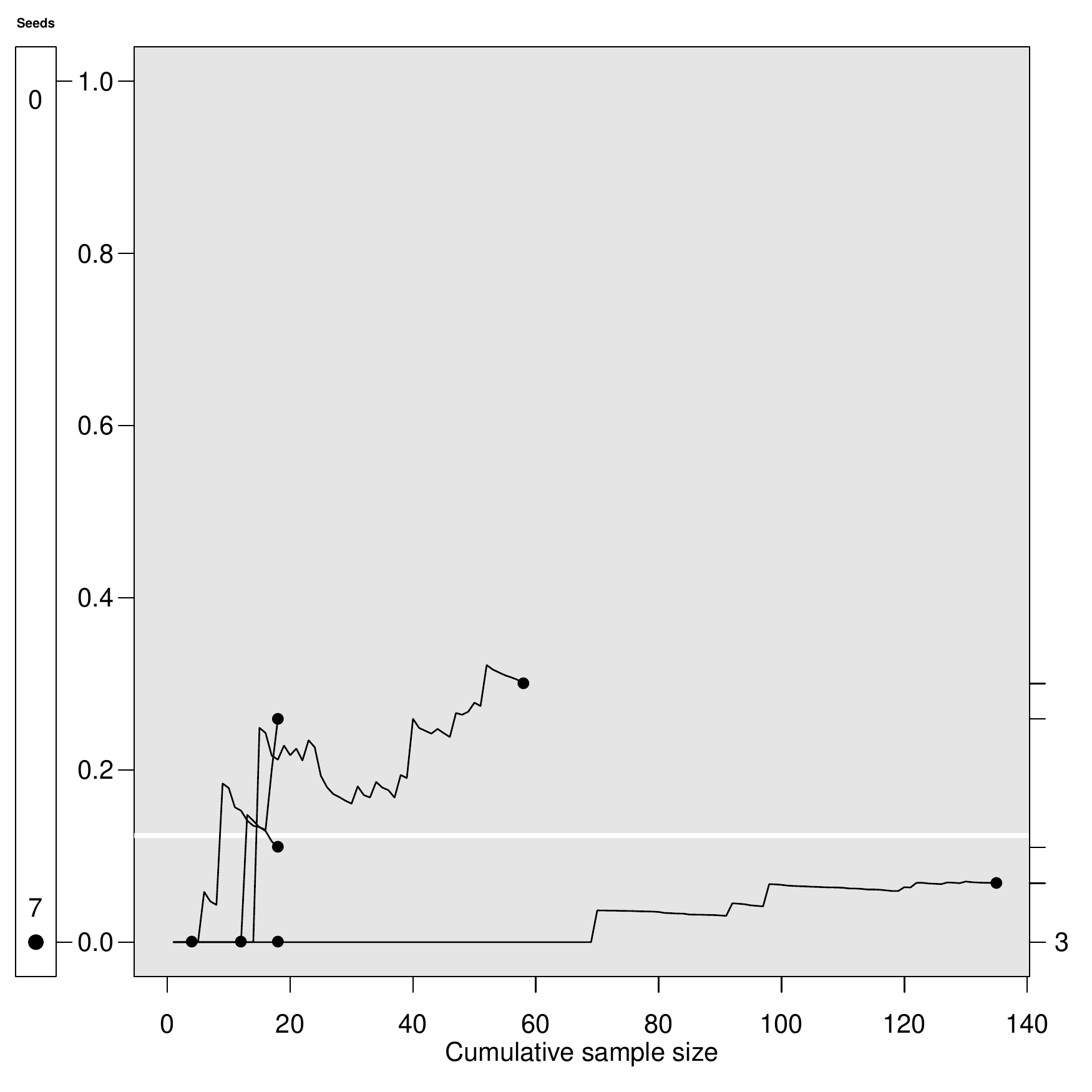}}
  \hspace{0in}
  \subfigure[All Points Plot]{
    \label{fig:MSM_HI_hetero_allpoints} 
     \includegraphics[width=0.3\textwidth]{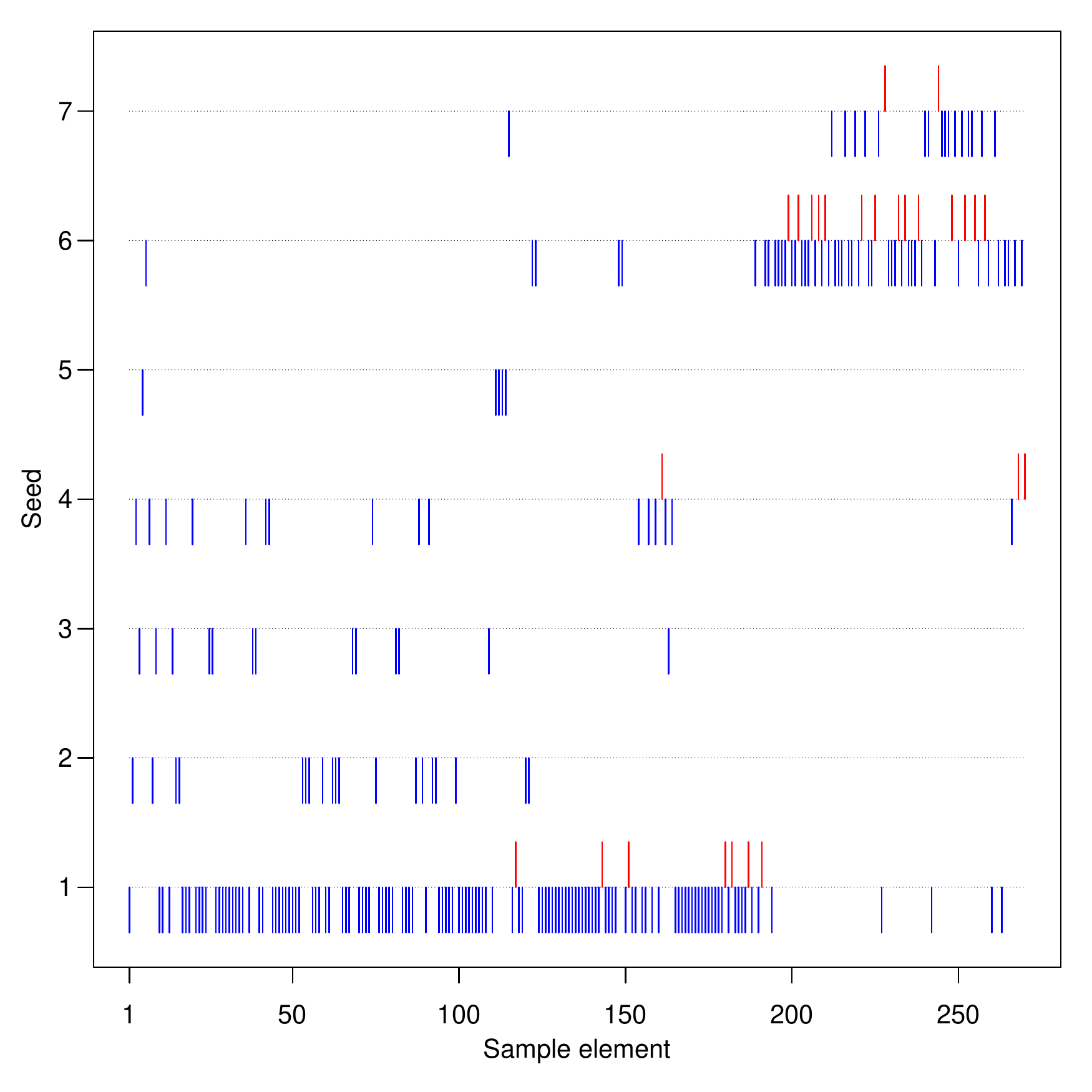}}
  \caption{Three diagnostic plots for estimates of MSM in Higuey that self-identify as heterosexual.  The Convergence Plot (a) shows that data collected late in the sample differs from data collected early in the sample.  The Bottleneck Plot (b) shows that the chains explored different subgroups suggesting a problem with bottlenecks.  The All Points Plots (c) shows that the self-identified heterosexuals (represented by up-ticks in the plot) were unusual in that they both arrived in the sample late and arrived from a small number of chains, a fact that is difficult to infer from the previous two plots.}
     \label{fig:allthree} 
\end{figure}

\section{Reciprocation}
\label{sec:reciprocationA}

In this section, we introduce a measure of reciprocation of all network ties, rather than just the ties associated with coupon-passing. Although the recall task associated with reporting these data is more complicated than asking only about the recruiter, it is the reciprocation of all ties, not just those involved in coupon-passing that is necessary for estimating sampling probabilities.  This is because the estimation of sampling probabilities in RDS relies on self-reported network connections.  If all relationships are symmetric or reciprocated, then the number of network connections is related to a respondent's sampling probability.  Otherwise, it is the respondent's in-degree, or number of incoming relations, that is related to sampling probability.  Unfortunately, reporting numbers of incoming relations is very difficult.   Current estimators for RDS data therefore require reciprocation for two reasons.  First, out-ties are easier to self-report and therefore more often recorded, while in-ties are more directly related to sampling probabilities.  If all ties are reciprocated, then self-reported out-degree is the same as in-degree.  Furthermore, if all ties are reciprocated, the sampling process more closely approximates a random walk on an undirected graph, a common assumption of estimators used.

During the initial visit, participants were asked the following questions about their alters (MSM versions;  other groups were analogous):
\bq
\item How many of them (repeat the number in \ref{sizeq3}) know you well enough that they could give you a coupon within a week if they had been in this study? \label{ques:ngetwk}
\item If we were to give you as many coupons as you wanted, how many of them (repeat the number in \ref{sizeq3}) could you give a coupon to? \label{ques:ngive}
\item If we were to give you as many coupons as you wanted, how many of these MSM (repeat the number in \ref{ques:ngive}) do you think you could give a coupon to by this time next week?\label{ques:ngivewk}
\eq

Among all 3,860 respondents who responded to all of these questions, 29.7\% gave the same answer for both questions \ref{ques:ngetwk} and \ref{ques:ngivewk}, 46.7\% reported they could give more coupons than they might receive, and 23.6\% reported the opposite.  The median difference between responses to these questions was 0 and the mean difference of 1.5 more coupons that could be given out than received.  Larger differences are positively associated with larger maximum response to either question. 
 For this reason, we also consider normalized difference values, computed as follows:
\[
     \frac{\vert \ref{ques:ngetwk}-\ref{ques:ngivewk} \vert}{max(\ref{ques:ngetwk},\ref{ques:ngivewk})}.
\]
Using these normalized values, the median difference is still 0, with mean 0.40 and third quartile 0.67.  This approach is conceptually closer to the full requirement of the reciprocity assumption, but it is also subject to larger concerns of reporting accuracy.  Therefore, we prefer the approaches described in Section \ref{sec:reciprocation}.

\section{Measurement of Degree}

\subsection{Time dynamics}

We conducted three analyses to check whether the one week time frame in question~\ref{sizeq4} was reasonable (see Sec.~\ref{sec:degree_time_window}).  First, for each respondent, we calculated the proportion of his or her alters (based on question \ref{sizeq3}) that could be reached in a specific time frame (based on questions \ref{ncupday} and \ref{ncupweek}).  Fig.~\ref{fig:propgiveboth} depicts the average proportion of alters reachable in each period, by site, with logically inconsistent results excluded.\footnote{Responses were deemed logically inconsistent and therefore were excluded if a respondent reported being able to reach more contacts than (s)he knew (\ref{sizeq3}).  Three sites had high numbers of logically inconsistent responses:  FSW-SD (56, 21) (7 days, 1 day), DU-SA (65, 39), MSM-SD (63, 40).  A total of 42 responses were inconsistent across the remaining 7 sites.}   With the exception of the DU in Santiago, almost all alters were reachable within seven days.  The average rate across sites was 92\% reachable within one week.  Within one day, the across-site-average percent reachable was 62\%.

\begin{figure}[h]
\begin{center}
    \includegraphics[width=3.5in]{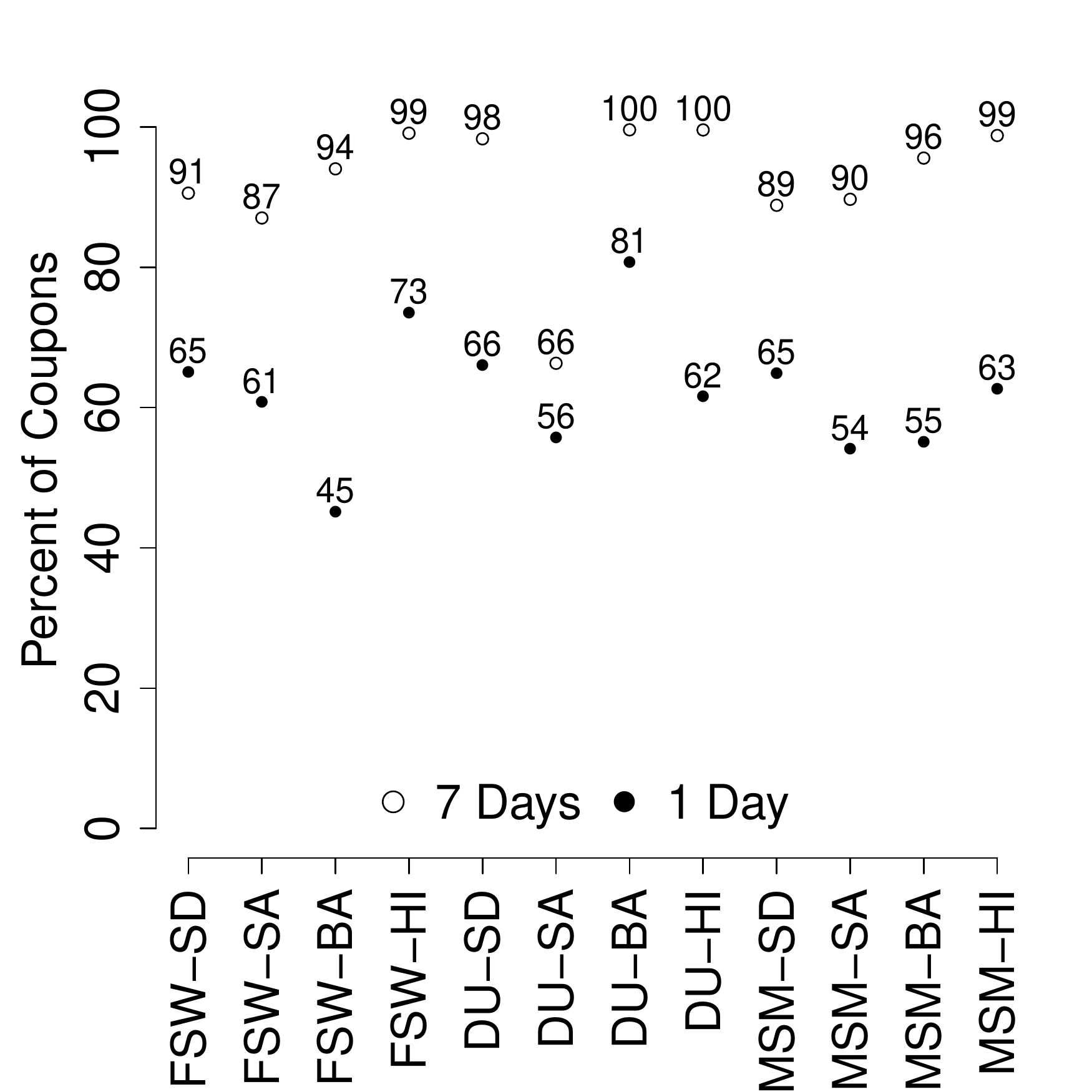}
\end{center}
\caption{Proportion of reported contacts respondent could get a coupon to in 1 or 7 days.} \label{fig:propgiveboth}
\end{figure}

Second, we considered the self-reported number of days each respondent took to distribute his or her coupons (asked at follow-up).  Fig.~\ref{fig:daysgiven} illustrates that across sites, over half (64\%) of coupons were distributed in one day and almost all (95\%) within seven days. 

\begin{figure}[h]
\begin{center}
\subfigure[All days]
{
    \label{daysgivena}
    \includegraphics[width=7cm]{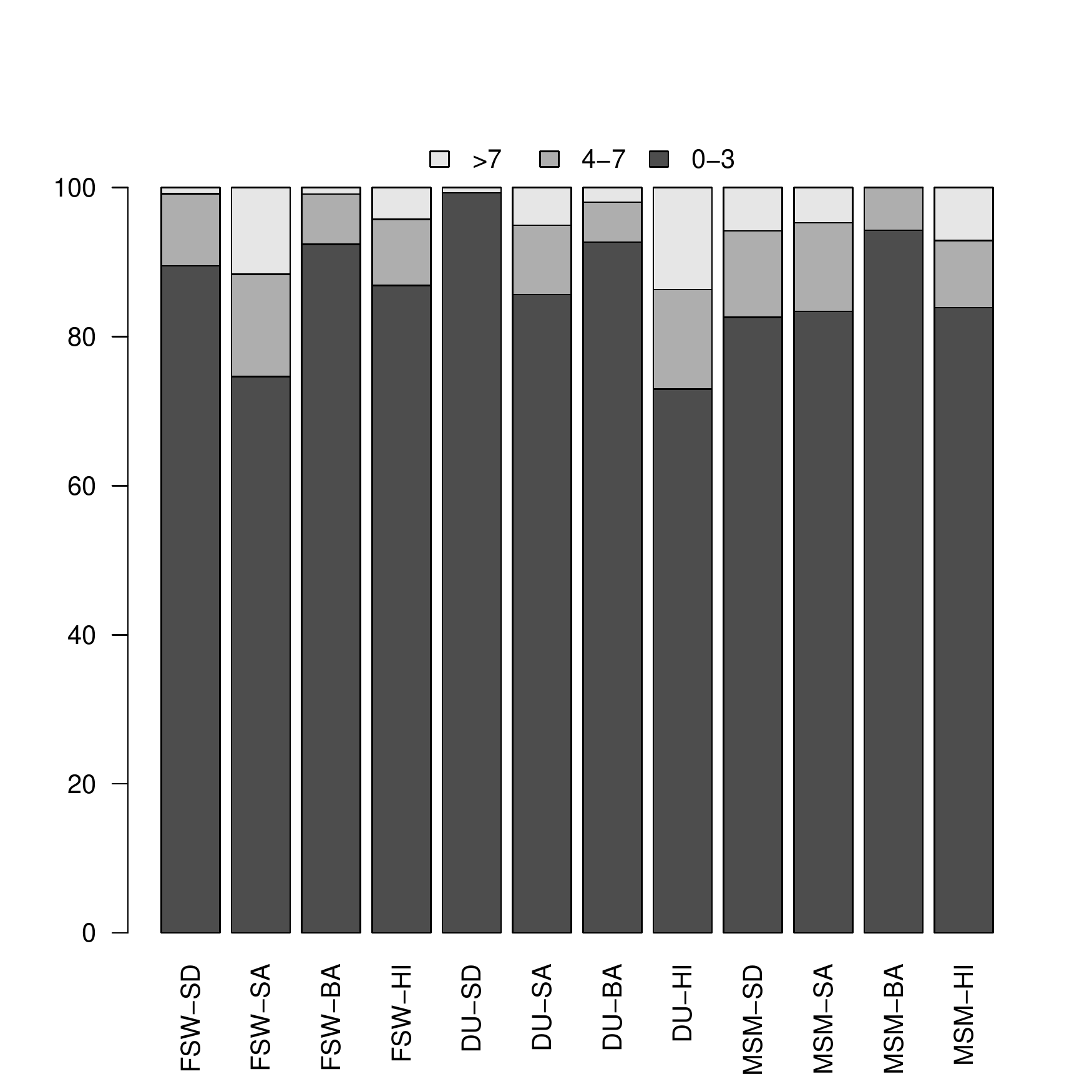}
} \hspace{-.3cm}
\subfigure[Days 0 - 3]
{
    \label{daysgivenb}
    \includegraphics[width=7cm]{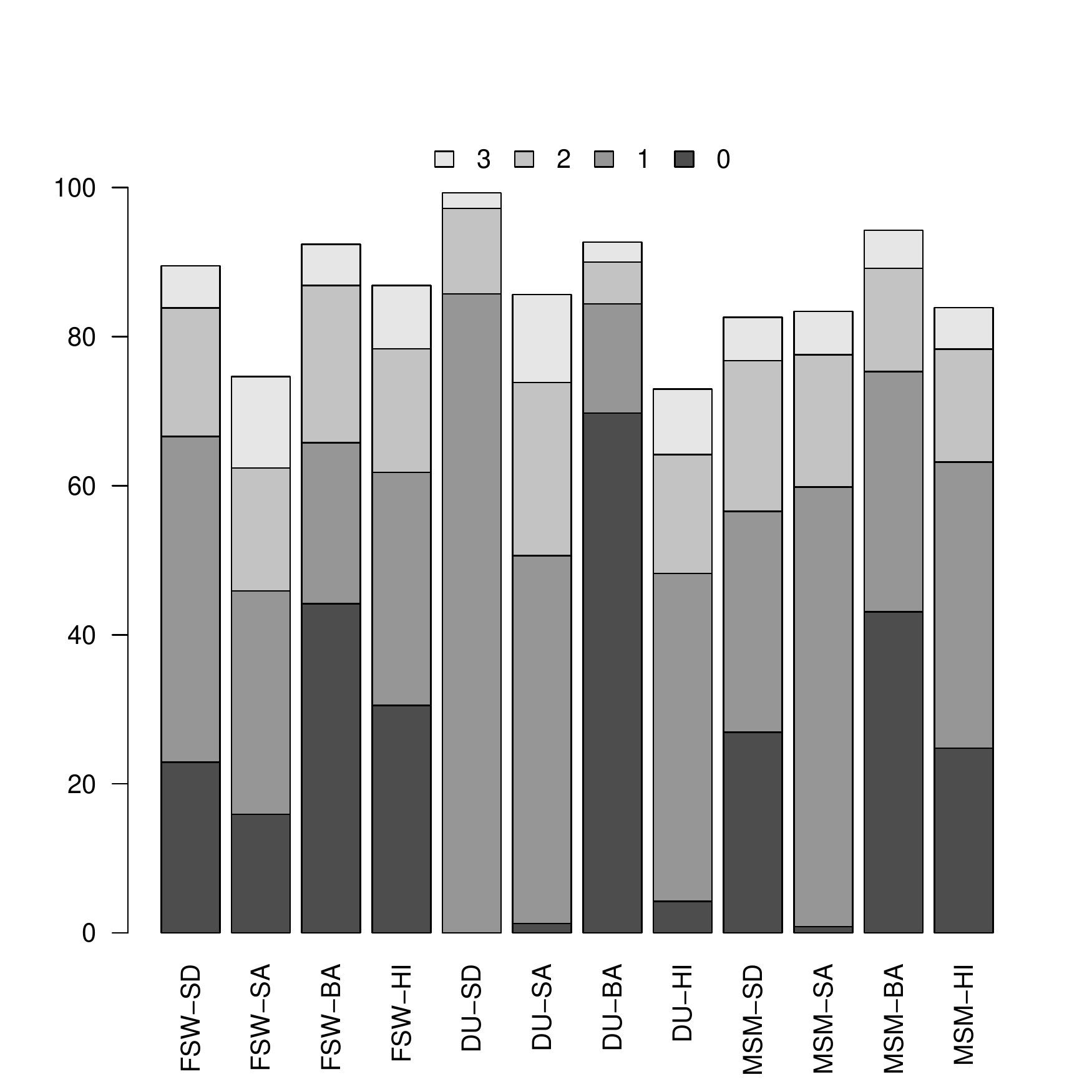}
}\end{center}  \caption{Percent of coupons distributed by number of days, by site.  Most coupons were distributed within 3 days, and nearly all within 7 (a).  Among DU in Barahona most coupons were distributed in one day (b).
}\label{fig:daysgiven}
\end{figure}

Finally, we examined the difference between the interview dates for each recruiter-recruit pair, a measure of time dynamics that does not rely on respondent's reports.\footnote{This measure may also be influenced by the capacity for survey sites to process interviews during high demand.}  Fig.~\ref{fig:bargapdata} shows that in each site, a substantial majority (79\% overall) 
of interviews occur within a week of the recruiter's interview.
\begin{figure}[h]
\begin{center}
    \includegraphics[width=3in]{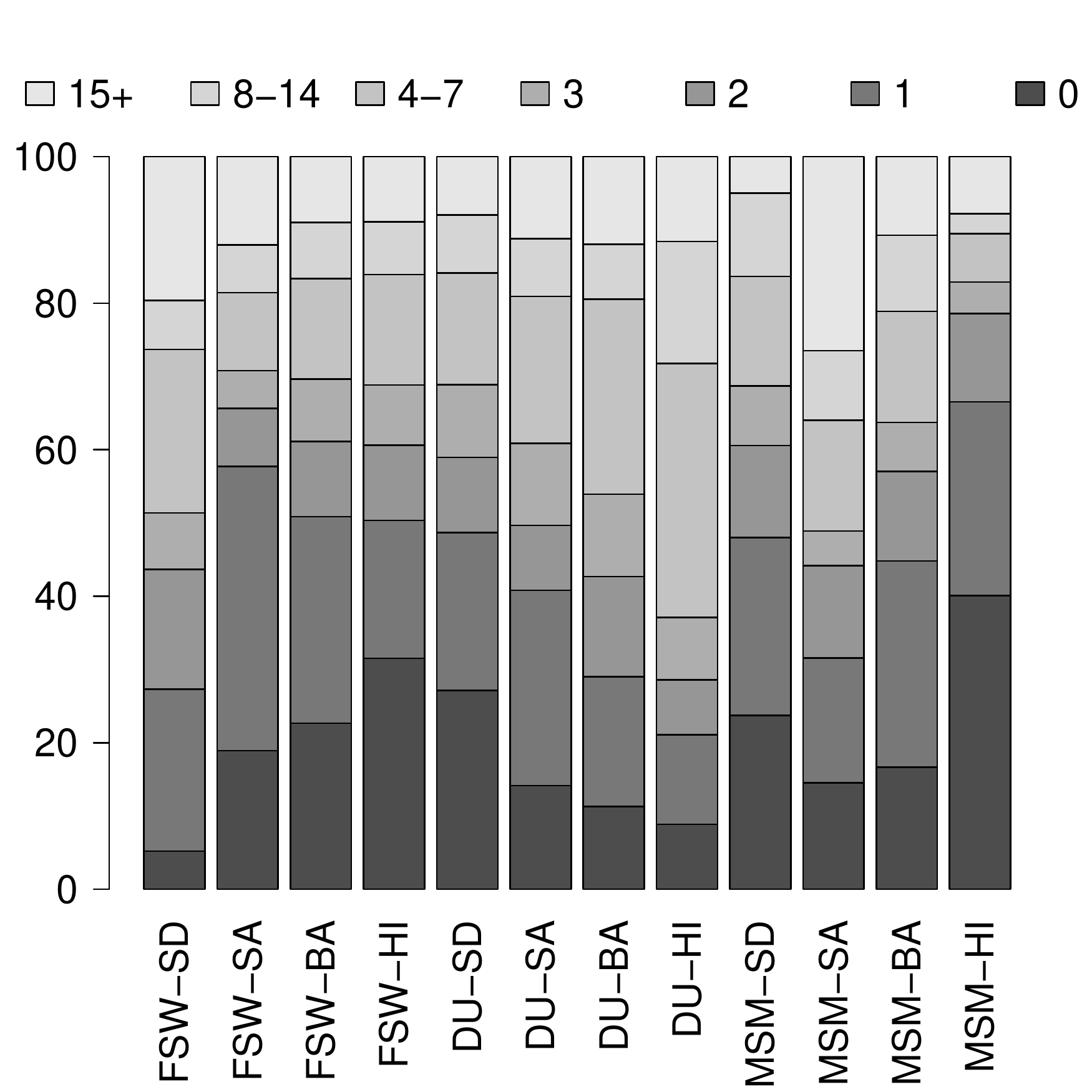}
\end{center}
\caption{Distribution of difference between recruiter's interview date and recruit's interview date, by site.}
\label{fig:bargapdata}
\end{figure}

Overall, these three results suggest that restricting social network recall to people a respondent has seen within the last week appears reasonable in this study.  Nearly all coupons were distributed within a week, and aside from the DU in Santiago, most respondents reported being able to reach nearly all social contacts within a week.  Because most coupons were distributed within a shorter period of a few days, it might even make sense to further restrict the recall period to two or three days.  Note that the validity of this measure, however, relies on the assumption that coupons were distributed to people incidentally encountered, rather than sought out.  Further study is necessary to determine whether respondents seek out their recruits, or select them from among incidentally encountered alters.

\subsection{Test-retest reliability}

In order to assess the test-retest reliability of the degree questions, questions \ref{sizeq1}-\ref{sizeq4} were included in both the initial and follow-up interviews.  The median difference in degree (\ref{sizeq4}) at interview and follow-up is 0 (25$^{th}$ percentile = -3, 75$^{th}$ percentile = 3).  Further, Fig.~\ref{fig:12sites_delta_degree_box} shows that results were similar across survey site and study population.

\begin{figure}
  \centering
  \includegraphics[width=0.5\textwidth]{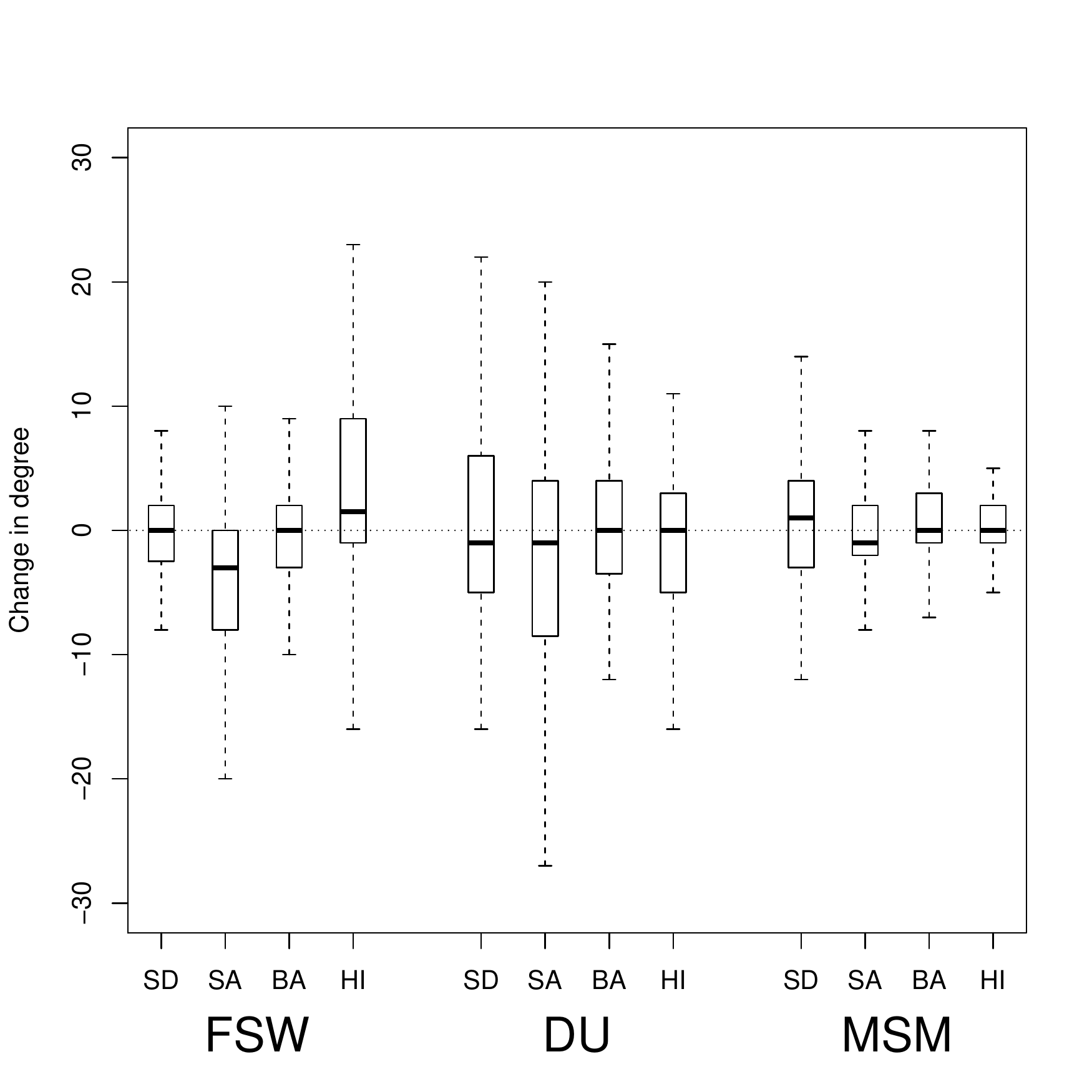}
   \caption{Boxplots of the difference between degree, measured by Question (\ref{sizeq4}), by group and city.  There is no general pattern of increase or decrease.  In order to show the median, 25th and 75th percentiles more clearly, this plot does not include points outside of the whiskers.}
   \label{fig:12sites_delta_degree_box} 
\end{figure}

\begin{figure}
  \centering
   \subfigure[]{
     \label{fig:12sites_spearman_P204} 
     \includegraphics[width=0.45\textwidth]{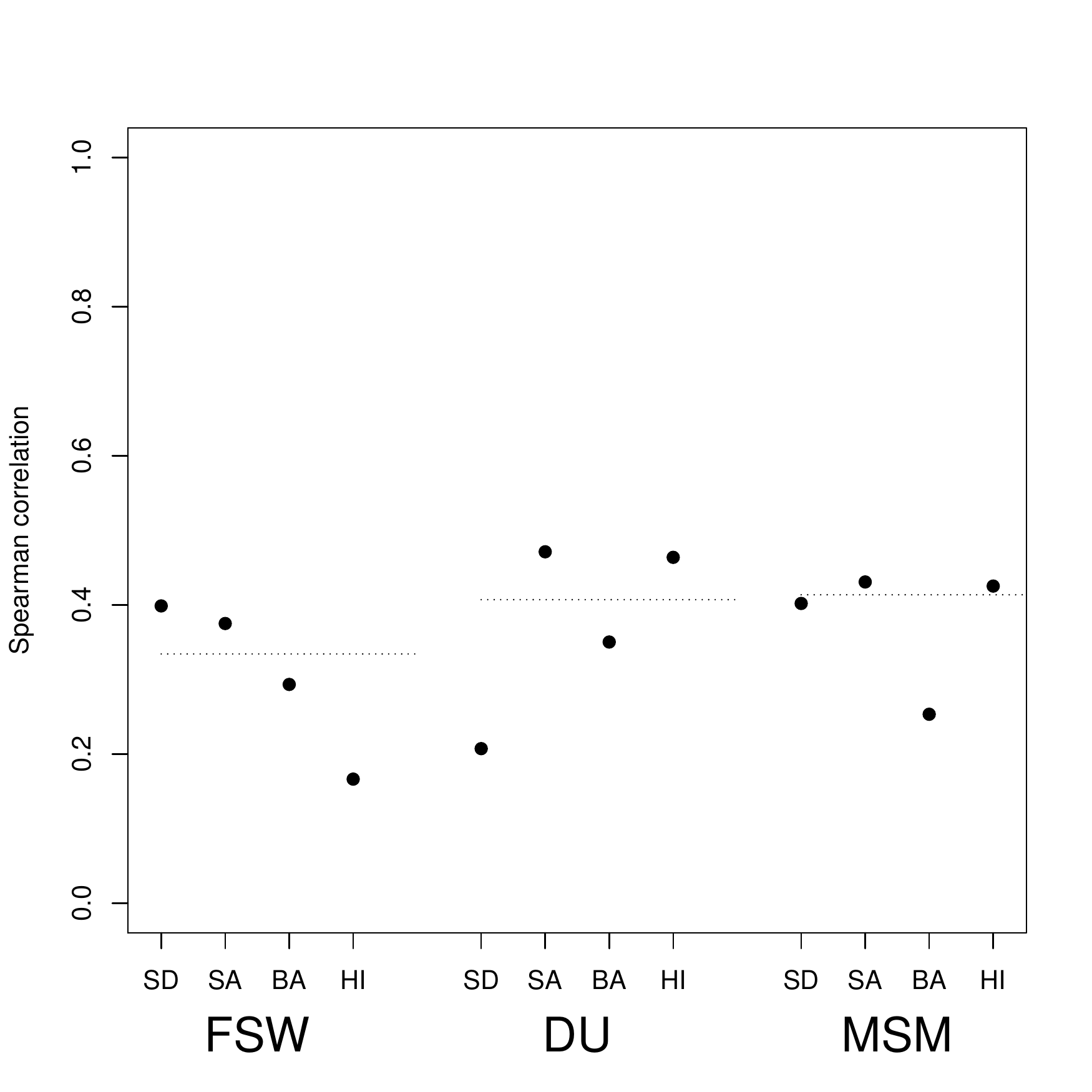}}
       \hspace{0in}
   \subfigure[]{
     \label{fig:12sites_spearman_correlation} 
     \includegraphics[width=0.45\textwidth]{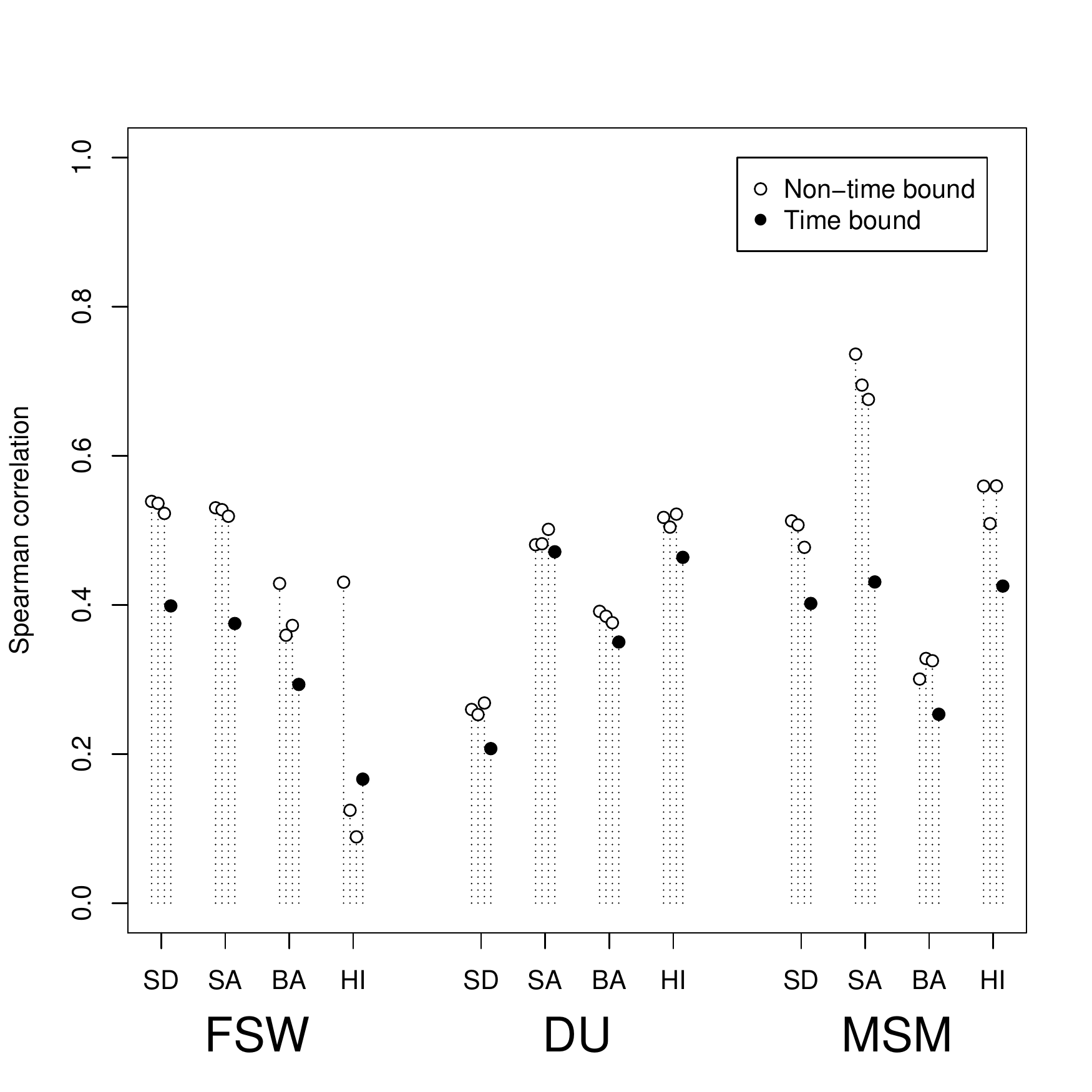}}     
     \caption{(a) Spearman rank correlation between test and retest measures for the main degree question (\ref{sizeq4}) with the median value for each group marked by the horizontal dotted line. (b) The measures that are not time-bounded (\ref{sizeq1}, \ref{sizeq2}, \ref{sizeq3}) have higher correlation.}
     \label{fig:correlation} 
\end{figure}

Fig.~S\ref{fig:12sites_spearman_P204} shows the test-retest reliability for the degree question used for estimation (question \ref{sizeq4}).  One potential reason for the low test-retest reliability of question \ref{sizeq4} is that it refers to a seven day time frame.  Therefore, even if respondents are perfectly accurate in their responses there could be test-retest variation because of week-to-week variation.  This issue of time-bounded questions has come up in other test-retest studies (e.g.,~\citet{van_groenou_test-retest_1990}), but is difficult to resolve because it is not reasonable to ask respondents at the follow-up interview about their experiences in the one week proceeding their initial interview as that is often about three weeks in the past.  However, one way to roughly gauge how much extra variability is introduced by this time frame is by examining the first three network size questions which are not time-bounded.  Fig.~S\ref{fig:12sites_spearman_correlation} shows that the test-retest reliability is higher for the non time-bounded questions, but only slightly so.   

Finally, we note that when considering measures of test-retest reliability, it is critical to consider any potential sources of dependence between the measures.  Here interviewers at the follow-up visit did not know respondent's answers from the initial visit.  Further, since the time period between interviews was generally around three weeks, it is extremely unlikely that respondents remembered their original responses to the degree questions.  One possible source of dependence that did exist in this study is that the respondents may have been interviewed by the same interviewer at the initial and follow-up visits, thus possibly increasing test-retest reliability.  

\section{Testing Recruitment Bias on Employment Status}

Most RDS inference relies on the assumption that recruits are selected at random from among the contacts of each recruiter.  Under this assumption, successful recruits should constitute a simple random sample of the personal networks of respondents.  In most cases, reviewing a Recruitment Bias Plot (e.g., Fig.~\ref{fig:refbiasplot}) should be sufficient to inform researchers' intuition about whether recruitment bias is a concern.  In some cases, researchers may want to test whether the observed recruitment patterns are consistent with random recruitment.  Researchers should also note that statistical significance is not the same as estimator bias, so even a perfect test would not be a good judge of whether a recruitment bias is strong enough to be of concern.

With no recruitment bias, the coupons should be passed to a simple random sample of the recruiter's contacts, and the coupons returned should be returned by a simple random sample of those receiving coupons. To test these assumptions non-parametrically, we compare the (unweighted) count of employed at each stage to a null distribution approximated by simulated simple random sampling from the reported composition of each recruiter's eligible alters.  To test for biased coupon passing, for example, we simulate the coupon-recipients of each recruiter by drawing $n_i^c$ samples from among $\ref{sizeq3}_i$ units, including $\ref{empl1}_i$ employed, where $n_i^c$ is the reported number of coupons distributed by $i$, $\ref{sizeq3}_i$ is $i$'s reported number of contacts, and $\ref{empl1}_i$ is $i$'s reported number of employed alters.   Non-parametric null distributions for returning coupons and for overall recruitment were constructed similarly, with test statistics and reference distributions described in Table \ref{tab:recruitmenttestdescription}.

\begin{table}[ht]\caption{\small Test statistics and reference distributions for testing for Recruitment Bias at three levels:  in the passing of coupons, in returning coupons, and overall.}
\begin{center}
\begin{small}
\begin{tabular}{lll}
Test & Test statistic & Reference Distribution \\
\hline
Coupon Passing & Count of employed  & SRS from contact composition\\
& coupon-recipients & of each recruiter\\
\hline
Returning Coupons &  Count of employed recruits & SRS from composition of  \\ 
& recruits & coupon recipients of each recruiter \\
\hline
Overall  &  Count of employed & SRS from contact composition \\
& recruits & of each recruiter \\
\end{tabular}
\end{small}
\label{tab:recruitmenttestdescription}
\end{center}
\end{table}

  Our tests show very small p-values, suggesting the reported recruitment patterns are very unlikely absent recruitment bias (see Table \ref{tab:recruitmenttest}).  However, one reason for these extreme findings could be poor data quality, perhaps due to a desirability bias of employment status reporting.  Many data points are logically inconsistent, with more employed alters receiving coupons than were originally reported known, or with more employed recruits than coupons given to employed people.  The percents of inconsistent reports are also given in Table \ref{tab:recruitmenttest}.  In addition, there is no evidence that those with more reported employed contacts tend to recruit more employed people (the correlation between these proportions is negative in many of the samples).  Therefore, while we feel this test is mathematically appropriate, we suggest caution in its use, or the use of earlier tests relying on self-reported network compositions. 

Finally, we note that other approaches have been used previously to compare reported network composition to actual sample recruits.  \citet{heckathorn_extensions_2002} look at the correlation between implied population proportions across several groups under random sampling and observed cross-group recruitment patterns.  This approach is not ideal because we would like to test whether the compositions are the same, not just correlated.   \citet{wang_respondent-driven_2005} therefore extend this approach by using a t-test to compare the sample proportion of the observed data to the proportion of reported degree.  This approach does compare the quantities of interest, but relies on an unrealistic binomial approximation to the distribution of the estimated proportions. \citet{wejnert_web-based_2008} introduce a chi square test to compare the expected referral matrix under random referral to the observed referral matrix.   This approach also relies on distributional assumptions, in particular an assumption of independence of observations.  It is unclear from the \citet{wang_respondent-driven_2005} and \citet{wejnert_web-based_2008} papers which form of weighting is used to estimate the composite degree characteristics in the latter two approaches.  

\begin{table}[ht]\caption{\small P-values for non-parametric tests of recruitment bias based on employment status on three levels:  Which contacts are given coupons, which coupon recipients return coupons to become recruits, and overall, which contacts become recruits.  P-values suggest the reported recruitment patterns are very unlikely absent recruitment bias.  Proportion inconsistent records the proportion of cases in each setting in which the number of employed persons selected was larger than the number available.  This suggests the apparently large effects may be due to data quality issues.}
\begin{center}
\begin{tabular}{l|cccc||cccc|}
& \multicolumn{4}{c}{P-value} & \multicolumn{4}{c}{Proportion Inconsistent} \\
& SD & SA & BA & HI & SD & SA & BA & HI \\
\hline
Coupon Passing & 0.369 & $< .0001$ &  $< .0001$ &  $< .0001$ & 0.094 & 0.137 & 0.201 & 0.216 \\ 
Returning Coupons &  $< .0001$ &  $< .0001$&  $< .0001$ &  $< .0001$ & 0.519 & 0.286 & 0.352 & 0.243 \\ 
\hline
Overall  &  $< .0001$ &  $< .0001$ &  $< .0001$ &  $< .0001$ & 0.202 & 0.143 & 0.192 & 0.206 \\ 
\end{tabular}
\label{tab:recruitmenttest}
\end{center}
\end{table}

\end{document}